\documentclass[3p,onecolumn,sort&compress]{article}
\usepackage[utf8]{inputenc}
\usepackage{bm}
\usepackage{amsmath,amsfonts,amssymb,amsthm}
\usepackage{graphicx}
\usepackage{hyperref}
\usepackage{cleveref}
\usepackage{enumerate}
\usepackage{caption}
\usepackage{subfig}
\usepackage{algorithm,algorithmicx,algpseudocode}
\usepackage{tikz}
\usepackage{color} 

\newcommand{\angstrom}{\mbox{\normalfont\AA}}

\newcommand{\bs}[1]{\boldsymbol{#1}}

\newtheorem{remark}{Remark}
\newtheorem*{notation}{Notation}
\newtheorem{proposition}{Proposition} 

\graphicspath{{Figs/}}

\makeatletter
\def\ps@pprintTitle{%
  \let\@oddhead\@empty
  \let\@evenhead\@empty
  \let\@oddfoot\@empty
  \let\@evenfoot\@oddfoot
}
\makeatother

\title{A Riemannian Stochastic Representation for Quantifying Model Uncertainties in Molecular Dynamics Simulations}
\author{Hao Zhang and Johann Guilleminot\\
   Department of Civil and Environmental Engineering\\Duke University\\
   Durham, NC 27708, USA\\
   \texttt{johann.guilleminot@duke.edu}}
\date{July 13, 2022}

\begin{document}

\maketitle

\begin{abstract}
A Riemannian stochastic representation of model uncertainties in molecular dynamics is proposed. The approach relies on a reduced-order model, the projection basis of which is randomized on a subset of the Stiefel manifold characterized by a set of linear constraints defining, \textit{e.g.}, Dirichlet boundary conditions in the physical space. We first show that these constraints are, indeed, preserved through Riemannian pushforward and pullback actions to, and from, the tangent space to the manifold at any admissible point. This fundamental property is subsequently exploited to derive a probabilistic model that leverages the multimodel nature of the atomistic setting. The proposed formulation offers several advantages, including a simple and interpretable low-dimensional parameterization, the ability to constraint the Fr\'{e}chet mean on the manifold, and ease of implementation and propagation. The relevance of the proposed modeling framework is finally demonstrated on various applications including multiscale simulations on graphene-based systems.
\end{abstract}

\subsubsection*{Keywords:}
Model uncertainty; molecular dynamics; reduced-order modeling; Stiefel manifold; uncertainty quantification

\section{Introduction}
Molecular dynamic (MD) simulations are widely employed to study microscopic processes and predict macroscopic thermodynamical properties in various science and engineering fields, including biophysics, computational chemistry, and materials science. The accuracy of MD simulations heavily depends on the interatomic potentials that are used to evaluate the force fields governing the interactions within the dynamical system. Such functions are usually designed and calibrated according to experimental data or first-principles calculations, leading to a myriad of models for the same atomic system. 

For example, models for water-biomolecular interactions include 3-site potential models such as TIPS \cite{TIP}, SPC \cite{SPC}, TIP3P \cite{TIP3P4P}, SPC/E \cite{SPCE}, as well as 4-site models such as BF \cite{BF}, TIP4P \cite{TIP3P4P}, OPC \cite{OPC}, to name a few. The coarse-grained Optimized Potential for Liquid Simulation (OPLS) potential model and its variants \cite{opls,pe,pe2,thermal-pe} can be employed for MD simulations of polyethylene, which is one of the simplest polymer systems. In the case of graphene sheets (which are composed of carbon atoms only), many models were proposed in the literature, including AIREBO \cite{airebo}, BOP \cite{bop}, REBO-2 \cite{reboII}, LCBOP\cite{lcbop}, ReaxFF \cite{reaxff}, and Tersoff-type potentials \cite{tersoff}. For many systems of interest, and most critically for newly developed materials, the appropriate choice of interatomic potentials is rarely known in advance, and a detailed analysis must be conducted before MD simulations are performed, according to specific simulation targets and environment \cite{roadmap}.

The selection and calibration of interatomic potentials can induce substantial uncertainties across scales. Several approaches have been pursued in the literature to propagate various types of uncertainties from fine to coarser atomistic scales (see, \textit{e.g.}, \cite{gaussian-uq,interphases,wang2019modeling}), as well as in MD-to-continuum coupling methods \cite{plasticity-UQ, MD-continuum-uq} where intermediate coarse-scale results are estimated through fine-scale MD simulations. There is also a very substantial number of papers that have reported on the impact of parametric uncertainties in a continuum multiscale setting; see, \textit{e.g.}, \cite{micro-UQ,fiber-uq,ghanem}, in a non-exhaustive manner. Restricting the discussion to uncertainty quantification for MD simulations, most of the works have focused on the propagation of parametric uncertainties in potential parameters \cite{bayesian-md,bayesian-md2,uq-ionicflow1, uq-ionicflow2}, sampling-induced uncertainties \cite{sampling-md-uq}, and uncertainties for models built using machine learning techniques \cite{mlpotential-uq}.

In contrast, the consideration of model-form uncertainties has received relatively little attention to date. The issue of adaptive model selection under uncertainty, considering candidates that are obtained through different coarse-graining strategies for example, was addressed in \cite{bayesian-selection, bayesian-selection2} using Bayesian formalism. A framework enabling extrapolation from one model to another, using functional perturbations, was proposed in \cite{functional-uq}. In the recent work \cite{wang2019modeling}, the probabilistic approach developed in \cite{soize2017nonparametric} to capture model uncertainties through the construction of a probability measure for the projection basis of a reduced-order model was applied to identify, and investigate the effect of, model uncertainties in MD simulations. While the methodology was shown to perform satisfactorily in terms of variability representation, the parameterization to pullback from the tangent space to the manifold used in \cite{soize2017nonparametric,wang2019modeling} does not allow for the mean of the fluctuations to be constrained. Furthermore, hyperparameter identification suffers from a curse of dimensionality unless assumptions about the structure of the statistical fluctuations are made---the number of hyperparameters scaling as $\mathcal{O}(n^2)$ for a reduced-order model of dimension $n$ \cite{soize2017nonparametric,Farhat2019}. The aim of this work is to propose a new formulation that circumvents these limitations and fully leverages the multimodel nature of the atomistic setting, using \textit{ad hoc} Riemannian projection and retraction operators. Note that this setting is intrinsically different from the one considered in \cite{soize2017nonparametric}, where only one model (about which statistical fluctuations are to be prescribed) is assumed to be available. The use of Riemannian operators is motivated by the preservation of linear constraints onto the tangent space, which enables the derivation of a particularly simple and easily implementable model, and allows the mean of the fluctuations to be constrained in the Fr\'{e}chet sense.

The remainder of this paper is organized as follows. The stochastic modeling framework is introduced in Section \ref{sec:overall-framework}. We present the reduced-order model, as well as strategies to parameterize the formulation on the tangent space to the Stiefel manifold. We subsequently define the probabilistic model and propose a strategy to constrain the Fr\'{e}chet mean on the manifold. Section \ref{sec:applications} is dedicated to various applications, including a toy example, related to sampling on the unit sphere, and multiscale MD-based predictions on a graphene system. Conclusion is finally presented in Section \ref{sec:conclusion}.

\section{Riemannian Stochastic Representation of Model Uncertainties}
\label{sec:overall-framework}
\subsection{Reduced-Order Modeling for Molecular Dynamics Simulations}
The evolution of the whole system, composed of $N_a$ atoms in $\mathbb{R}^d$, is described by Newton's second law of motion 
\begin{equation}
	[M] \bs{\ddot{q}}(t) = \bs{f}(t; \bs{q}(t))\,,
\end{equation}
where $[M] \in \mathbb{R}^{N \times N}$, $\bs{q}\in \mathbb{R}^{N}$, and $\bs{f} \in \mathbb{R}^{N}$ denotes the mass matrix, the position vector, and the force vector, respectively (with $N = d \times N_a$). Appropriate initial conditions are assumed and left unspecified throughout this section. In addition, we assume that $\bs{q}$ satisfies a set of linear constraints, written as
\begin{equation}\label{eq:def-linear-constraints}
[B]^T\bs{q}(t)=[0_{N_{\mathrm{CD}},n}]\,, \quad \forall t \geq 0\,,
\end{equation}
where $N_{\mathrm{CD}}$ is the number of constrained degrees of freedom in the system and $[B] \in \mathbb{R}^{N \times N_{\mathrm{CD}}}$ satisfies 
\begin{equation}\label{eq:B-is-orthogonal}
[B]^T[B] = [I_{N_{CD}}]\,.
\end{equation}
The above constraint can be used to specify homogeneous Dirichlet boundary condition, for example. 

To fix ideas, assume that the system is defined for a \textit{given} choice of interatomic potential (describing all types of interactions). A reduced-order model (ROM) can then be constructed by using a proper orthogonal decomposition (POD). To that end, we consider the state variable
\begin{equation}
    \label{eq:displacement}
    \bs{z}(t) = \bs{q}(t) - \bs{q}(0)\,, \quad \forall t \in [0, T]\,,
\end{equation}
which satisfies the equilibrium equation
\begin{equation}\label{eq:Newton-z}
[M]\ddot{\bs{z}}(t) = \bs{\widetilde{f}}(t) \,,
\end{equation}
supplemented with appropriately modified boundary conditions, and $\bs{\widetilde{f}}(t) = \bs{f}(\bs{z}(t)+ \bs{q}(0),t)$. 

Let $t_0 = 0 < t_1 < \cdots < t_{N_t} = T$ be a discretization of the time interval $[0,T]$, where $t_j = j \Delta t$ and $\Delta t$ is the time step. Let $\mathcal{J} = \{j_1,\ldots,j_{N_s}\} \subset \{1,\cdots,N_t\}$, with $1 \leqslant N_s \leqslant N_t$, not necessarily ordered and with distinct elements, and consider the sequence of snapshots $\{\bs{\mu}^{(k)}\}_{k = 1}^{N_s}$ such that $\bs{\mu}^{(k)} = \bs{q}(t_{j_k})$. Let 
\begin{equation}
[X] = [\underline{\bs{\mu}}^{(1)} \ldots \underline{\bs{\mu}}^{(N_s)}]\,, 
\end{equation}
where  $\underline{\bs{\mu}}^{(k)} = \bs{\mu}^{(k)} - \bs{q}(0)$ for $1 \leqslant k \leqslant N_s$, and introduce the singular value decomposition
\begin{equation}\label{eq:eigenvalue-X}
[X]=[U][S][V]^T\,,
\end{equation}
where the sequence of singular values is nonincreasing. A reduced-order basis (ROB) $[\Phi]$ can be classically obtained by retaining the $n$ first columns (which are referred to as POD modes) of $[U]$. The number $n$ of modes can be determined through a convergence analysis enabling a tradeoff between dimensionality reduction and projection error.

Now consider the linear mapping
\begin{equation}\label{eq:projection-phys-to-reduc}
\bs{z}(t) = [\Phi] \bs{y}(t)\,,
\end{equation}
where $\bs{y}$ is the reduced variable with values in $\mathbb{R}^{n}$. The matrix $[\Phi]$ satisfies the orthogonality property
\begin{equation}\label{eq:W-is-orthogonal}
[\Phi]^T[\Phi] = [I_{n}]\,,
\end{equation}
as well as the boundary condition
\begin{equation}\label{eq:DBC}
[B]^T[\Phi]=[0_{N_{\mathrm{CD}},n}]\,,
\end{equation}
where $[B]^T$ defines a boundary condition operator and $N_{\mathrm{CD}}$ is the number of constrained degrees of freedom in the system. The Galerkin projection of Eq.\,\eqref{eq:Newton-z} expressed in terms of atom displacements (with physical variable $\bs{z}$) reads as
\begin{equation}
\label{eq:rom}
[\mathcal{M}] \ddot{\bs{y}}(t) = \mathcal{F}(t)\,,
\end{equation}
where $[\mathcal{M}]$ and $\mathcal{F}$ are the projected mass matrix and reduced force vector, respectively:
\begin{equation}
\label{eq:10}
[\mathcal{M}] = [\Phi]^T[M][\Phi]\,, \quad \mathcal{F}(t)=[\Phi]^T\,\bs{\widetilde{f}}(t)\,.
\end{equation}
Note at this stage that the previous reduced-order formulation is not introduced to accelerate simulations since most of the computational time is spent into assembling procedures. This numerical burden may be circumvented, in practice, by bypassing back-and-forth projections between the physical and reduced space, using, \textit{e.g.}, machine-learning-based surrogates for forces in the reduced space. This aspect is out of the scope of the present work.

Since the reduced-order basis $[\Phi]$ satisfies the orthogonality property stated in Eq.\,\eqref{eq:W-is-orthogonal}, it is necessary to introduce the set of orthogonal matrices 
\begin{equation}\label{eq:stiefel-manifold}
St(N,n) = \{[Y] \in \mathbb{R}^{N \times n}\textnormal{~such that~}[Y]^T[Y]=[I_{n}]\}\,,
\end{equation}
called the compact Stiefel manifold, where $\mathbb{R}^{N \times n}$ is the set of all $N \times n$ real matrices. Owing to the constraint given by Eq.~\eqref{eq:DBC}, the matrix $[\Phi]$ then belongs to the subset 
$\mathbb{S}_{N,n} \subset St(N,n)$ defined as
\begin{equation}\label{eq:state-space-W}
\mathbb{S}_{N,n} = \{[Y] \in \mathbb{R}^{N \times n}\textnormal{~such that~}[Y]^T[Y]=[I_{n}]\,, ~ [B]^T[Y]=[0_{N_{\mathrm{CD}},n}]\}\,.
\end{equation}
The set $\mathbb{S}_{N,n}$ constitutes the admissible space for $[\Phi]$, and can therefore be interpreted as the support of the probability measure for the stochastic counterpart of $[\Phi]$, denoted by $[\bs{\Phi}]$. The main challenge then lies in the construction of a proper probabilistic model for $[\bs{\Phi}]$.

\begin{notation}
In the following, we denote by $[W^{(1)}], \ldots, [W^{(m)}]$ the reduced-order bases in $\mathbb{S}_{N,n}$ obtained by considering each interatomic potential separately, assuming here that $m$ candidate models are available. The global reduced-order basis obtained by concatenating snapshots obtained for all considered potentials is denoted by $[W]$.
\end{notation}

\subsection{Construction of the Stochastic Reduced-Order Model}
\subsubsection{Problem Statement}
We consider the stochastic modeling of the random matrix $[\bs{\Phi}]$, defined on a probability space $(\Theta, \mathcal{T}, P)$ and taking values in the subset $\mathbb{S}_{N,n}$ of the Stiefel manifold $St(N, n)$. We assume that the collection $\{[W^{(1)}], \ldots, [W^{(m)}], [W]\}$ of reduced-order bases is given, and that $[W]$ belongs to the convex hull of $[W^{(1)}], \ldots, [W^{(m)}]$. The tangent space of $St(N,n)$ at $[Y]$ is defined as 
\begin{equation}
    T_{[Y]} St(N,n) = \{[\Delta] \in \mathbb{R}^{N \times n}~|~[Y]^{T}[\Delta] + [\Delta]^T [Y] = [0_n]\}\subset \mathbb{R}^{N \times n}\,,
\end{equation}
where $[Y] \in St(N,n)$ is called the base (or reference) point on the Stiefel manifold, and $[0_n]$ is the null matrix of size $n \times n$. The projection onto the tangent space to the Stiefel manifold at $[Y]$ (push-forward operation) is denoted by
\begin{equation}
\label{eq:proj}
    P_{[Y]}: St(N,n) \to T_{[Y]} St(N,n)\,,
\end{equation}
while the retraction (pull-back operation) is denoted by
\begin{equation}
    \label{eq:retraction}
    R_{[Y]}: T_{[Y]} St(N,n) \to St(N,n)\,.
\end{equation}
There are several ways to define such projection and retraction operators; see, \textit{e.g.}, Chapter 4 in \cite{Absil-2007}. The retraction operator based on the polar decomposition (see Eq.~(4.7), p.~59, in \cite{Absil-2007}), namely
\begin{equation}
R_{[Y]}([\Delta]) = \left([Y] + [\Delta]\right) \left([I_{n}] + [\Delta]^T[\Delta]\right)^{-1/2}\,,    
\end{equation}
was used in \cite{soize2017nonparametric}, in particular (see \cite{wang2019modeling} for an application in a molecular dynamics setting). In fact, using the parameterization
\begin{equation}
    [\Delta] = [A] - [W][D]
\end{equation}
on $T_{[W]}St(N,n)$, where $[A] \in \mathbb{R}^{N \times n}$ is arbitrary and $[D] = \textnormal{Sym}([W]^T[A])$, it is seen that the pulled-back point 
\begin{equation}
[\widetilde{Y}] = R_{[Y]}(s[\Delta]) = \left([W] + s[\Delta]\right) \left([I_{n}] + s^2 [\Delta]^T[\Delta]\right)^{-1/2}\, \quad s \geq 0\,, 
\end{equation}
satisfies the Dirichlet boundary condition $[B]^T[\widetilde{Y}]=[0_{N_{\mathrm{CD}},n}]$ if $[A]$ also satisfies 
\begin{equation}\label{eq:BC-on-A}
    [B]^T[A]=[0_{N_{\mathrm{CD}},n}]\,.
\end{equation}
The representation 
\begin{equation}
    [A] = ([I] - [B][B]^T)[U]\,, \quad [U] \in \mathbb{R}^{N \times n} \textnormal{ arbitrary}\,,
\end{equation}
trivially satisfies Eq.~\eqref{eq:BC-on-A} (see Eq.~\eqref{eq:B-is-orthogonal}) and was introduced in \cite{soize2017nonparametric} to model uncertainties through the randomization of $[U]$. The main advantage of this approach is that the model ensures admissibility of samples by construction, since the stochastic reduced-order basis belongs to $\mathbb{S}$ almost surely. The complexity of (and nonlinearity in) the retraction operator, however, makes statistical inference intricate, since the mean of the stochastic model cannot be enforced for instance. In addition, the formulation introduced in \cite{soize2017nonparametric} to model the stochastic version of $[U]$ introduces a curse of dimensionality in terms of hyperparameters, with a number of parameters that scales as $\mathcal{O}(n^2)$; see \cite{Farhat2019} for a discussion. In the following section, we propose a new representation that fully takes advantage of the \textit{multimodel} molecular dynamics setting and in particular, of the dataset $\{[W^{(1)}], \ldots, [W^{(m)}], [W]\}$.

\subsubsection{Riemannian Stochastic Modeling}
Let $[Y]$ be a reference point on the Stiefel manifold $St(N,n)$, and consider two points, denoted by $[\Delta]$ and $[\widetilde{\Delta}]$, on the tangent space $T_{[Y]} St(N,n)$. The canonical inner product associated with the tangent space $T_{[Y]} St(N,n)$ is then given by 
\begin{equation}
    \langle [\Delta], [\widetilde{\Delta}] \rangle_{[Y]} = \text{tr}([\Delta]^T ([I_N] - \frac{1}{2} [Y][Y]^T)[\widetilde{\Delta}])
\end{equation}
and induces the (canonical) metric $\|[\Delta]\|_{[Y]} = \langle [\Delta], [\Delta] \rangle_{[Y]}^{1/2}$. Note that $\|[\Delta]\|_{[Y]}$ is the length of tangent vector $[\Delta]$ on the tangent space at the base point $[Y]$ and corresponds to the arc length between $[\Delta]$ and $[Y]$ on the Stiefel manifold. 

A Riemannian projection operator $P_{[Y]}: St(N,n) \ni [\widetilde{Y}] \mapsto [\Delta] \in T_{[Y]} St(N,n)$ can be obtained as 
\begin{equation}
    [\Delta] = \log_{[Y]}^{St}([\widetilde{Y}])\,,
\end{equation}
where $\log_{[Y]}^{St}$ is the Riemannian Stiefel logarithm at point $[Y]$, defined such that $||\Delta||_{[Y]}=\langle [\Delta], [\Delta] \rangle_{[Y]}^{1/2}$ represents the geodesic distance between $[\widetilde{Y}]$ and $[Y]$. The retraction operator $R_{[Y]}: T_{[Y]} St(N,n) \ni [\Delta] \mapsto [\widetilde{Y}] \in St(N,n)$ is defined as
\begin{equation}
    [\widetilde{Y}] = \exp_{[Y]}^{St}([\Delta])\,,
\end{equation}
where $\exp_{[Y]}^{St}$ is the Riemannian Stiefel exponential at $[Y] \in St(N,n)$; see Chapter 5 in \cite{Absil-2007} for a review. 

No closed-form results exist for the computation of the Riemannian Stiefel logarithm, which must be evaluated numerically. An optimization-based approach was proposed in \cite{Rentmeesters}, while iterative algorithms based on matrix-algebraic representations geodesic can be found in \cite{zimmermann2019manifold} (see Algorithms 7 and 8 therein for the computation of the Stiefel exponential and logarithm, respectively); see also \cite{zimmermann2017matrix}. Note that there exists an empirical condition, given by $||[Y] - [\widetilde{Y}]||_2 \leq 2$, that ensures that the Stiefel logarithm algorithm converges. 

In this work, we rely on the algorithms proposed in \cite{zimmermann2017matrix,zimmermann2019manifold} and utilize the iterative matrix construction to demonstrate important results related to the constraint given in Eq.~\eqref{eq:DBC}. These results are presented in the form of propositions below. Note that $n \leq N/2$ and that for most dynamical systems of interest, the condition $n \ll N$ is met. 

\begin{proposition}\label{prop-log}
Let $[Y]$ and $[\widetilde{Y}]$ be two points belonging to $\mathbb{S}_{N,n} \subset St(N,n)$. Then $[\Delta] = \log_{[Y]}^{St}([\widetilde{Y}]) \in T_{[Y]} St(N,n)$, where the Riemannian Stiefel logarithm is defined through the matrix-algebraic representation proposed in \cite{zimmermann2019manifold}, satisfies the linear constraint $$[B]^T [\Delta] = [0_{N_{\mathrm{CD}}\times n]}\,,$$ where $[B]$ is defined by Eq.~\eqref{eq:DBC}.
\end{proposition}

\begin{proof} Using the matrix-algebraic representation derived in \cite{zimmermann2019manifold} (see Algorithm 8 therein, as well as \cite{zimmermann2017matrix}), the Stiefel logarithm can be computed as
\begin{equation}
    [\Delta] = \log_{[Y]}^{St}([\widetilde{Y}]) = [Y][A_\tau] + [Q_L][B_\tau]\,,
\end{equation}
where $[Q_L] \in \mathbb{R}^{N \times n}$ stems from the compact (thin) QR decomposition
\begin{equation}
    ([I_N] - [Y][Y]^T)[\widetilde{Y}] = [Q_L][N_L]\,.
\end{equation}
The matrices $[A_k] \in \mathbb{R}^{n \times n}$ and $[B_k] \in \mathbb{R}^{n \times n}$ are associated with the sequence of matrices $\{[A_k], [B_k]\}_{k \geq 0}$ satisfying the system of nonlinear algebraic equations
\begin{equation}
    \left[ \begin{matrix}
    [A_{k+1}] & -[B_{k+1}]^T \\
    [B_{k+1}] & [C_{k+1}]
    \end{matrix}
    \right] = \log \left\{ \exp \left\{ \left[ \begin{matrix}
    [A_{k}] & -[B_{k}]^T \\
    [B_{k}] & [C_{k}]
    \end{matrix}
    \right] \right\} \exp \left\{ \left[ \begin{matrix}
    [0_n] & [0_n] \\
    [0_n] & -[C_{k}]
    \end{matrix}
    \right] \right\} \right\}\,, 
\end{equation}
with
\begin{equation}
    \left[ \begin{matrix}
    [A_{0}] & -[B_{0}]^T \\
    [B_{0}] & [C_{0}]
    \end{matrix}
    \right] = \log \left\{ \left[ \begin{matrix}
    [Y]^T[\widetilde{Y}] & [X_0] \\
    [N_L] & [Y_0]
    \end{matrix}
    \right] \right\}\,.
\end{equation}
The matrices $\{[X_0],[Y_0]\}$ are obtained by completion, and $\tau$ is the smallest integer such that $\|[C_\tau]\|_2 \leq \epsilon$, with $\epsilon$ a given threshold parameter. Assuming the invertibility of $[N_L] \in \mathbb{R}^{n \times n}$ (which follows when $\textnormal{rank}(([I_N] - [Y][Y]^T)[\widetilde{Y}]) = n$), we have that
\begin{equation}
\begin{aligned}
    [B]^T [\Delta] & = [B]^T \left( [Y][A_\tau] + [Q_L][B_\tau] \right)\,,\\
    & = [B]^T [Y][A_\tau] + [B]^T ([I_N] - [Y][Y]^T)[\widetilde{Y}][N_L]^{-1} [B_\tau]\,, \\
    & = [0_{N_{\mathrm{CD}}\times n]}\,,
\end{aligned}
\end{equation}
since $[B]^T [Y] = [B]^T [\widetilde{Y}] = [0_{N_{\mathrm{CD}}\times n]}$.
\end{proof}

\begin{proposition}\label{prop-exp}
Let $[Y] \in \mathbb{S}_{N,n} \subset St(N,n)$ and consider $[\Delta] \in T_{[Y]} St(N,n)$ satisfying $[B]^T [\Delta] = [0_{N_{\mathrm{CD}}\times n]}$, where $[B]$ is defined by Eq.~\eqref{eq:DBC}. Then $[\widetilde{Y}] = \exp_{[Y]}^{St}([\Delta]) \in St(N,n)$ (Riemannian Stiefel exponential) satisfies the linear constraint $$[B]^T [\widetilde{Y}] = [0_{N_{\mathrm{CD}}\times n]}\,,$$ that is, $[\widetilde{Y}] \in \mathbb{S}_{N,n} \subset St(N,n)$.
\end{proposition}
\begin{proof} Following \cite{edelman1998geometry}, the Riemannian exponential is evaluated as
\begin{equation}
    St(N,n) \ni [\widetilde{Y}] = \exp_{[Y]}^{St}([\Delta]) = [Y][M] + [Q_E][N_E]\,,
\end{equation}
where $[M],[N_E] \in \mathbb{R}^{n\times n}$ are defined as
\begin{equation}
    \left[ \begin{matrix}
    [M]\\
    [N_E]
    \end{matrix}
    \right] = \exp \left\{\left[ \begin{matrix}
    [Y]^T [\Delta] & -[R_E]^T \\
    [R_E] & [0_n]
    \end{matrix}
    \right]\right\} \left[ \begin{matrix}
    [I_n]\\
    [0_{n}]
    \end{matrix}
    \right]\,.
\end{equation}
The matrices $[Q_E] \in \mathbb{R}^{N \times n}$ and $[R_E] \in \mathbb{R}^{n \times n}$ arise in the compact (thin) QR decomposition
\begin{equation}
    ([I_N] - [Y][Y]^T)[\Delta] = [Q_E][R_E]\,.
\end{equation}
Assuming that $[R_E]$ is invertible (which follows when $\textnormal{rank}(([I_N] - [Y][Y]^T)[\Delta]) = n$), we deduce
\begin{equation}
\begin{aligned}
    [B]^T [\widetilde{Y}] & = [B]^T \left( [Y][M] + [Q_E][N_E] \right)\,,\\
    & = [B]^T [Y][M] + [B]^T ([I_N] - [Y][Y]^T)[\Delta][R_E]^{-1} [N_E]\,, \\
    & = [0_{N_{\mathrm{CD}}\times n]}\,,
\end{aligned}
\end{equation}
as $[Y]$ and $[\Delta]$ satisfy $[B]^T [Y] = [B]^T [\Delta] = [0_{N_{\mathrm{CD}}\times n]}$.
\end{proof}
Propositions \ref{prop-log} and \ref{prop-exp} imply that the satisfaction of the linear constraint is preserved through the pushforward and pullback actions defined by the Riemannian Stiefel logarithm and exponential. Applying these results to the proposed framework, we can now derive
\begin{proposition}
Let $\{[\Delta^{(i)}]\}_{i = 1}^m$ be the projections of the reduced-order bases
$\{[W^{(i)}] \in \mathbb{S}_{N,n}\}_{i = 1}^m$ onto the tangent space $T_{[W]} St(N,n)$ at $[W] \in \mathbb{S}_{N,n}$, $[\Delta^{(i)}] = \log_{[W]}^{St}([W^{(i)}])$ for $1 \leq i \leq m$. Then the linear combination $[\widetilde{\Delta}] = \sum_{i = 1}^{m} p_i [\Delta^{(i)}]$, with $(p_1, \ldots, p_m) \in \mathbb{R}^m$, satisfies the property $$\exp_{[W]}^{St}([\widetilde{\Delta}]) \in \mathbb{S}_{N,n} \subset St(N,n)\,.$$ 
\end{proposition}
\begin{proof}
The result is immediate using Propositions \ref{prop-log} and \ref{prop-exp}.
\end{proof}

The above proposition suggests to seek the stochastic representation as
\begin{equation}\label{eq:PhiConvexCombi}
    [\bs{\Phi}] := \exp_{[W]}^{St}\left\{ \sum_{i = 1}^{m} P_i \log_{[W]}^{St}([W^{(i)}]) \right\}\,.
\end{equation}
This form ensures that $[\bs{\Phi}]$ takes values in the constrained set $\mathbb{S}_{N,n}$, by construction. 

In Eq.~\eqref{eq:PhiConvexCombi}, the random vector $\bs{P} = (P_1, \ldots, P_m)^T$ is defined on a probability space $(\Theta, \mathcal{T}, P)$. A natural choice for the probability measure of $\bs{P}$ is the Dirichlet distribution with concentration parameter $\bs{\alpha}$, $\bs{P} \sim \mathcal{D}(\bs{\alpha})$. This choice ensures that $P_i \geq 0$ and $\sum_{i = 1}^{m} P_i = 1$ almost surely, and therefore defines a stochastic Riemannian convex combination on the Stiefel manifold. In practice, this construction leads to samples that belong to the convex hull defined by the reduced-order bases dataset $\{[W^{(i)}]\}_{i = 1}^m$ (see \cite{afsari2013convergence} for an analysis in a deterministic setting).

\begin{remark}
With the proposed formulation, uncertainty propagation can be achieved through Monte Carlo simulations, as well as by using state-of-the-art stochastic collocation methods. More specifically, let $\bs{Y}$ be the random variable with values in $\mathbb{R}_{>0}^m$ and with independent components, such that $Y_i \sim \mathcal{G}(\alpha_i, 1)$. In this case, $\bs{P}$ and $\bs{Y}$ are related through
\begin{equation}
    P_i = \frac{Y_i}{\sum_{j = 1}^{m} Y_j}\,, \quad 1 \leq i \leq m\,.
\end{equation}
It follows that $[\bs{\Phi}]$ can equivalently be viewed as a function of $\bs{Y}$ (that is, $[\bs{\Phi}] = [\bs{\Phi}(\bs{P})] = [\bs{\Phi}(\bs{Y})]$), which enables the use of, \textit{e.g.}, polynomial chaos expansions in terms of Laguerre polynomials \cite{Xiu-2002,Soize-2004} to represent, and efficiently identify, stochastic quantities of interest defined through a multiscale operator; see \cite{ghanem2017handbook,le2010spectral} for reviews regarding representations and stochastic solvers.
\end{remark}

\subsubsection{Integrating a Constraint on Fr\'echet Mean}\label{subsubsec:alpha-with-frechet}
The aim of this section is to derive a formulation that allows the empirical mean model associated with the representation \eqref{eq:PhiConvexCombi} to be prescribed. To this end, we assume that the global reduced-order basis $[W]$ belongs to the convex hull of $\{[W^{(i)}]\}_{i = 1}^m$, and consider the identification of the concentration parameter $\bs{\alpha}$ such that 
\begin{equation}
    \mathbb{E}\{[\bs{\Phi}]\} \approx [W]\,,
\end{equation}
where the mean holds in the Fr\'echet sense. Recall that the Rienmannian $L^2$ center of mass of a dataset $\{[\Phi^{(1)}], \ldots, [\Phi^{(q)}]\}$ composed of $q$ samples of $[\bs{\Phi}]$ (in $St(N, n)$) is defined as the minimizer of 
\begin{equation}
    h([V]) = \frac{1}{2} \sum_{i = 1}^{q} w_i\, d([V],[\Phi^{(i)}])^2\,,
\end{equation}
where $\{w_i\}_{i = 1}^{q}$ are scalar weights in the $(q-1)$-dimensional simplex and $d$ is the Riemannian canonical distance. Imposing that the gradient of the objective function vanishes at $[W]$ then yields 
\begin{equation}
    \sum_{i = 1}^{q} w_i \log_{[W]}^{St} \left\{ [\Phi^{(i)}] \right\} = [0_{N \times n}]\,.
\end{equation}
Using the definition \eqref{eq:PhiConvexCombi} and taking $w_i = 1/q$ for all weights then implies
\begin{equation}
    \sum_{j = 1}^{m} \left( \frac{1}{q} \sum_{i = 1}^{q} p_j^{(i)} \right) \log_{[W]}^{St}([W^{(j)}]) = [0_{N \times n}]\,,
\end{equation} 
where $p_j^{(i)}$ denotes the i-th realization of the component $P_j$ of $\bs{P} \sim \mathcal{D}(\boldsymbol{\alpha})$ ($p_j^{(i)} = p_j(\theta_i)$, $\theta_i \in \Theta$). Since
\begin{equation}
    \frac{1}{q} \sum_{i = 1}^{q} p_j^{(i)} \approx \frac{\alpha_j}{\sum_{i = 1}^{m} \alpha_i}
\end{equation}
for $q$ sufficiently large, it can be deduced that the concentration parameters must satisfy
\begin{equation}
    T_{[W]} St(N,n) \ni \sum_{j = 1}^{m} \alpha_j \log_{[W]}^{St}([W^{(j)}]) = [0_{N \times n}]\,.
\end{equation}
The above property can hence be enforced by imposing the constraint
\begin{equation}
    \|\sum_{j = 1}^{m} \alpha_j \log_{[W]}^{St}([W^{(j)}])\|_F = 0\,.
\end{equation}
In practice, $\bs{\alpha}$ can be evaluated as
\begin{equation}
    \label{eq:qp}
    \bs{\alpha} = \textnormal{argmin}_{\bs{a} \in \mathbb{R}_{> 0}^m}~ \|\sum_{i=1}^{m} a_i \log_{[W]}^{St}([W^{(j)}])\|_F^2\,,
\end{equation}
which is recast, for implementation purposes, as 
\begin{equation}\label{eq:quadprogpb}
    \bs{\alpha} = \textnormal{argmin}_{\bs{a} \in \mathbb{R}_{> 0}^m}~ \bs{a}^T [H] \bs{a}\,,
\end{equation}
where $[H]$ is the symmetric positive-definite matrix in $\mathbb{R}^{m \times m}$, the entries of which are given by
\begin{equation}
    H_{ij} = \textnormal{tr}\left(\log_{[W]}^{St}([W^{(i)}])^T \log_{[W]}^{St}([W^{(j)}])\right)\,.
\end{equation}
This problem can be solved by any conventional quadratic programming algorithm. In this work, the built-in MATLAB function $\verb+quadprog+$ is used for the sake of illustration.

\subsubsection{Scaling Fluctuations on the Tangent Space}\label{subsubsec:scaling-tangent-space}
Defining the stochastic reduced-order basis as
\begin{equation}
\label{eq:PhiConvexCombi-unscaled}
    [\bs{\Phi}] = \exp_{[W]}^{St}\left\{ \sum_{i = 1}^{m} P_i \log_{[W]}^{St}([W^{(i)}]) \right\}\,, \quad \bs{P} \sim \mathcal{D}(\bs{\alpha})\,,
\end{equation}
restricts statistical fluctuations in the convex hull of $\{[W^{(i)}]\}_{i = 1}^m$. In order to increase fluctuations, a scaling parameter $c \geq 1$ is introduced to scale variations on the tangent space:
\begin{equation}\label{eq:PhiConvexCombi-scaled}
    [\bs{\Phi}] = \exp_{[W]}^{St}\left\{c \sum_{i = 1}^{m} P_i \log_{[W]}^{St}([W^{(i)}]) \right\}\,.
\end{equation}
It should be noticed that the calibration strategy for the concentration parameter $\bs{\alpha}$ is insensitive to multiplicative scaling (see Section \ref{subsubsec:alpha-with-frechet}). Consequently, considering $c > 1$ may lead to a shift in the Fr\'{e}chet mean that is all the more pronounced that the distance between the Fr\'echet mean taken over the dataset $\{[W^{(i)}]\}_{i = 1}^m$ and the global reduced-order basis $[W]$ is important. The value of $c$ may be calibrated in practice solving a statistical inverse problems on microscopic or macroscopic quantities of interest (see Section \ref{subsubsec:sto-model-tension} for an example).

\subsection{Summary of the Proposed Approach}\label{subsec:summary}
\begin{figure}[htbp]
    \centering
    \includegraphics[width=\textwidth]{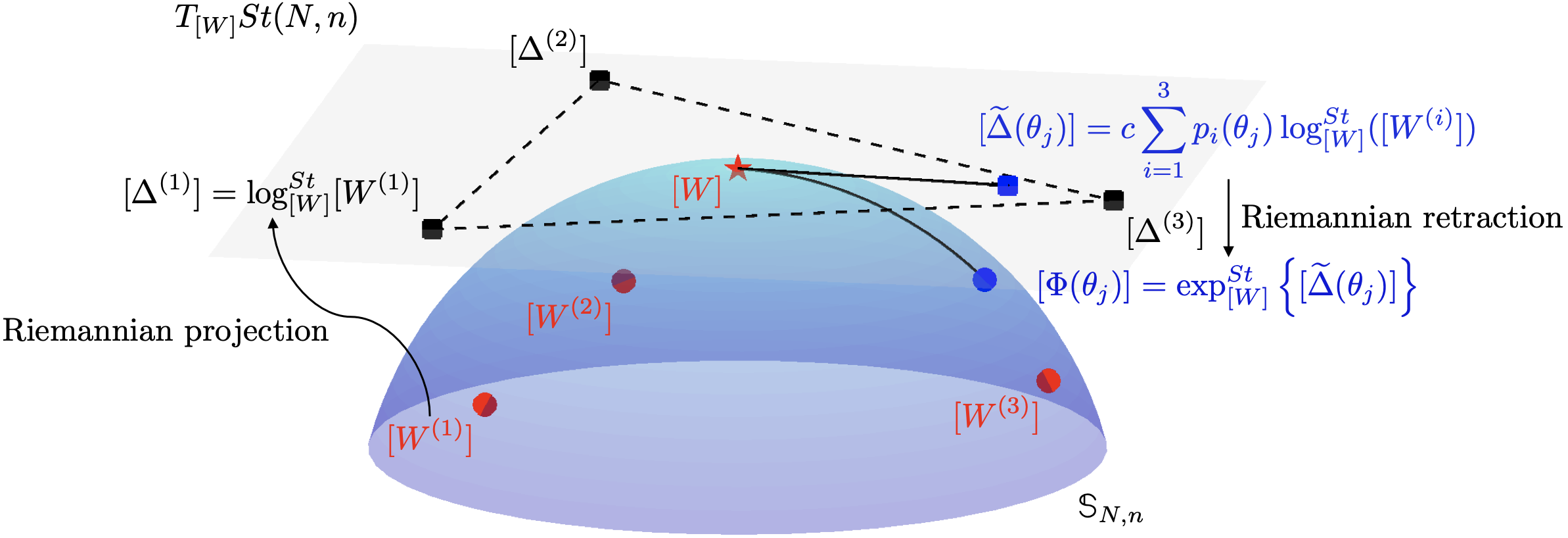}
    \caption{Schematic view of the proposed stochastic modeling strategy on $\mathbb{S}_{N,n}$, with $m = 3$. Riemannian operators are defined with respect to the canonical metric. Note that geometries of sets are illustrative and may not reflect actual structures.}
    \label{fig:overall-procedure}
\end{figure}

The main steps of the proposed modeling framework are listed below and are schematically illustrated in Fig.~\ref{fig:overall-procedure}. 
Recall that $\{[W^{(1)}],\ldots,[W^{(m)]}\}$ are the reduced-order bases in $\mathbb{S}_{N,n}$ computed through a proper orthogonal decomposition with snapshots associated with the full dynamical system for $m$ different input candidate models, and that $[W]$ denotes the global reduced-order basis obtained by gathering all snapshots for all models. The global ROB is taken as base point and target Fr\'{e}chet mean in the sampling procedure. 
\begin{enumerate}[Step 1:]
\item Compute the tangent vectors $\{[\Delta^{(1)}],\dots,[\Delta^{(m)}]\}$ using the Riemannian projection operator, $[\Delta^{(i)}]=\text{log}_{[W]}^{St} [W^{(i)}]$ with $[\Delta^{(i)}] \in T_{[W]}St(N,n)$.
\item Compute the concentration parameters $\bs{\alpha} = (\alpha_1,\ldots,\alpha_m)$ by solving the quadratic programming problem defined in Eq.~\eqref{eq:quadprogpb}.
\item Draw $\nu$ samples $\{\bs{p}(\theta_j)\}_{j = 1}^{\nu}$ of $\bs{P} \sim \mathcal{D}(\bs{\alpha})$, $\theta_j \in \Theta$ for $1 \leq j \leq \nu$.
\item Compute the associated samples $\{[\Phi(\theta_j)]\}_{j = 1}^{\nu}$ of $[\bs{\Phi}]$ as 
\begin{equation}
    [\Phi(\theta_j)] = \exp_{[W]}^{St}\left\{c \sum_{i = 1}^{m} p_i(\theta_j) \log_{[W]}^{St}([W^{(i)}]) \right\} \in \mathbb{S}_{N,n} \subset St(N,n)\,,
\end{equation}
where $c = 1$ for stochastic Riemannian convex combinations or $c \geq 1$ to enforce fluctuations beyond the convex hull of the dataset.
\end{enumerate}
In the next section, we deploy the proposed approach on a variety of applications. The case of the unit sphere is first presented in Section \ref{sec:simple-sphere} to illustrate the approach with standard visualization in $\mathbb{R}^3$. Applications to molecular dynamics simulations on graphene-based systems are then discussed in Sections \ref{sec:harmonic} and \ref{sec:multiscale}, with focus on microscopic and macroscopic responses respectively. The open-source package LAMMPS \cite{LAMMPS} is used for both full-order and reduced-order MD simulations.

\section{Applications}
\label{sec:applications}
\subsection{Illustrative Example: Sampling on (a Subset of) the Unit Sphere $St(3,1)$}
\label{sec:simple-sphere}
\subsubsection{Sampling Without Linear Constraints}
\begin{figure}[htbp]
	\centering
	\subfloat[2D view]{\includegraphics[width = 0.45\textwidth]{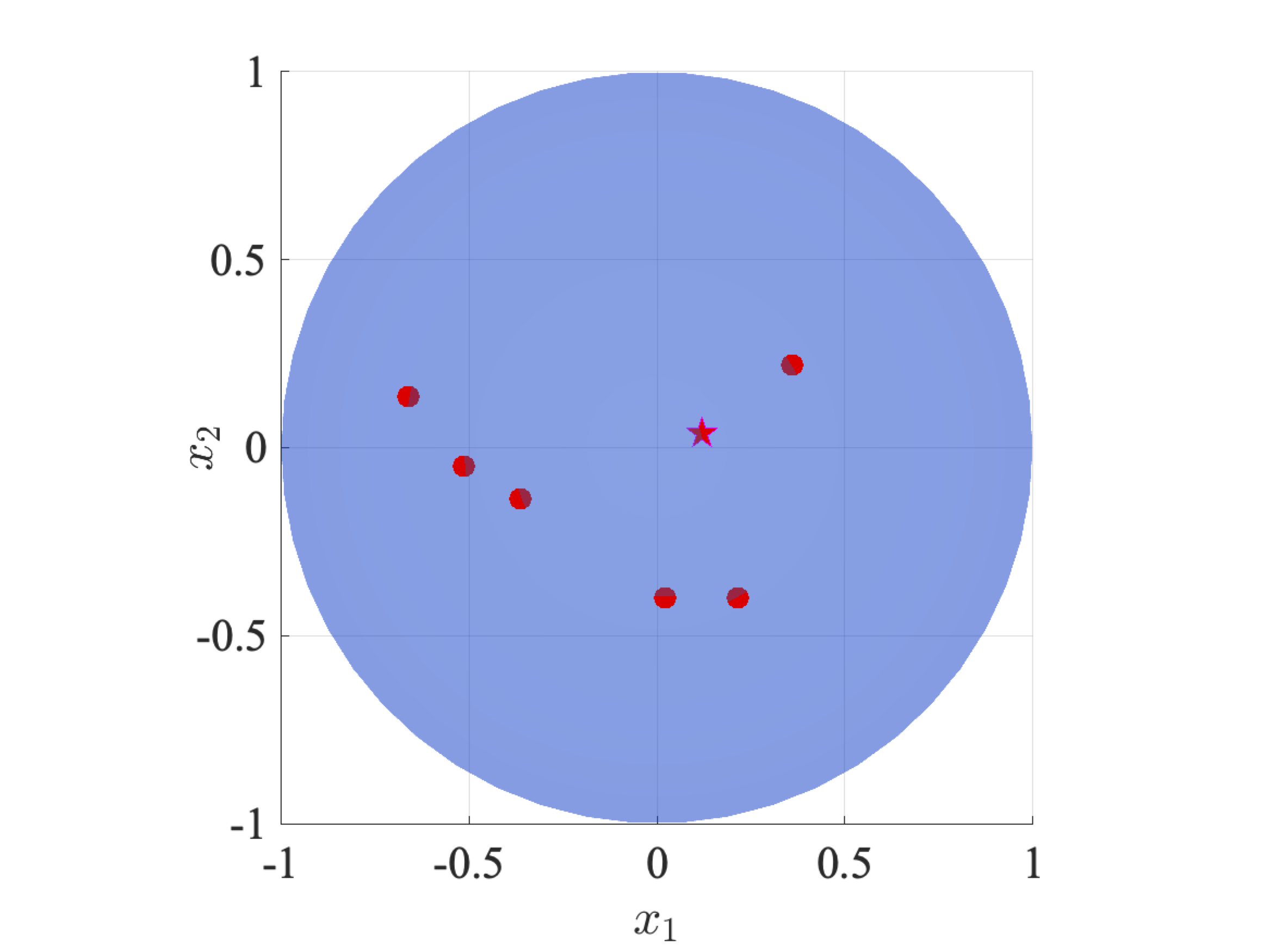}}\;
	\subfloat[3D view]{\includegraphics[width = 0.45\textwidth]{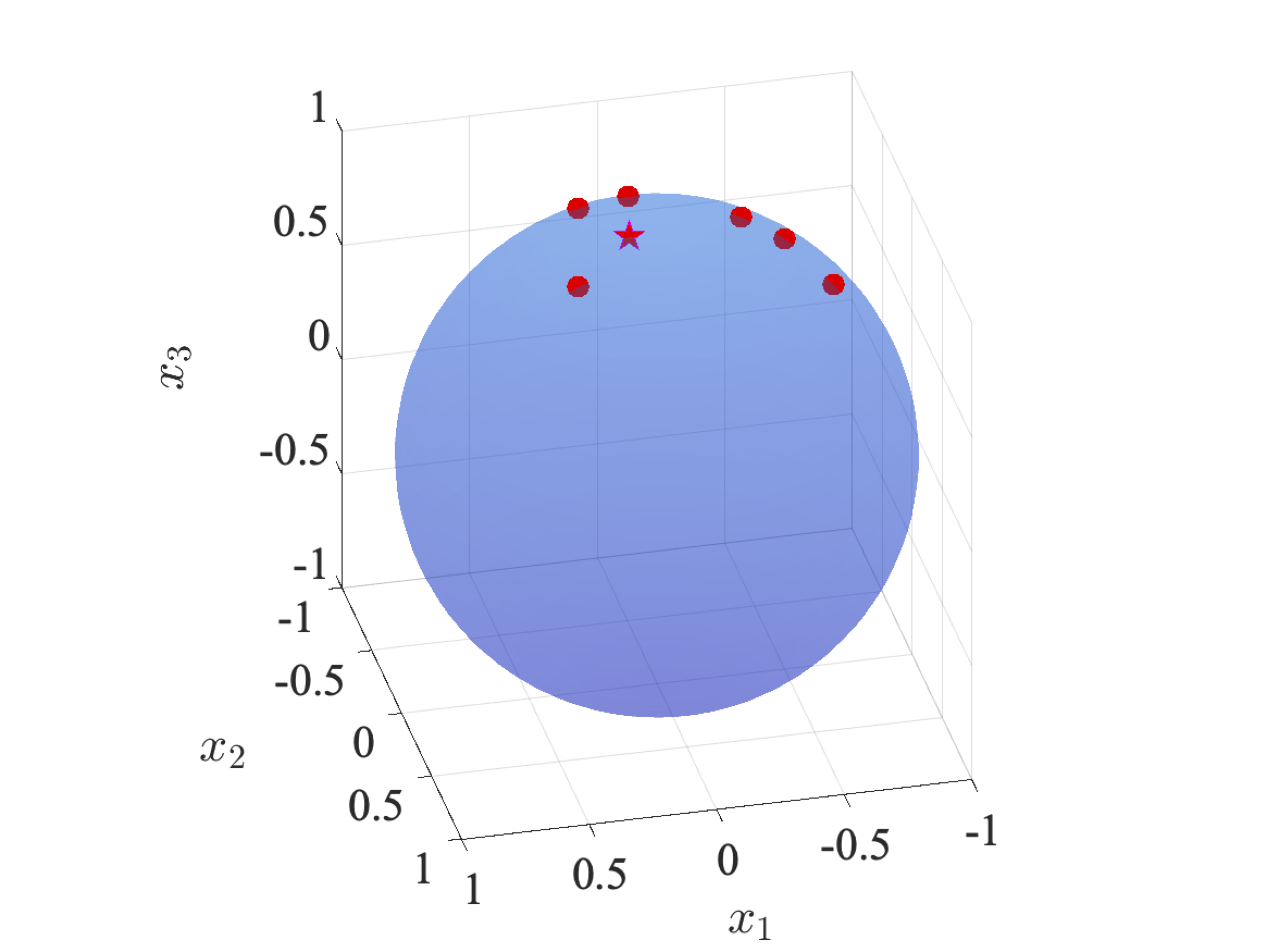}}\;
	\caption{Dataset on the unit sphere in $\mathbb{R}^3$: base point (red star) and vertices (red points).}
    	\label{fig:original-sphere}
\end{figure}

In this first example, we consider sampling on the half unit sphere (that is, without the linear constraints defined by the matrix $[B]$, see Eq.~\eqref{eq:def-linear-constraints}). The dataset consists of seven points randomly distributed on the sphere ($m = 6$), with one base point included in the convex hull defined by the remaining points; see Fig. \ref{fig:original-sphere}.

\begin{figure}[htbp]
	\centering
    	\subfloat[2D view, unconstrained]{\includegraphics[width = 0.45\textwidth]{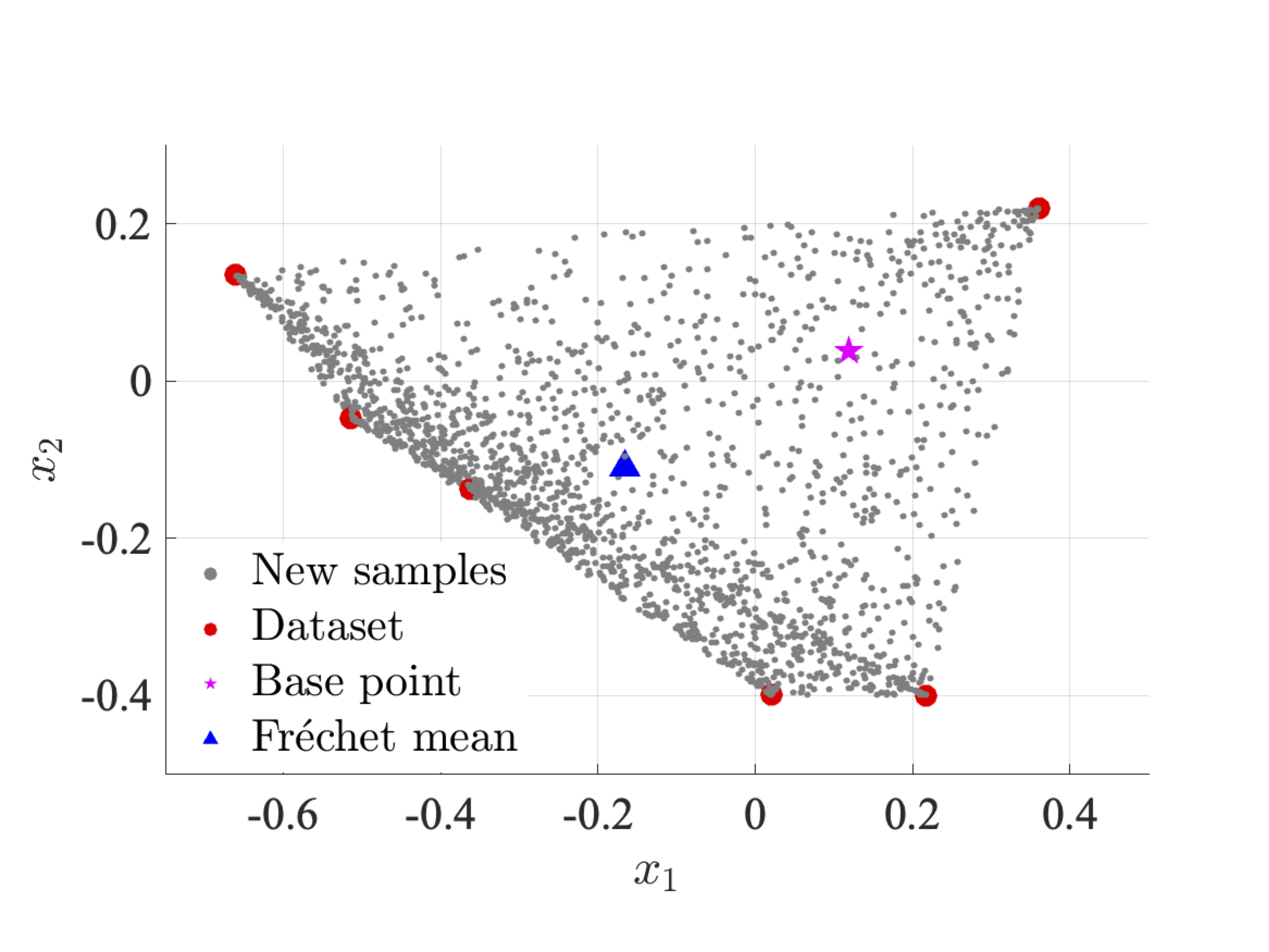}}\;
    	\subfloat[2D view, constrained]{\includegraphics[width = 0.45\textwidth]{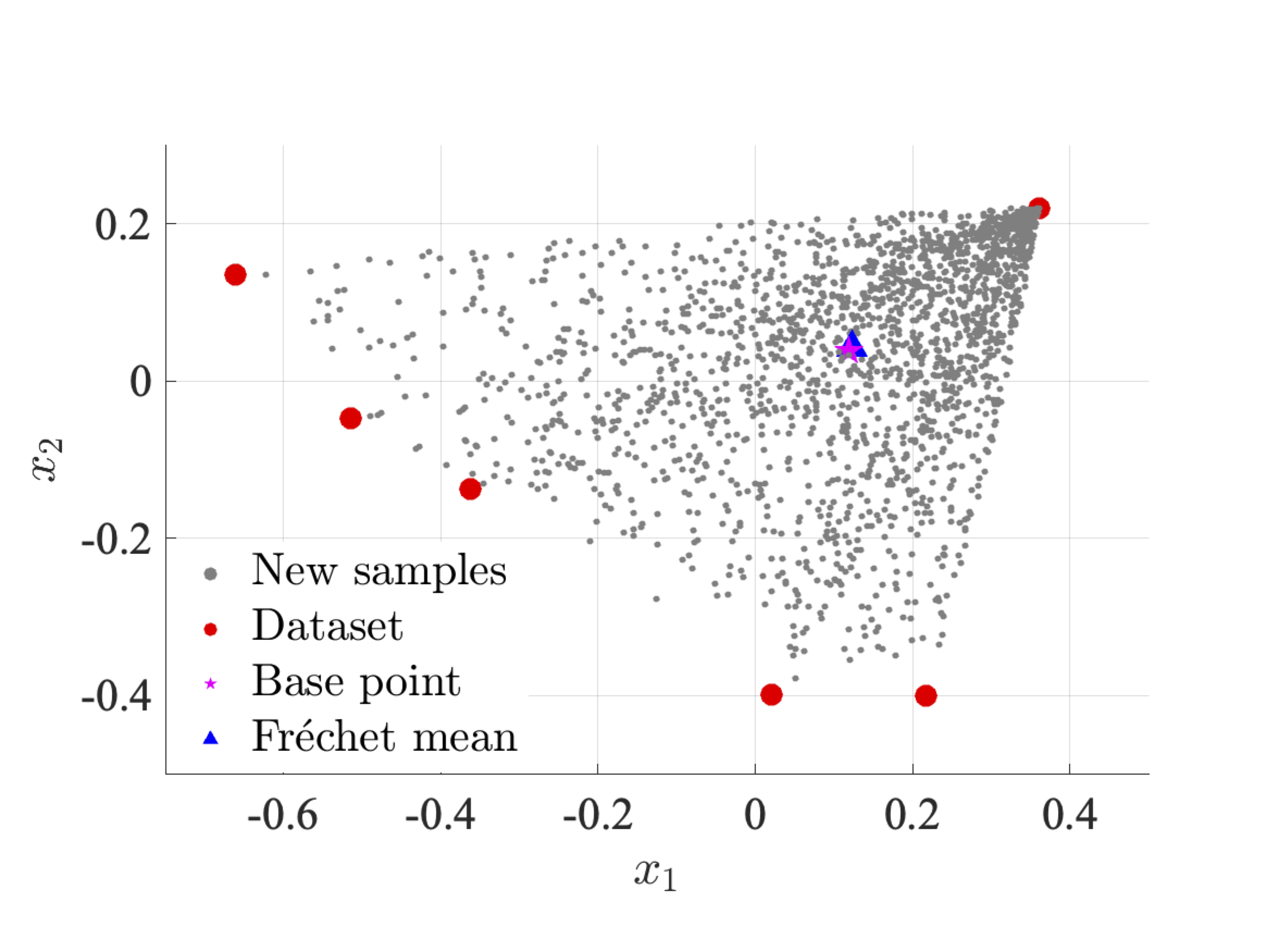}}\;\\
    	\subfloat[3D view, unconstrained]{\includegraphics[width = 0.45\textwidth]{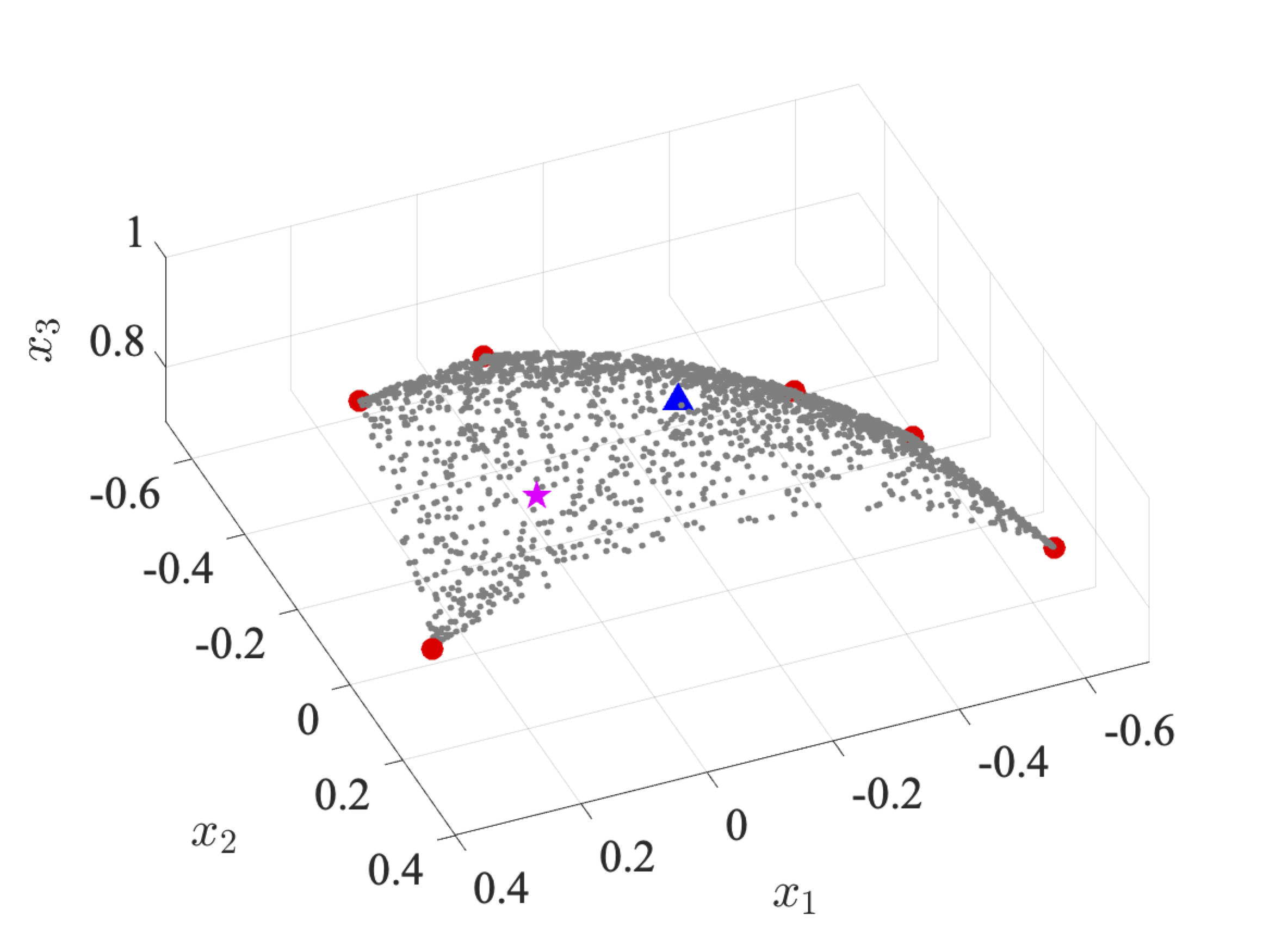}}\;
    	\subfloat[3D view, constrained]{\includegraphics[width = 0.45\textwidth]{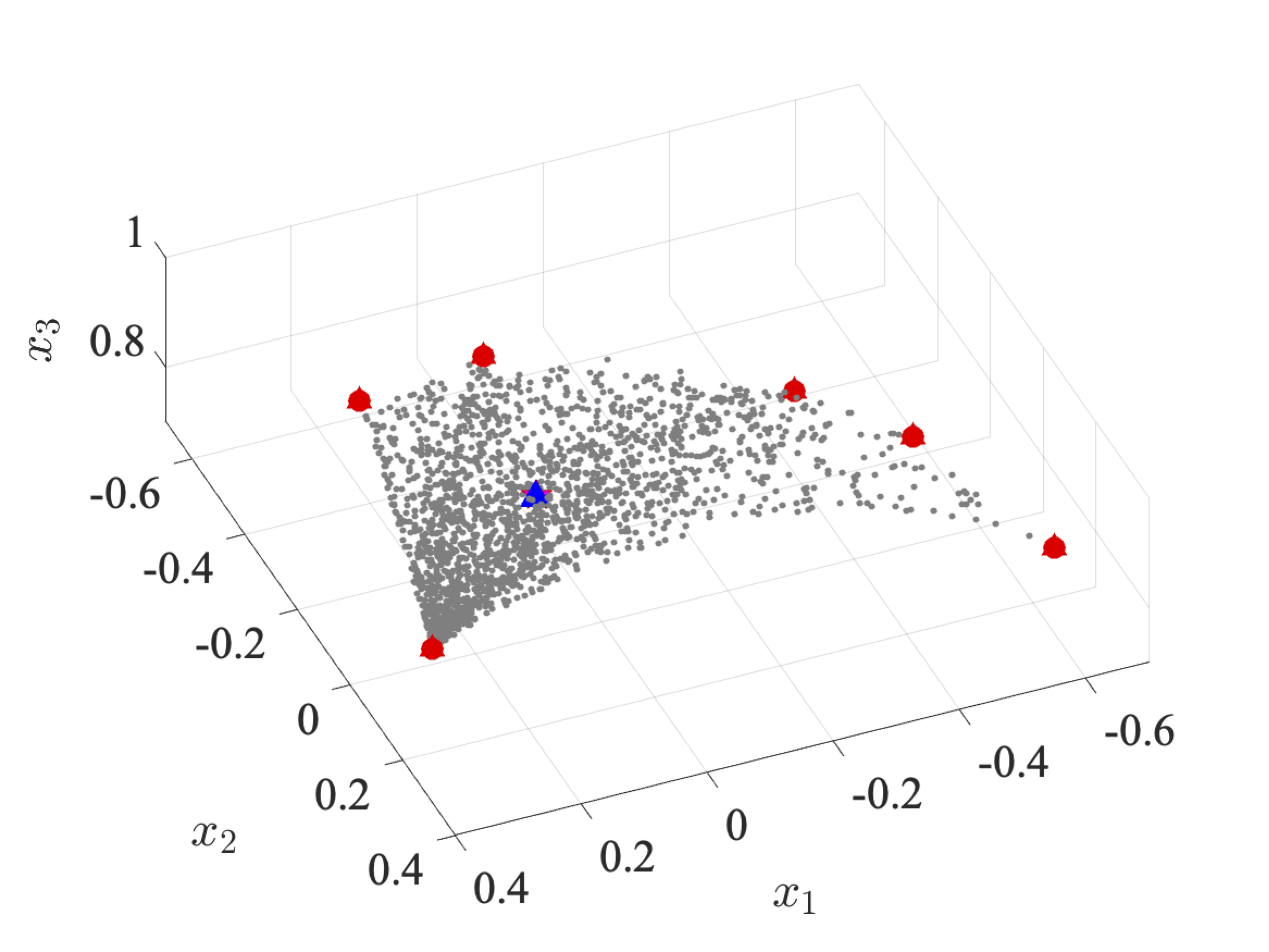}}\;
	\caption{Left panel (unconstrained sampling): samples obtained for $c = 1$ and $\alpha_i = 0.2$ for $i \in \{1, \ldots, 6\}$. Right panel (constrained sampling): samples obtained for $c = 1$ and $\bs{\alpha} = (0.2357, 0.2000, 0.1820, 0.1805, 1.4151,0.2866)^T$.}
    	\label{fig:sphere_unscaled}
\end{figure}

Two specific choices are made at this point. In a first setting, all concentration parameters are set to the same value, $\alpha_i = 0.2$ for $i \in \{1, \ldots, 6\}$. In the second configuration, concentration parameters are calibrated such that the Fr\'{e}chet mean is as close as possible to the aforementioned base point, following the strategy proposed in Section \ref{subsubsec:alpha-with-frechet}. Here, $c$ is set to 1 so that only stochastic Riemannian convex combinations are used. Fig.~\ref{fig:sphere_unscaled} shows a set of 2,000 samples for both cases.
It is seen that uniform sampling in the convex hull can be achieved by setting all concentration parameters equal to a small value (see the left panel in Fig.~\ref{fig:sphere_unscaled}). The Fr\'{e}chet mean of the samples, computed using the algorithm detailed in Appendix \ref{app:frechet}, then lies far away from the chosen base point. In contrast, determining the concentration parameters by solving the quadratic programming problem defined in Eq.~\eqref{eq:quadprogpb} allows for the Fr\'{e}chet mean to be constrained to the neighborhood of the base point.

\begin{figure}[htbp]
	\centering
	\subfloat[2D view, unconstrained]{\includegraphics[width = 0.45\textwidth]{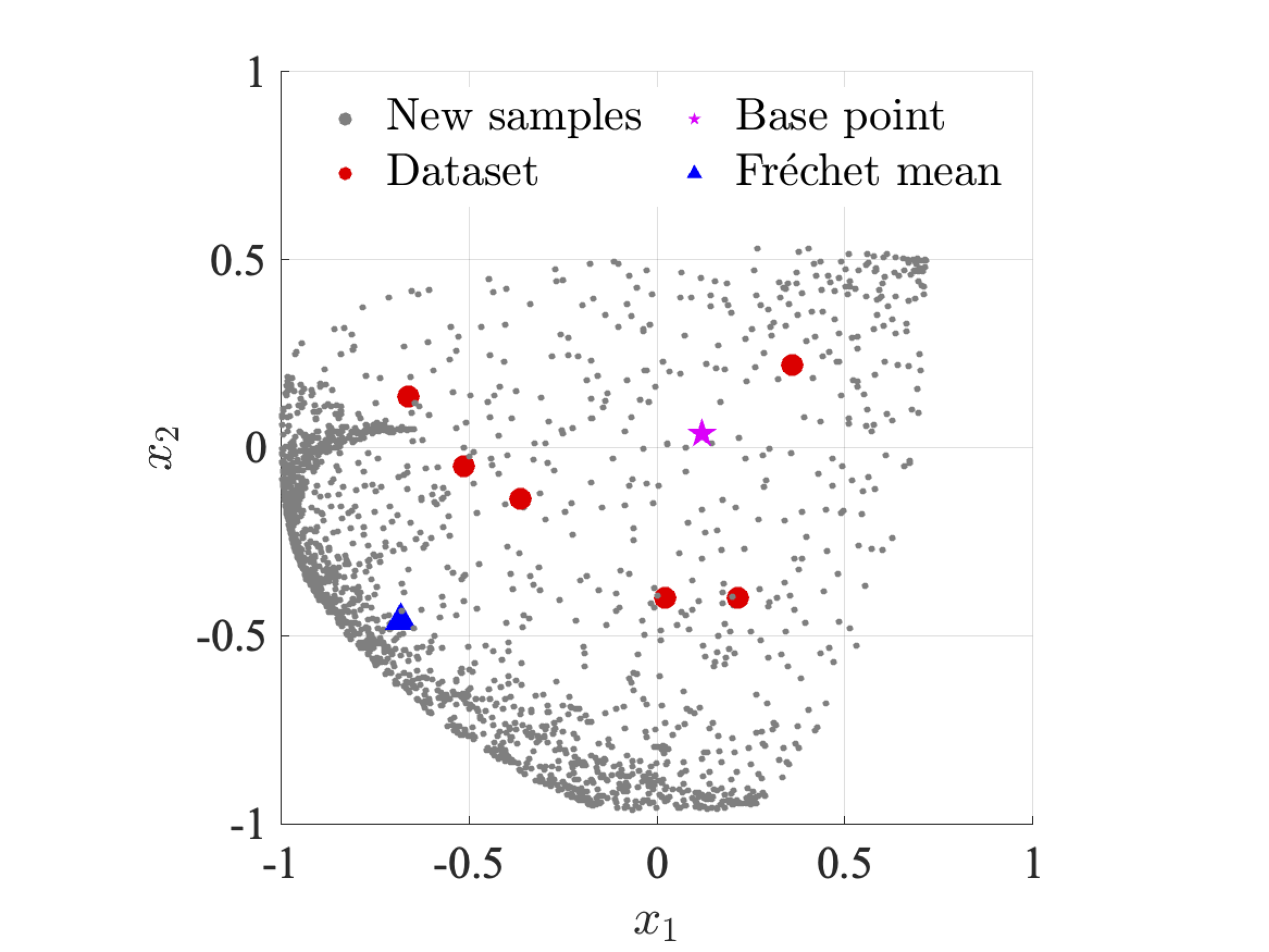}}\;
    	\subfloat[2D view, constrained]{\includegraphics[width = 0.45\textwidth]{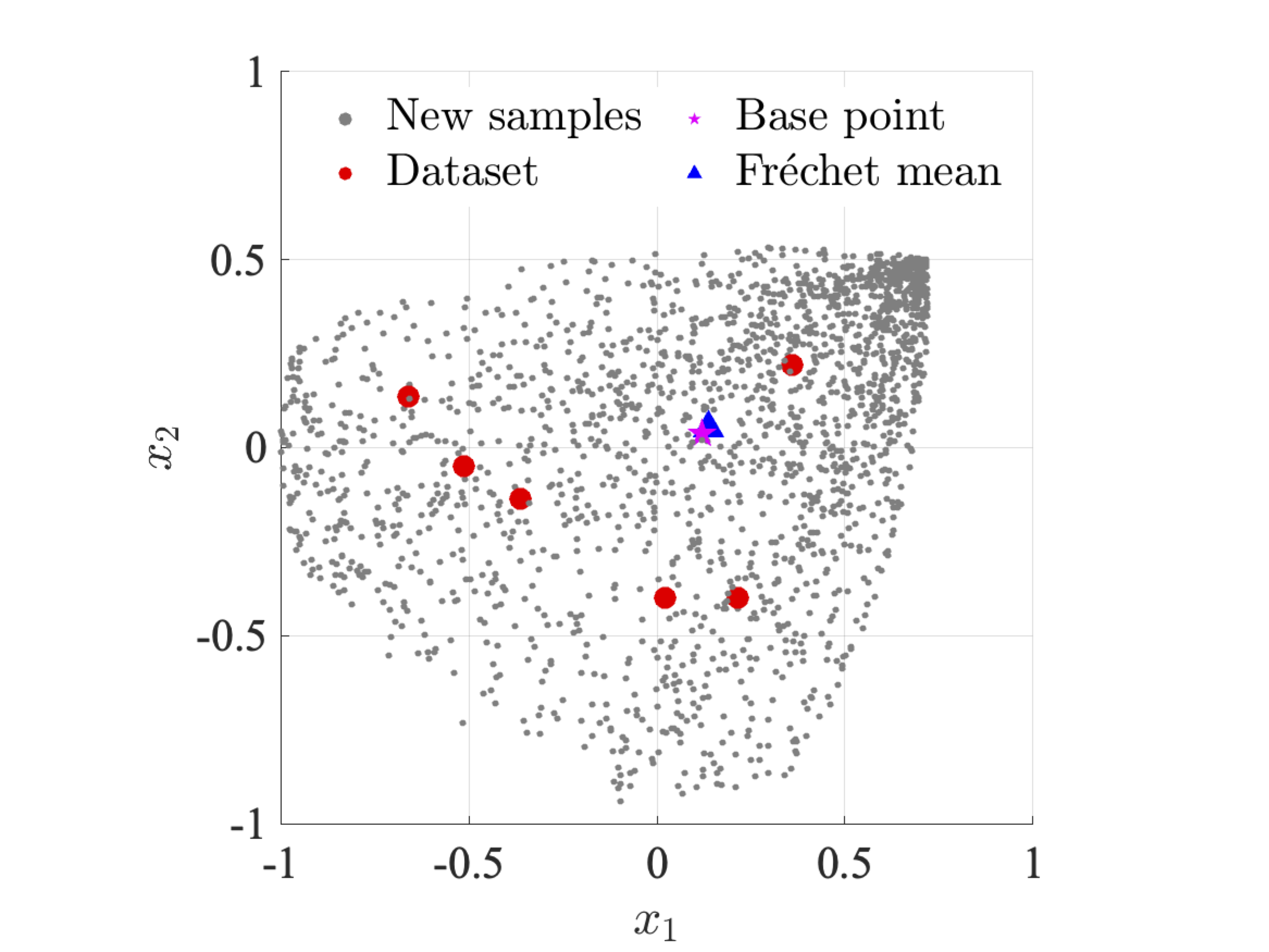}}\;\\
    	\subfloat[3D view, unconstrained]{\includegraphics[width = 0.45\textwidth]{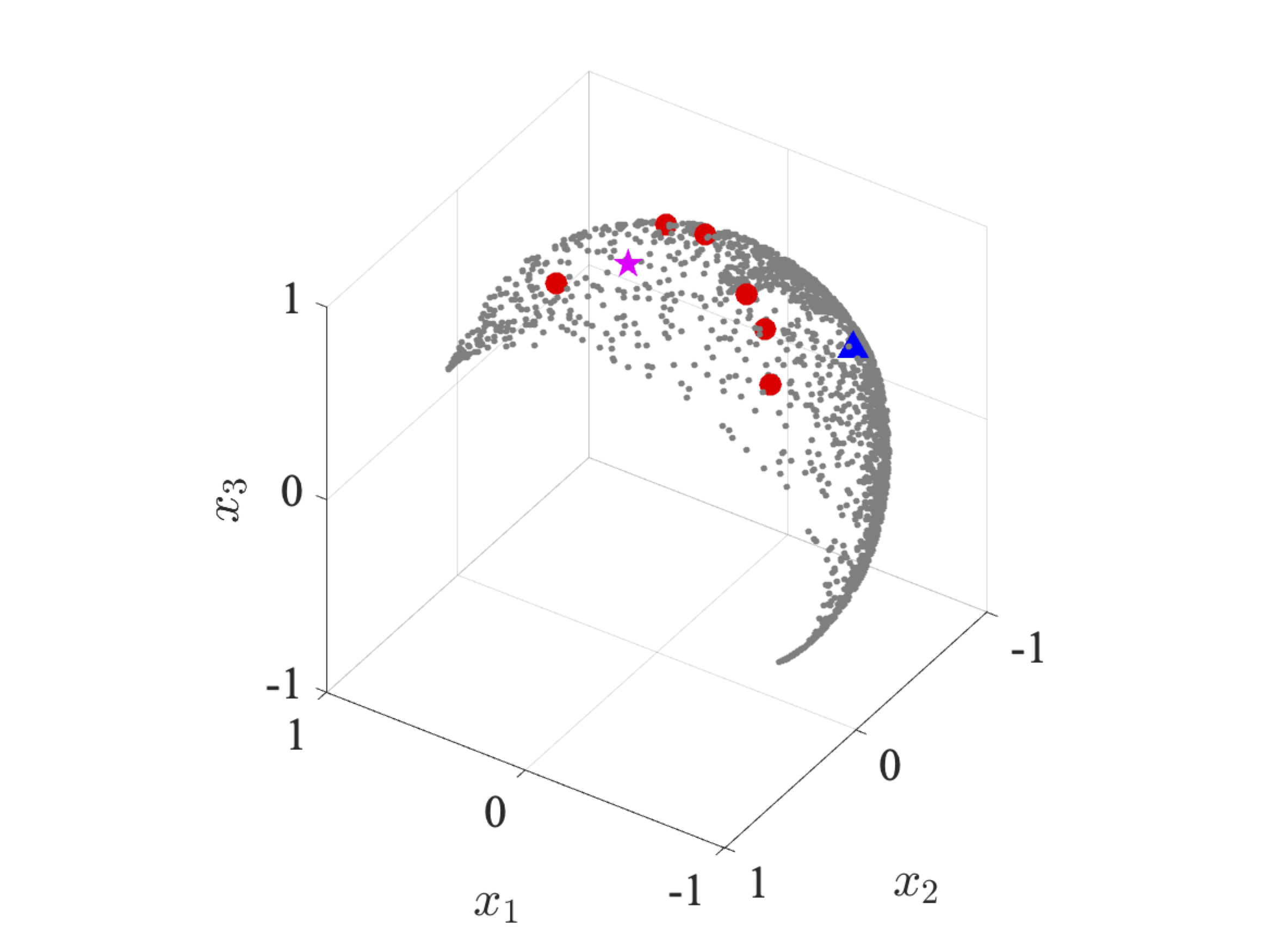}}\;
    	\subfloat[3D view, constrained]{\includegraphics[width = 0.45\textwidth]{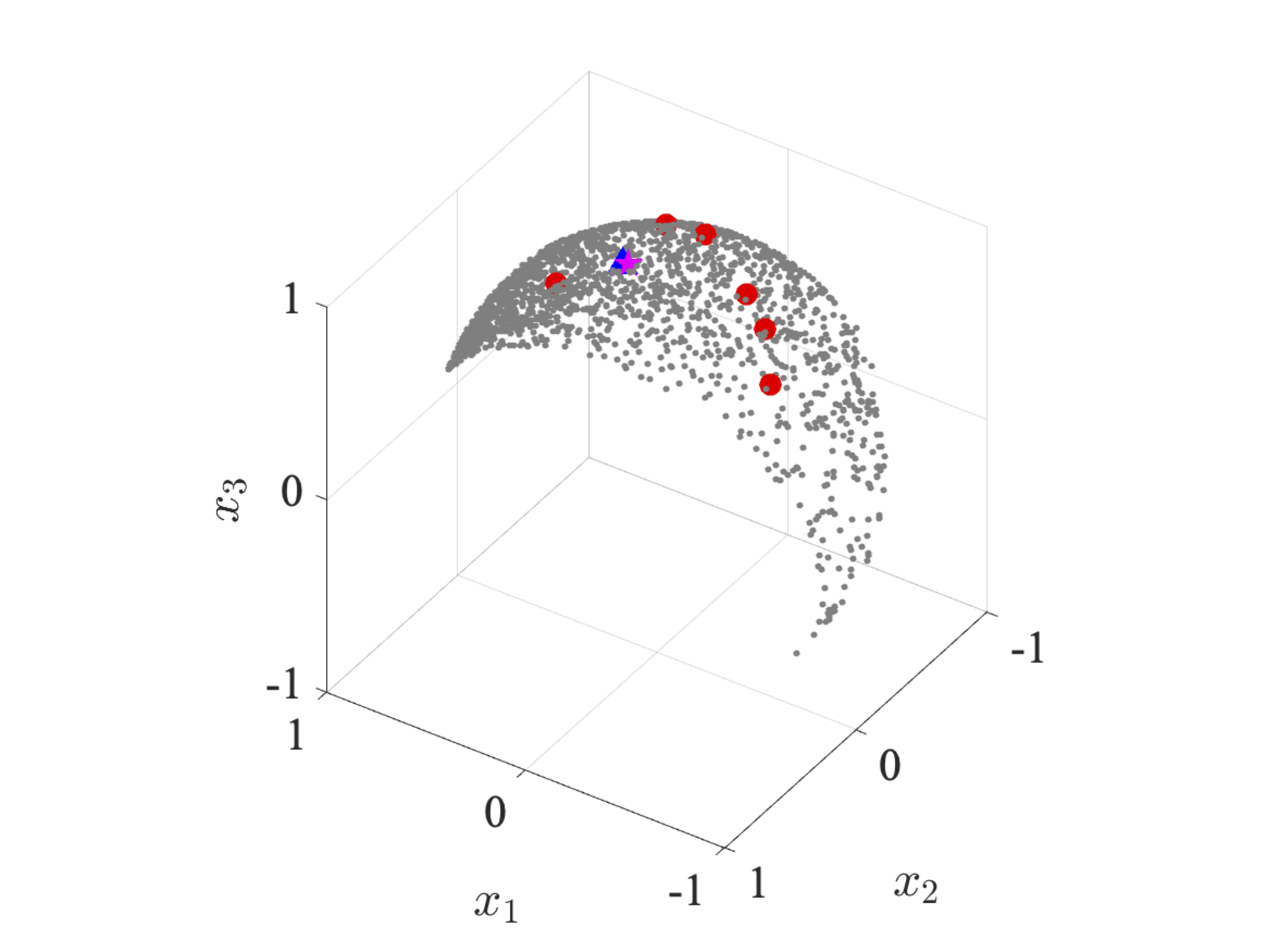}}\;
    	\caption{Left panel (unconstrained sampling): samples obtained for $c = 3$ and $\alpha_i = 0.2$ for $i \in \{1, \ldots, 6\}$. Right panel (constrained sampling): samples obtained for $c = 3$ and $\bs{\alpha} = (0.24, 0.20, 0.18, 0.18, 1.42,0.27)^T$.}
    	\label{fig:sphere_scaled}
\end{figure}

We next consider $c = 3$, using the same two configurations for the concentration parameters. Samples can be seen in Fig.~\ref{fig:sphere_scaled}.
As expected, the generated samples are distributed beyond the convex hull defined by the given vertices (red dots). In addition, the distance between the Fr\'{e}chet mean and the base point substantially increases for unconstrained sampling (left panel in Fig.~\ref{fig:sphere_scaled}), while remaining small for constrained sampling (recall that the concentration parameters are not adjusted a posteriori since the calibration formulation is insensitive to multiplicative scaling). In fact, the later observation strongly depends on the positions of the vertices relative to the target mean: when the distance between the center of mass of the vertices and the target mean is sufficiently small, scaling on the tangent space generally leads to a small drift in the Fr\'{e}chet mean. On the contrary, a large distance implies a ``lack of symmetry'' in the definition of the sampling domain, in which case the mean is affected more significantly. 

\subsubsection{Sampling with Linear Constraints}
\begin{figure}[htbp]
	\centering
	\subfloat[2D view]{\includegraphics[width = 0.45\textwidth]{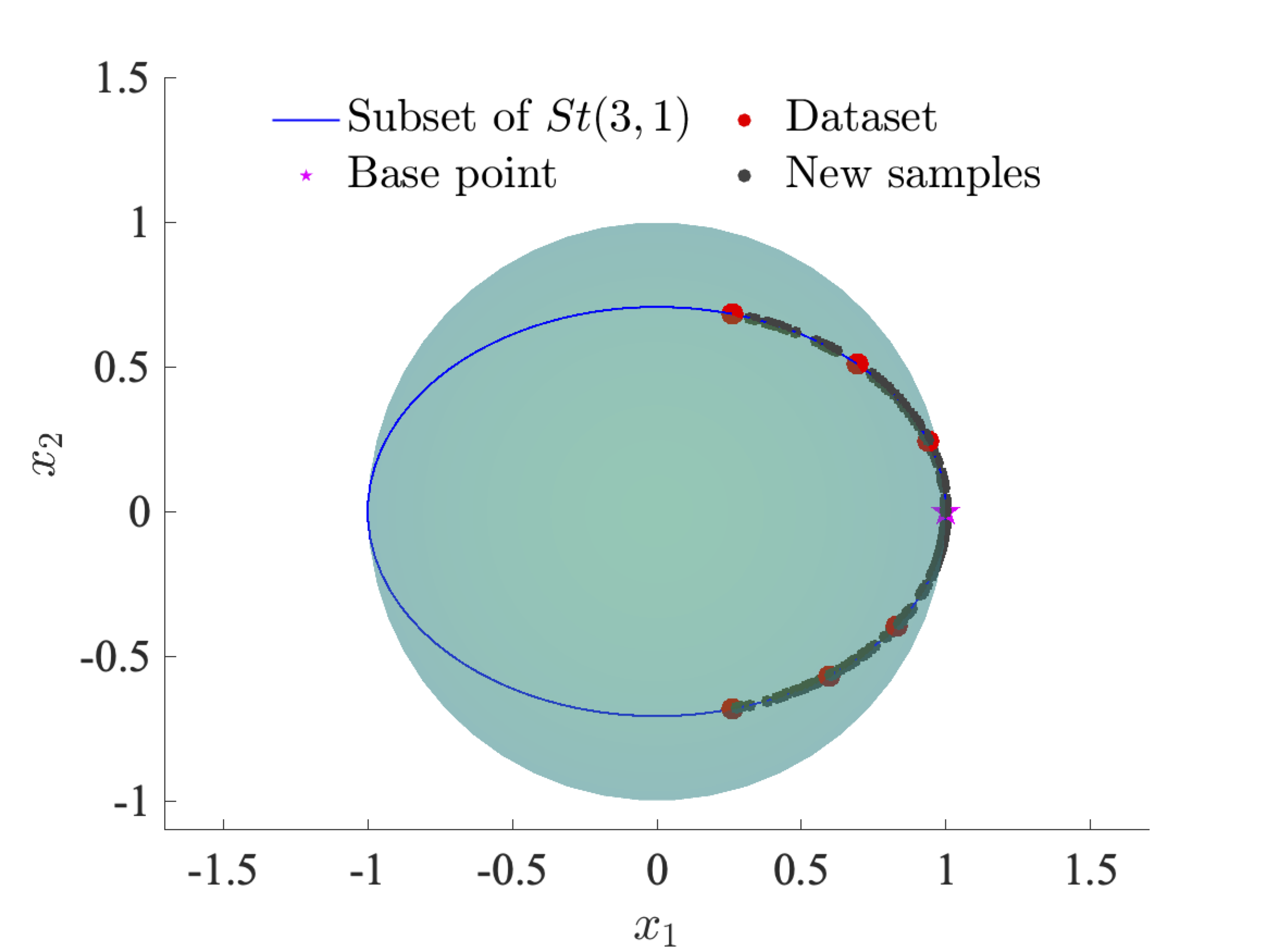}}\;
    	\subfloat[3D view]{\includegraphics[width = 0.45\textwidth]{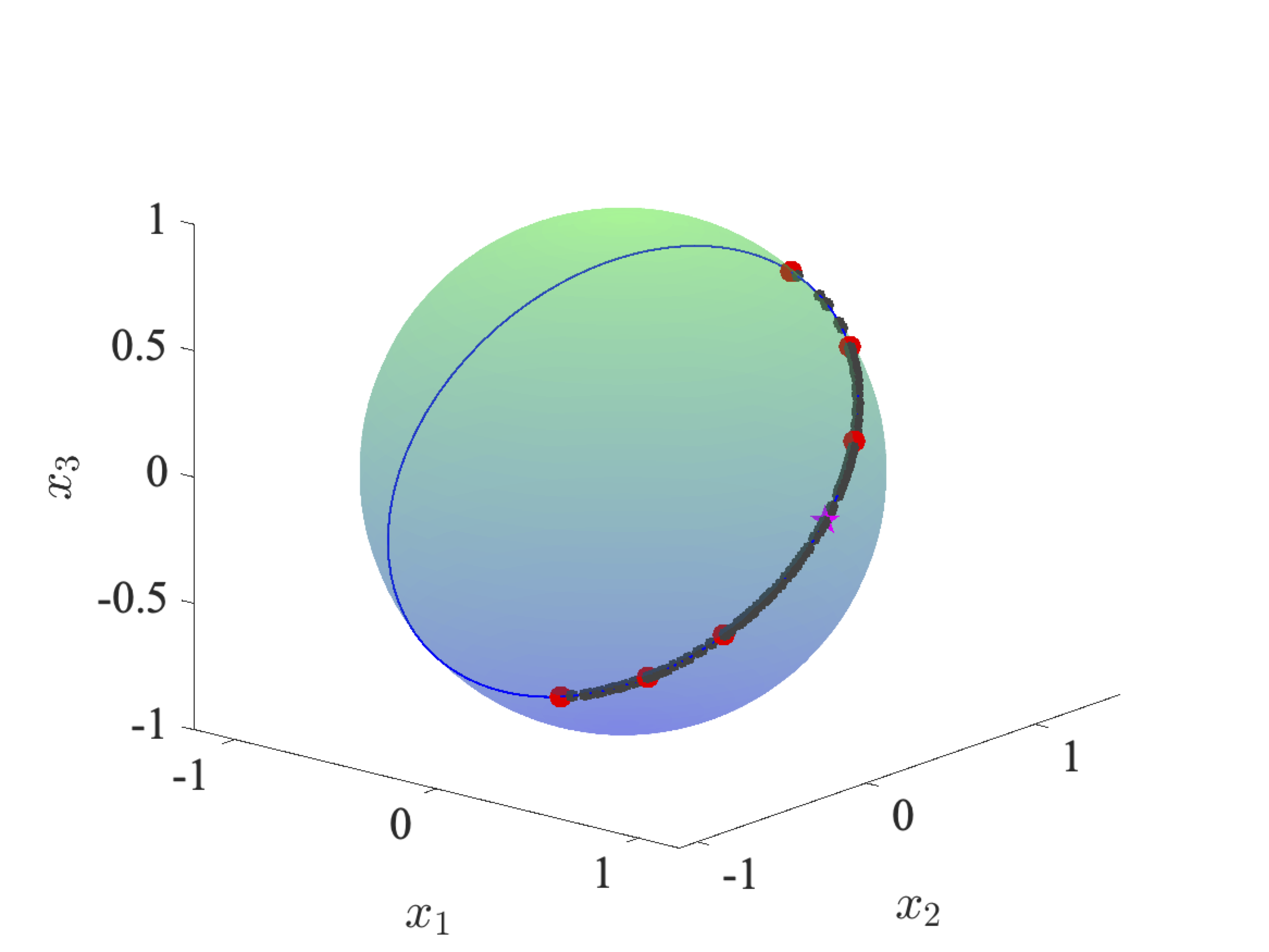}}\;
	\caption{Samples obtained for $c = 1$ and $\alpha_i = 0.2$ for $i \in \{1, \ldots, 6\}$ (no constraint on Fr\'{e}chet mean). Observe that all samples properly belong to $\mathbb{S}_{3, 1}$, owing to the use of the Riemannian projection and retraction operators.}
    	\label{fig:linear-constraint}
\end{figure}

We now turn to the proper integration of linear constraints on the half unit sphere. The only relevant case corresponds to $N_{\mathrm{CD}} = 1$, other values leading to overconstrained problems that are not appropriate in terms of sampling. Let $[B] = [b_{1},b_{2}, b_{3}]^{T} \in \mathbb{R}^{3 \times  1}$, with $\|[B]\| = 1$ (see Eq.~\eqref{eq:B-is-orthogonal}), and consider $[\Phi] = [\Phi_{1}, \Phi_{2}, \Phi_{3}]^{T} \in \mathbb{S}_{3 \times  1} \subset St(3,1)$. In this case, $\mathbb{S}_{3, 1}$ defines a semi-ellipse (as the intersection of the unit sphere and an arbitrary plane) embedded in $\mathbb{R}^3$. Without loss of generality, six points are randomly chosen through uniform sampling on a semi-ellipse ($m = 6$), and one base point is selected near the middle of the curve defined by these points. Concentration parameters are chosen as $\alpha_i=0.2$ for $i = 1, \ldots, 6$, and samples are shown in Fig.~\ref{fig:linear-constraint} for $c = 1$.
It is seen that all samples are distributed on the ellipse in a uniform manner (given the choice of the concentration parameters), which qualitatively shows that the linear constraint is properly satisfied.

\subsection{Single Graphene Sheet Subjected to Harmonic Excitation}
\label{sec:harmonic}
\subsubsection{System Description} \label{subsubsec:harmonic-system-desc}
\begin{figure}[htbp]
	\centering
    	\includegraphics[width=0.5\textwidth]{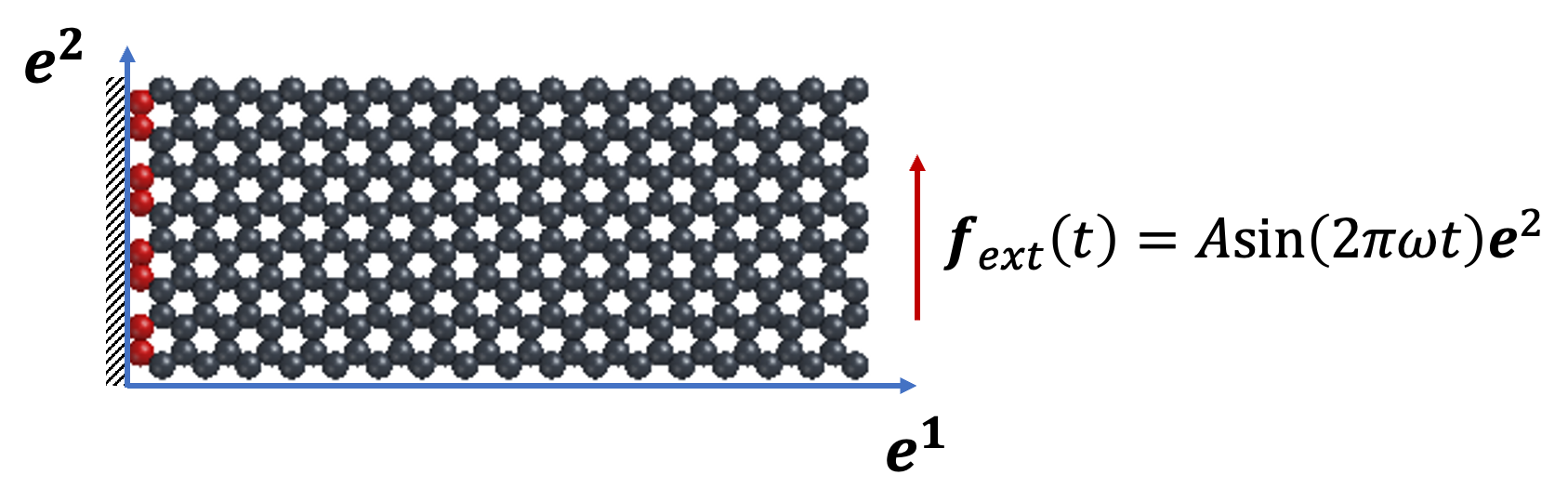}
    	\caption{Single-layer graphene sheet model composed of 272 atoms. A zero homogeneous Dirichlet boundary condition is applied on the left edge, and a harmonic excitation force is applied on the right edge.}
    	\label{fig:graphene_model}
\end{figure}

In this section, we apply the approach to molecular dynamics simulations on a single graphene sheet (in $\mathbb{R}^3$), composed of 272 carbon atoms, see Fig.~\ref{fig:graphene_model}. A zero Dirichlet boundary condition is applied on the left side of the structure (hence defining the linear constraints and matrix $[B]$), while a harmonic excitation force is applied on the right side according to
\begin{equation}
    \bs{f}_{ext}(t) = A\sin(2\pi\omega t) \bs{e}^2\,, \quad t \geq 0\,,
\end{equation}
with $A = 6$ [kcal$\cdot$mol$^{-1}\cdot\angstrom^{-1}$] and $\omega = 20\times10^9$ [rad/s]. 
Model uncertainties arise from the selection of the interatomic potentials governing the evolution of the system, and six different potentials commonly employed to model graphene-based systems are considered, namely AIREBO \cite{airebo}, BOP \cite{bop}, LCBOP \cite{lcbop}, Modified-Morse \cite{M-Morse}, REBO-2 \cite{reboII}, and Tersoff-2010 \cite{tersoff2010}. Atom displacement is chosen as the quantity of interest to study the influence of model-form uncertainties in the graphene system. Relaxation is performed through energy minimization before the external force $\bs{f}_{ext}$ is applied. Sampling is conducted in the microcanonical ensemble (NVE), with a time step set to 1 [fs] ($1\times10^{-15}$ [s]).

\subsubsection{Forward Simulations and Model Reduction}
\begin{figure}[htbp]
	\centering
    	\includegraphics[width = 0.9\textwidth]{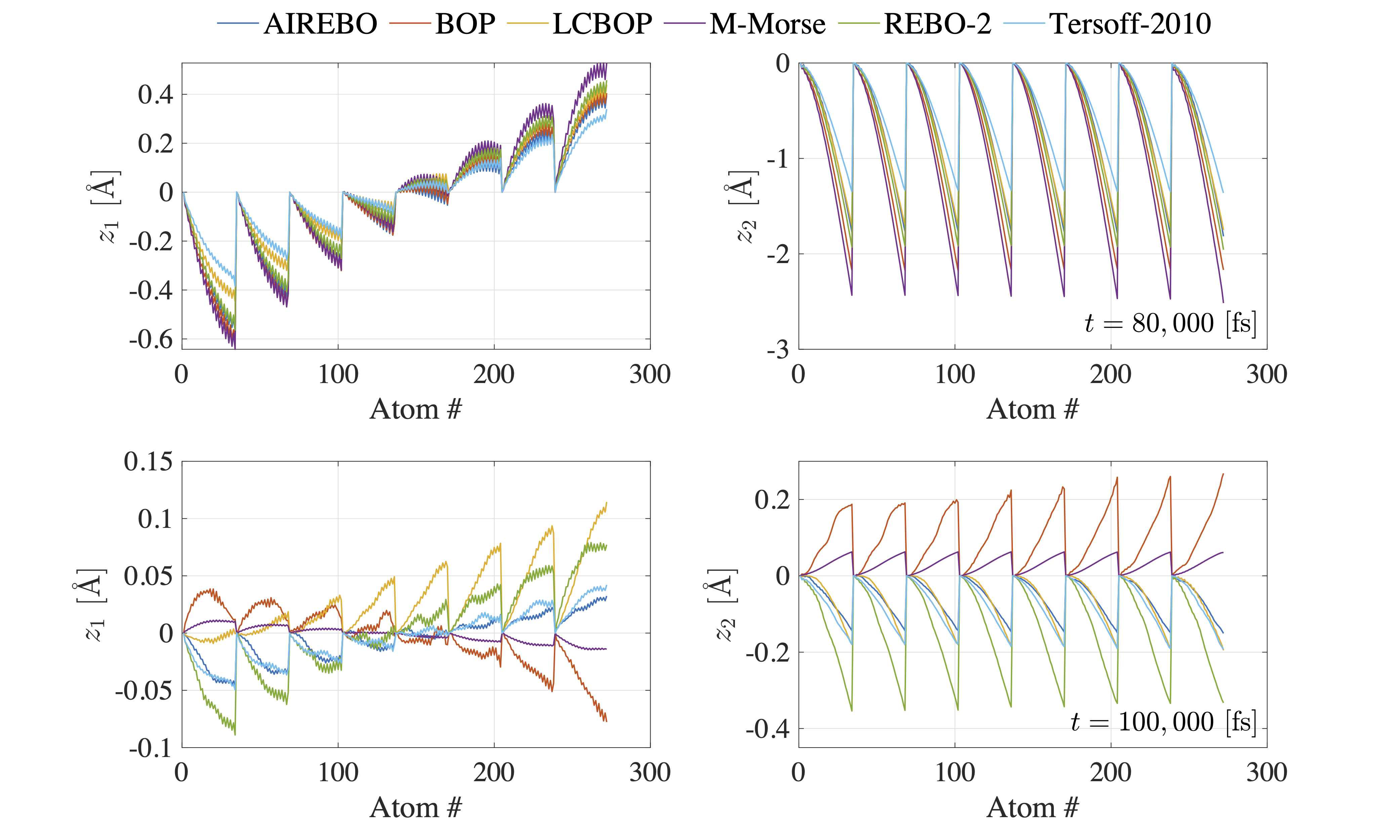}
    	\caption{Snapshots of horizontal and vertical displacements at $t=80$ [ps] (top row) and $100$ [ps] (bottom row), computed with six different interatomic potentials.}
    	\label{fig:traj6}
\end{figure}

The impact of model selection, viewed from the perspective of model uncertainties, is illustrated in Fig.~\ref{fig:traj6} where horizontal and vertical displacements for all atoms are displayed at $t = 80,000$ and $t = 100,000$ [fs], respectively, for the six considered potentials.
It is seen that the choice of the potential has a significant impact on the fine-scale dynamics of the graphene system, motivating the use of the proposed approach to quantify and propagate model-form uncertainties at relevant scales. 

\begin{figure}[htbp]
	\centering
	\includegraphics[width = 0.6\textwidth]{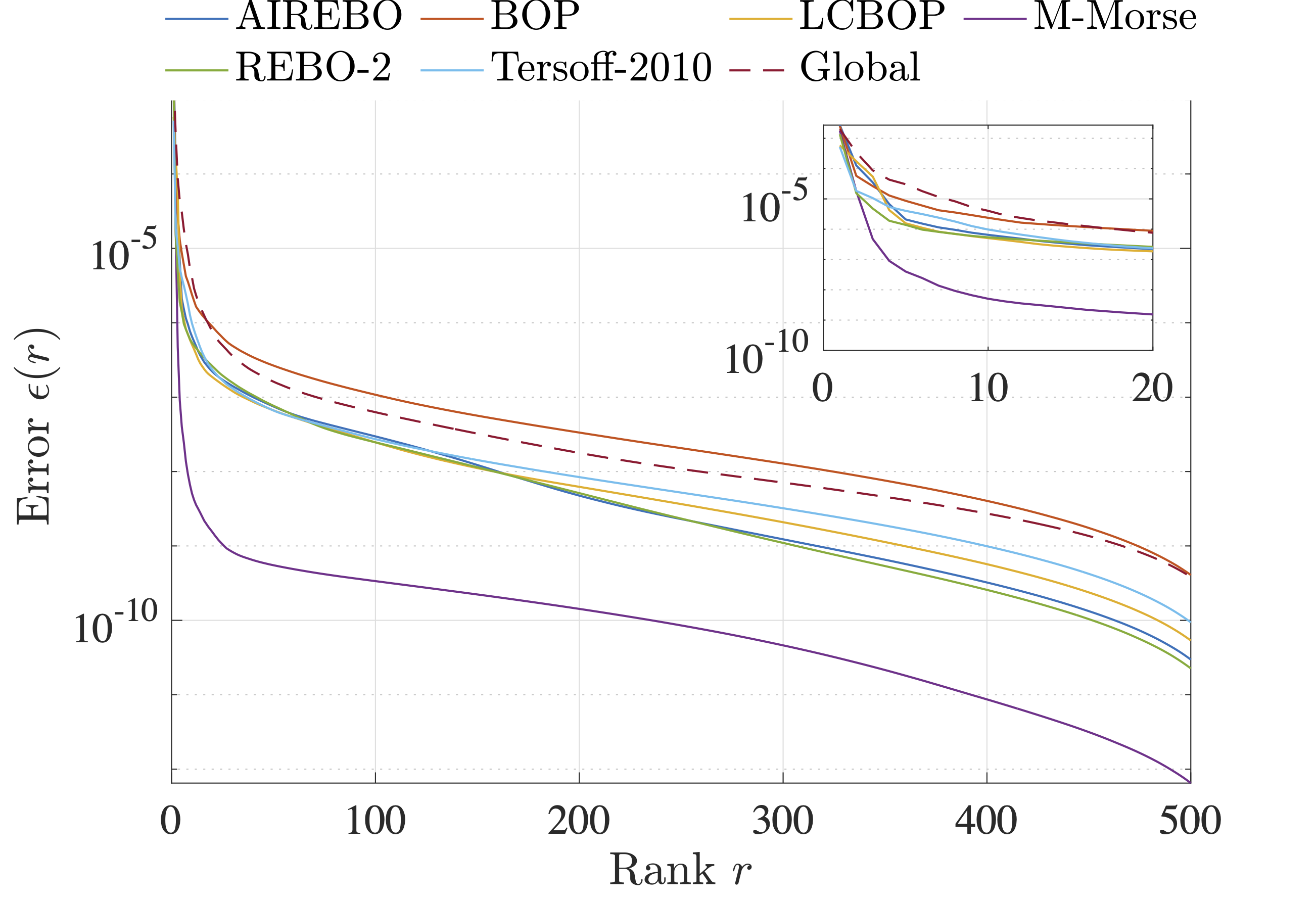}
	\caption{Graph of the $L^2$ error function for all scenarios.}
	\label{fig:err_function}
\end{figure}

The POD approach is next employed to construct the reduced-order bases $\{[W^{(i)}]\}_{i = 1}^{6}$ (the bases $\{[W^{(1)}],\ldots, [W^{(6)}]\}$ are associated with AIREBO, BOP, LCBOP, Modified-Morse, REBO-2, and Tersoff-2010 potentials, respectively). For each MD configuration (choice of interatomic potential), 1,000 displacement snapshots are collected with a time interval between consecutive snapshots set to 200 [fs] to promote independence (see Eq.~\ref{eq:displacement}). Recall that the global ROB $[W]$ is obtained by concatenating the displacement snapshots for all MD configurations. A singular value decomposition is used to identify the reduced dimension (taken as the minimum over all configurations) and the associated projection bases. Using a threshold of $1\times10^{-4}$, we identify the reduced dimension, $n = 5$ (see Fig.~\ref{fig:err_function}), and therefore consider stochastic modeling in $\mathbb{S}_{816,5} \subset St(816, 5)$. Note that the dimension of $St(816, 5)$ is: $816\times5-\frac{1}{2}\times 5\times (5+1) = 4,065$.

\subsubsection{Sampling Results}
\label{subsubsec-graphene-fine-scale-sampling}
The proposed stochastic model and sampling procedure are then deployed to generate reduced-order basis samples on $\mathbb{S}_{816,5}$. The scaling parameter $c$ is taken as 1, meaning that only Riemannian convex combinations are considered, and the concentration parameters are computed by solving the quadratic programming problem defined in Section \ref{eq:quadprogpb} (to reduce the distance between the global reduced-order basis $[W]$ and the center of mass of the generated samples):
\begin{equation}
    \bs{\alpha} = (0.494, 0.601,0.006,0.236,0.421,0.242)^T\,.
\end{equation}
In this example, the smallest eigenvalue of $[H]$ (in Eq.~\eqref{eq:quadprogpb}) is 0.482, which shows the well-posedness of the quadratic programming problem.

\begin{figure}[htbp]
	\centering
    	\includegraphics[width = 0.7\textwidth]{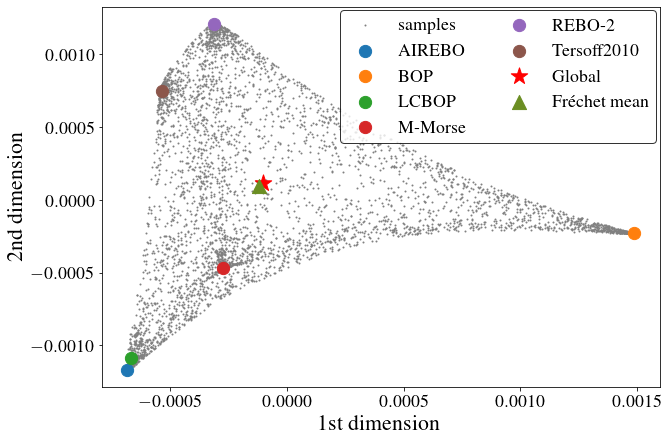}
    	\caption{First two dimensions after dimension reduction using the spectral embedding approach \cite{spectralembedding}. A total number of 5,000 samples are generated and shown, together with the original samples and the Fr\'{e}chet mean.}
    	\label{fig:spectral_embedding_harmonic}
\end{figure}

To visualize the dataset and the $2,500$ generated samples in a low-dimensional space (here, a two-dimensional space), several commonly used non-linear dimension reduction techniques were tested, including spectral embedding \cite{spectralembedding}, t-SNE \cite{tsne}, UMAP \cite{umap}, and PACMAP \cite{pacmap}. It was found through extensive numerical experiments that the spectral embedding approach typically delivers representations that can be interpreted more easily, in terms of structure; see Fig.~\ref{fig:spectral_embedding_harmonic}. 
This figure illustrates the fact that all samples are generated inside the convex hull defined by the dataset, with curved edges owing to the use of the non-linear reduction technique (data compression). It is also observed that the Fr\'{e}chet mean computed with the samples appears close to the global reduced-order basis $[W]$, demonstrating the efficiency of the proposed methodology to identify the concentration parameters based on a Fr\'{e}chet mean constraint in a molecular dynamics setting. 

Such visualization techniques and results should, however, be handled and interpreted with caution, due to the reduction process. A comparative study about such representations is beyond the scope of this work. Their use in the context of reduced-order modeling for dynamical systems, in particular, is an interesting topic that is left for future work.

\subsubsection{Forward Propagation of Model Uncertainties}
In this section, model-form uncertainties are propagated through Monte Carlo simulations with the stochastic reduced-order model corresponding to the graphene system subjected to harmonic excitation. This step necessitates the selection of the interatomic potential used after pullback in the physical space (to evaluate forces). Two strategies can be pursued at this point. In a first scenario, the same potential is used for all simulations, regardless of the reduced-order basis sample. This potential may be chosen, in practice, as the one minimizing the distance to the mean behavior. A second strategy consists in performing selection for each sample of the reduced-order basis, retaining the potential (in the physical space) that is the closest to the sample under consideration in the reduced-order space. In this case, the potential can be identified by computing relative distances between the sample $[\Phi(\theta_j)]$ and all elements in the dataset (i.e., $[W^{(1)}], \ldots, [W^{(m)}]$), using the canonical metric, or by leveraging the definition through a convex combination. Specifically, let $\mathcal{I}_j$, with $1 \leq \mathcal{I}_j \leq m$, be the integer such that $p_{\mathcal{I}_j}(\theta_j) = \max\{p_1(\theta_j), \ldots, p_m(\theta_j)\}$. The sample $[\Phi(\theta_j)]$ is then located closer to $[W^{(\mathcal{I}_j)}]$, so that the $(\mathcal{I}_j)$th potential may be used in the physical space.

\begin{figure}[htbp]
	\centering
	\subfloat[$t = 22$ {[ps]} (BOP)]{\includegraphics[width = 0.45\textwidth]{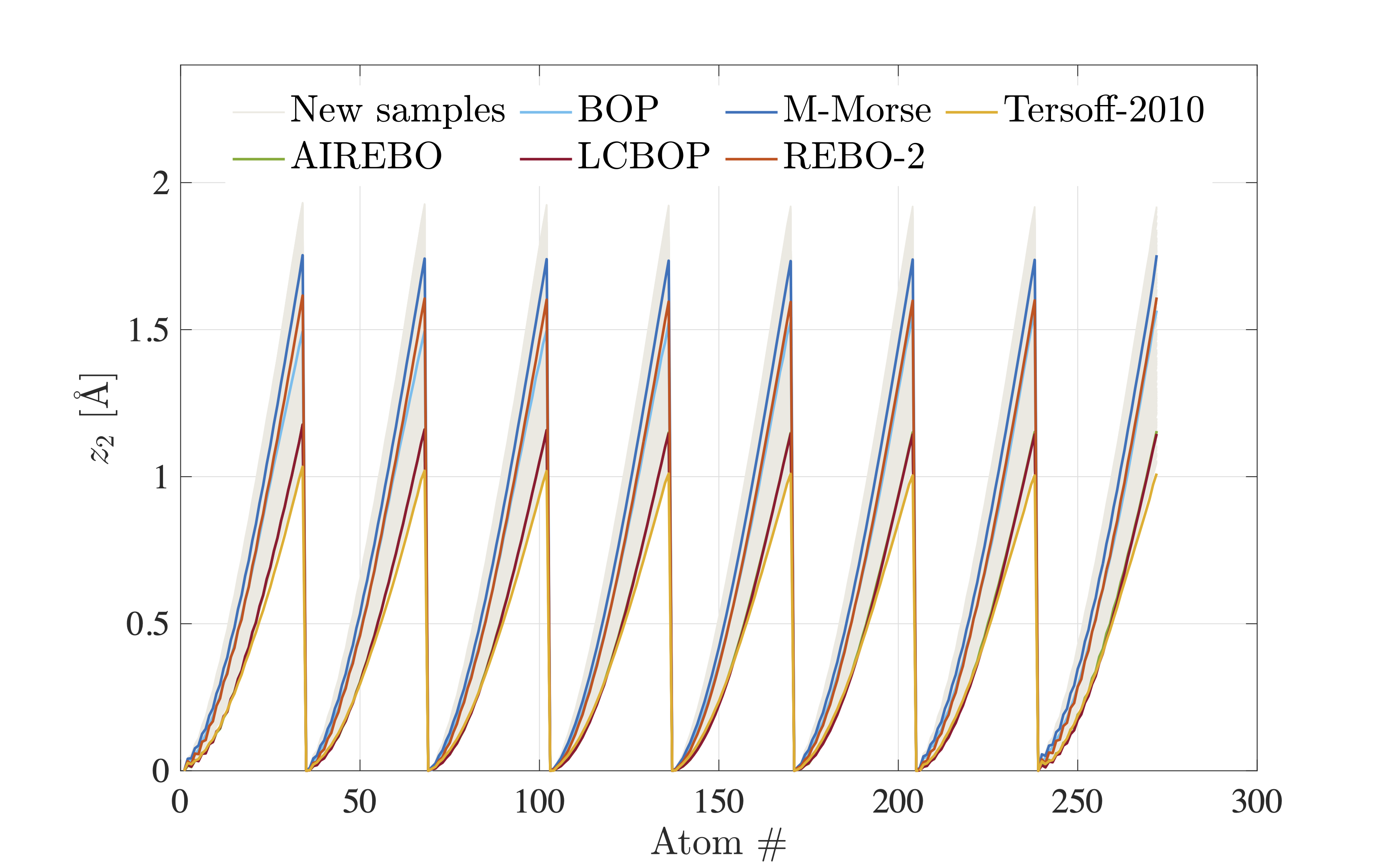}}\;
    	\subfloat[$t = 25$ {[ps]} (BOP)]{\includegraphics[width = 0.45\textwidth]{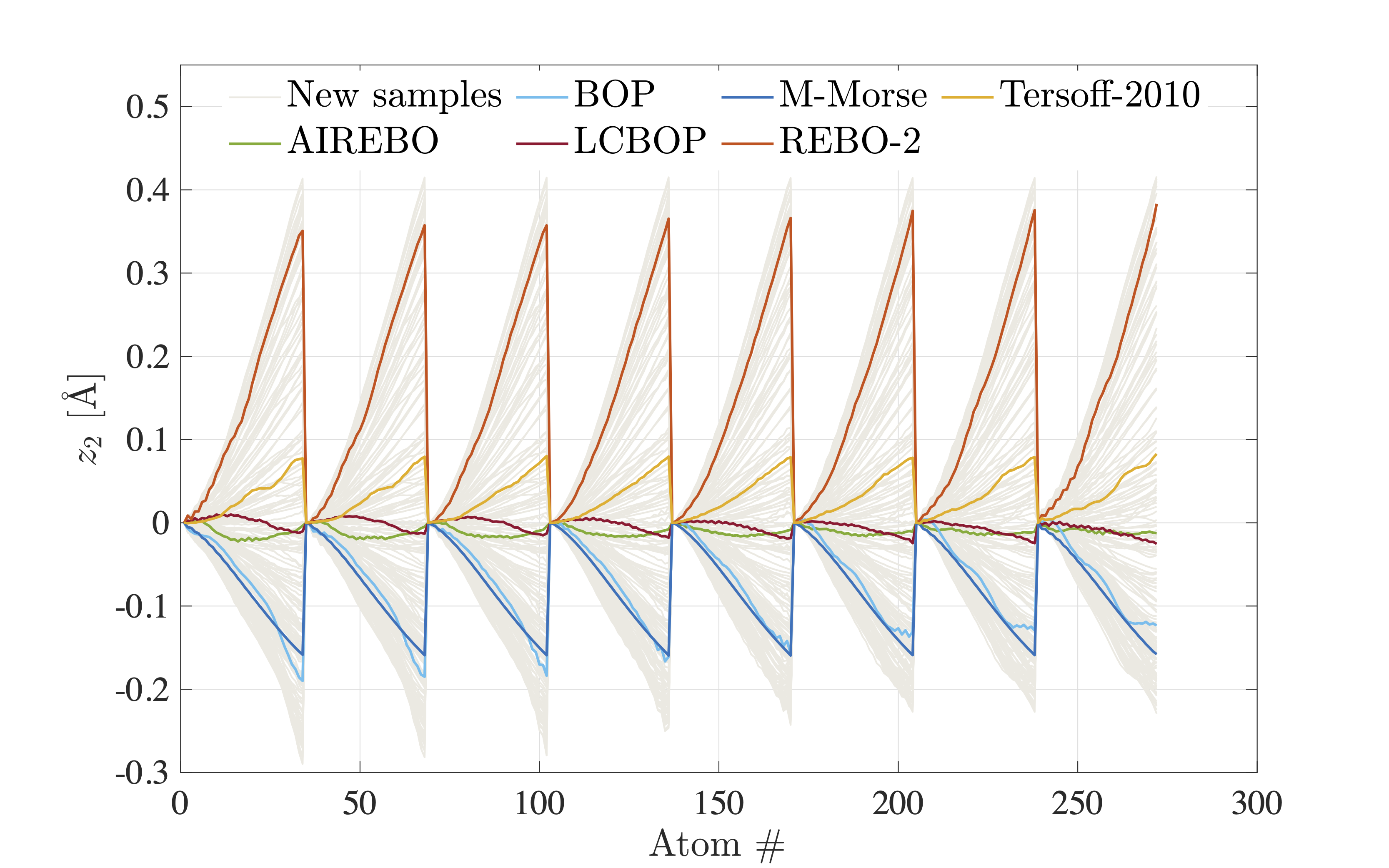}}\;\\
    	\subfloat[$t = 22$ {[ps]} (sample-based selection)]{\includegraphics[width = 0.45\textwidth]{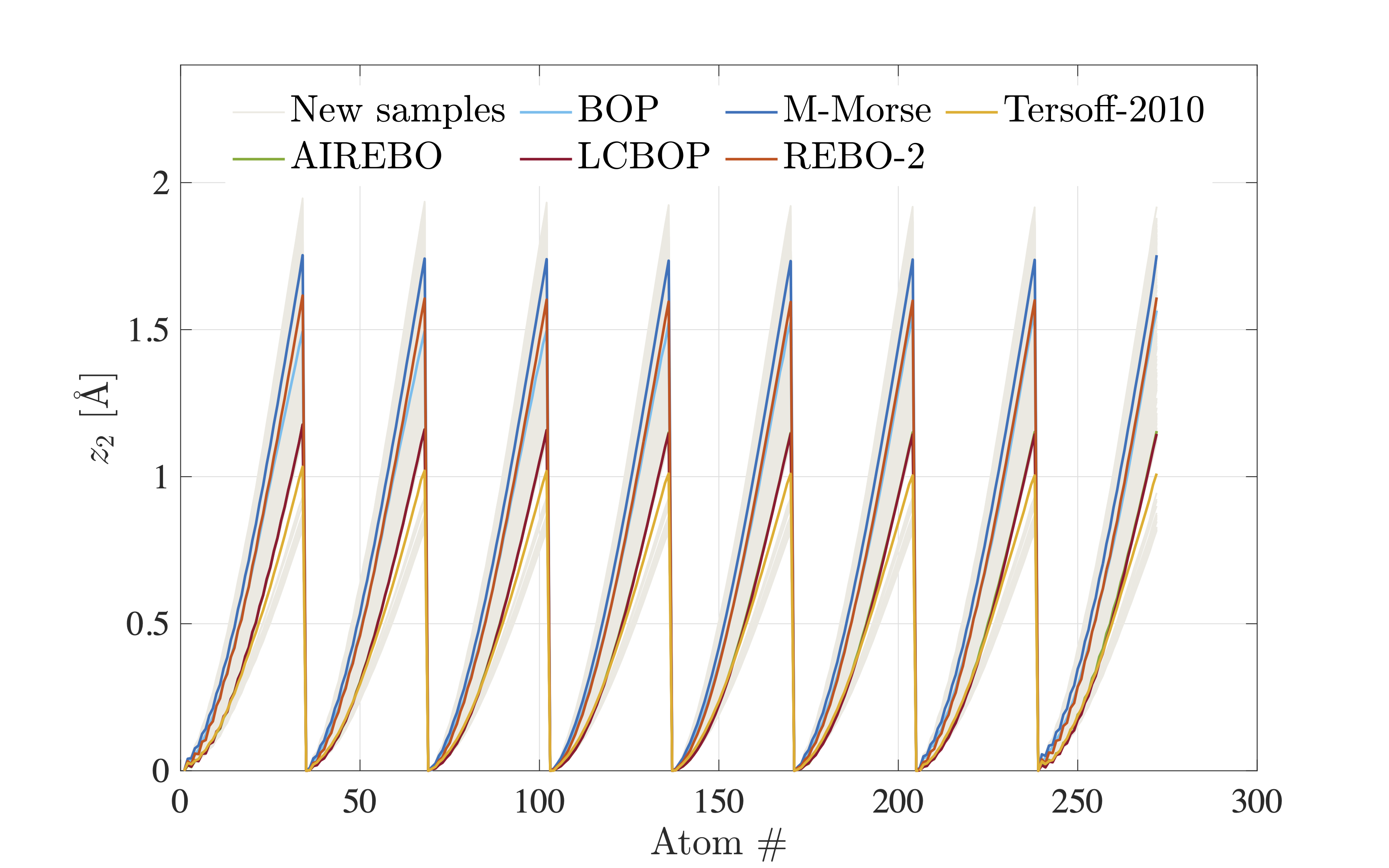}}\;
    	\subfloat[$t = 25$ {[ps]} (sample-based selection)]{\includegraphics[width = 0.45\textwidth]{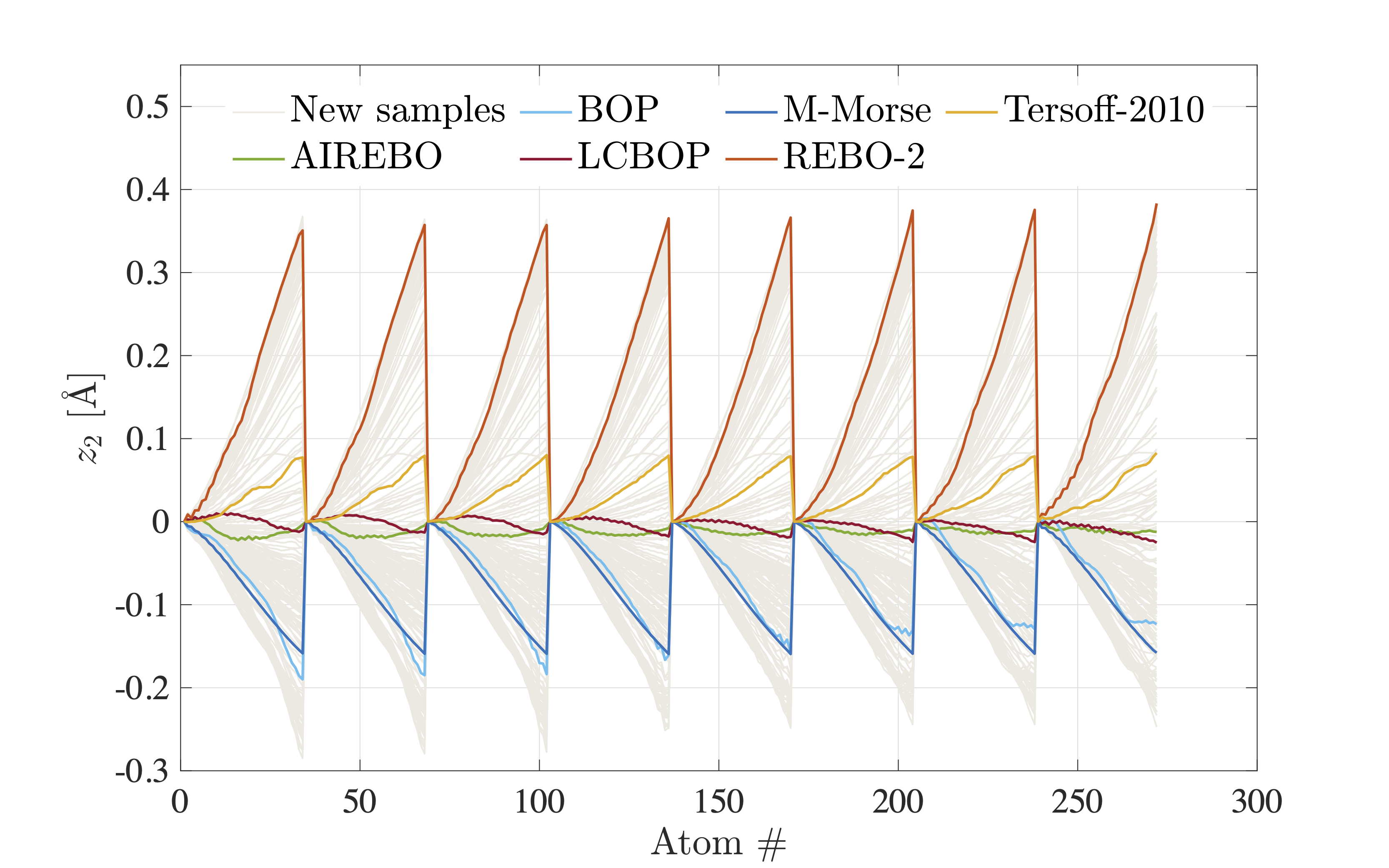}}\;
	\caption{Trajectories predicted by the reduced-order MD simulations using 200 ROB samples (grey solid lines), together with full-order MD simulation results (colored solid lines) at simulation times $t = 22$ [ps] (left panels) and $t = 25$ [ps] (right panels), respectively. In the top row, the BOP potential is used for all the reduced-order MD simulations, while sample-based selection is carried out in the bottom row.}
     	\label{fig:traj_harmonic}
\end{figure}

Results obtained with the above two strategies are shown in Fig.~\ref{fig:traj_harmonic}. In this example, the BOP potential \cite{bop} is used in the first strategy, and 200 samples are generated using the values given in Section \ref{subsubsec-graphene-fine-scale-sampling}. Snapshots of the vertical displacement (along $\bs{e}^2$) are displayed at $t = 22$ and $t = 25$ [ps]. Trajectories computed for the 200 samples of the stochastic reduced-order basis are shown, together with the trajectories corresponding to full-order MD simulations with all six reference potentials.
It is seen that both strategies yield fairly similar results in terms of spread. The domain defined by the set of full-order simulations is properly captured by the sampled trajectories, which indicates that model uncertainty has been successfully encoded into stochastic modelling process. It is worth mentioning that the zero Dirichlet boundary condition is also preserved across all samples and full-order models.

\begin{figure}[htbp]
	\centering
	\subfloat[$t=22$ {[ps]} (BOP)]{\includegraphics[width = 0.45\textwidth]{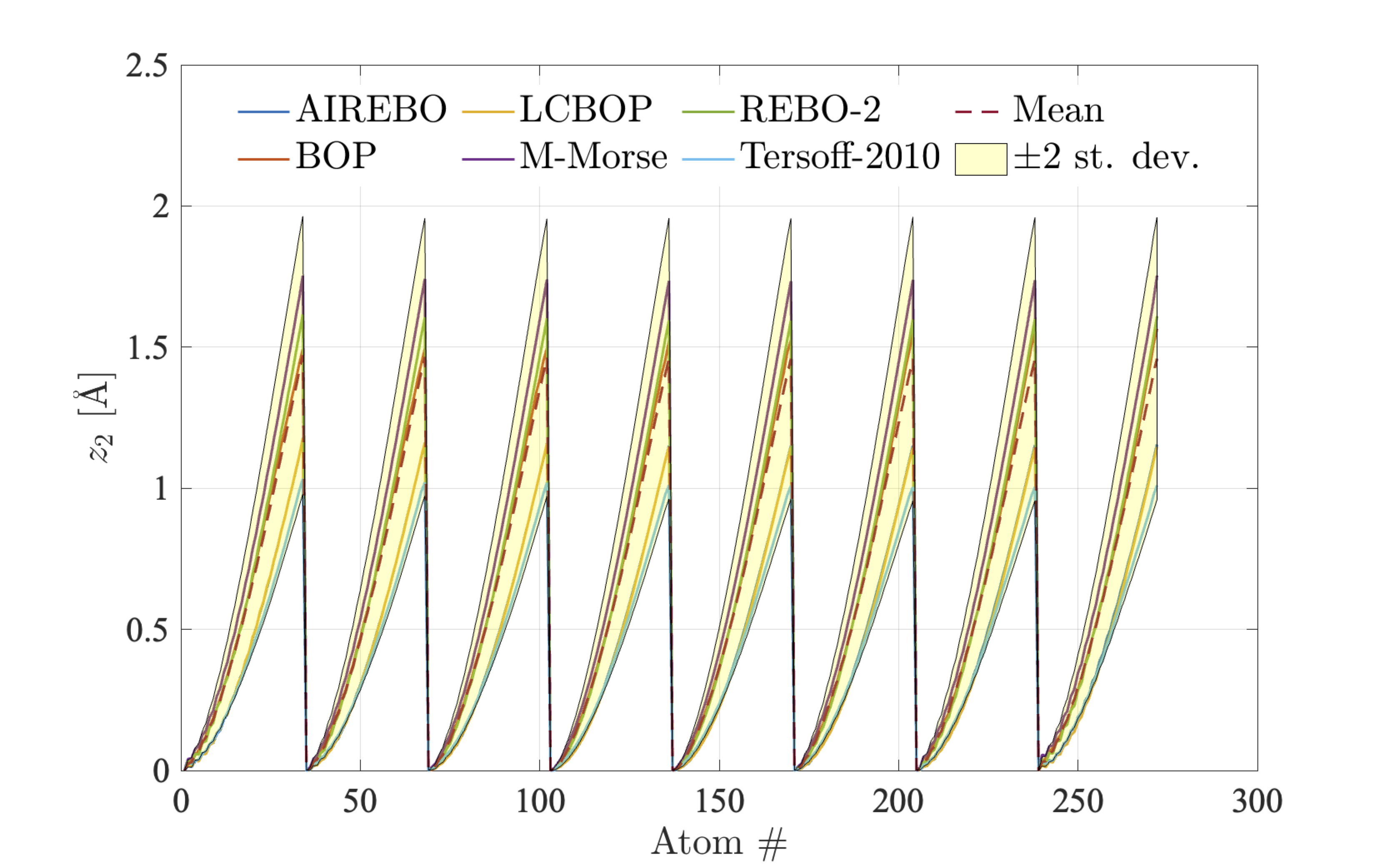}}\;
    	\subfloat[$t=25$ {[ps]} (BOP)]{\includegraphics[width = 0.45\textwidth]{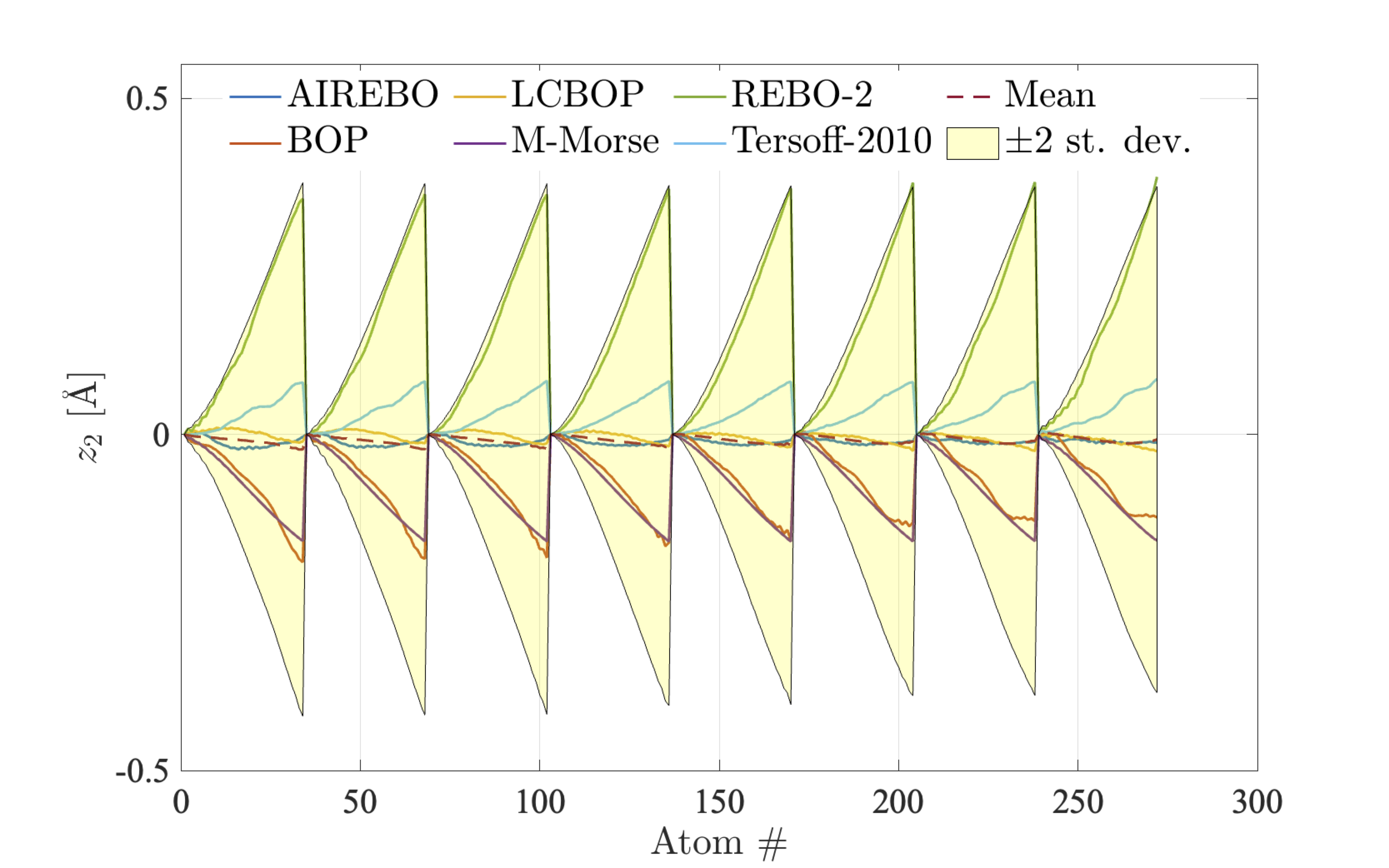}}\;\\
    	\subfloat[$t=22$ {[ps]} (sample-based selection)]{\includegraphics[width = 0.45\textwidth]{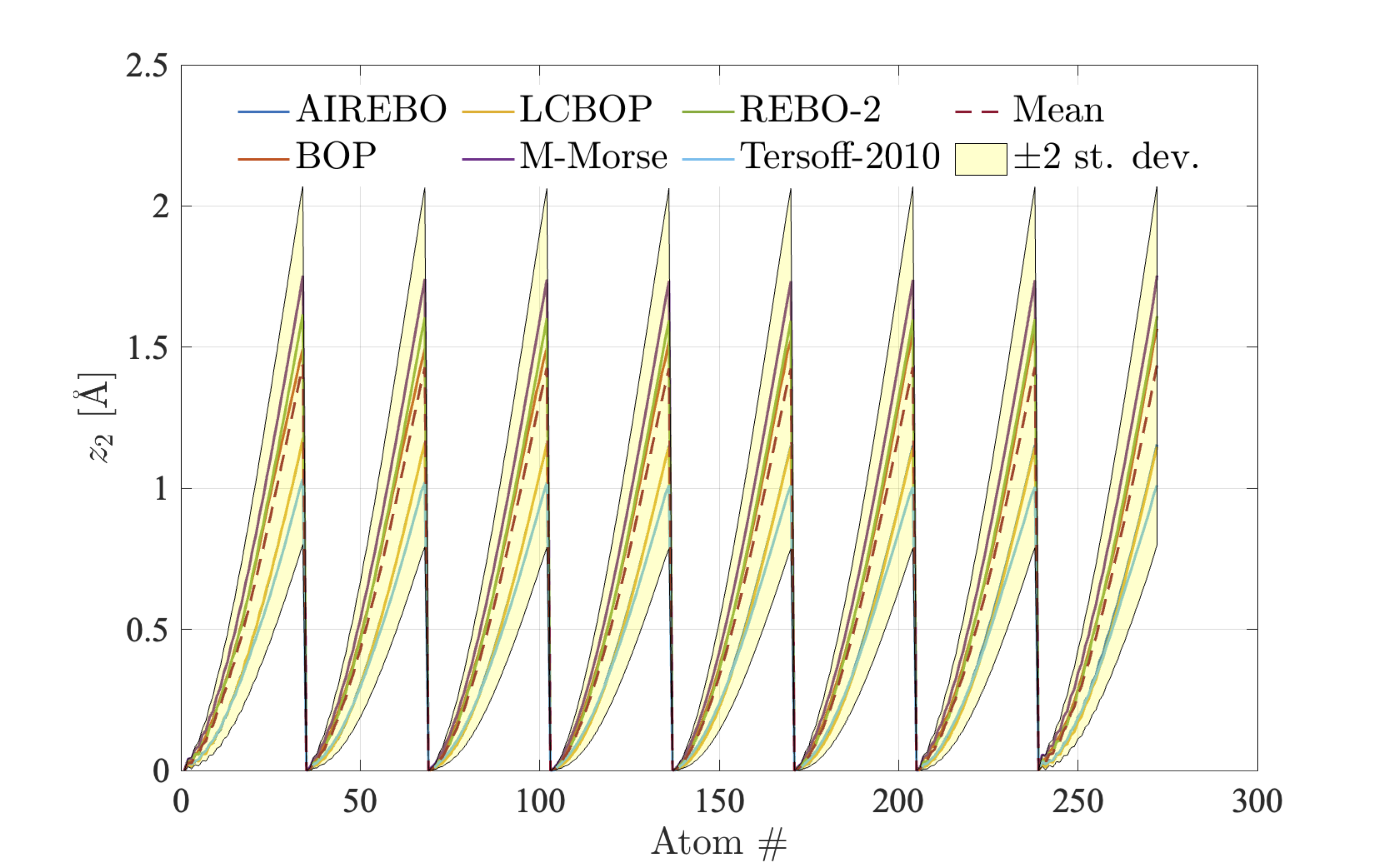}}\;
    	\subfloat[$t=25$ {[ps]} (sample-based selection)]{\includegraphics[width = 0.45\textwidth]{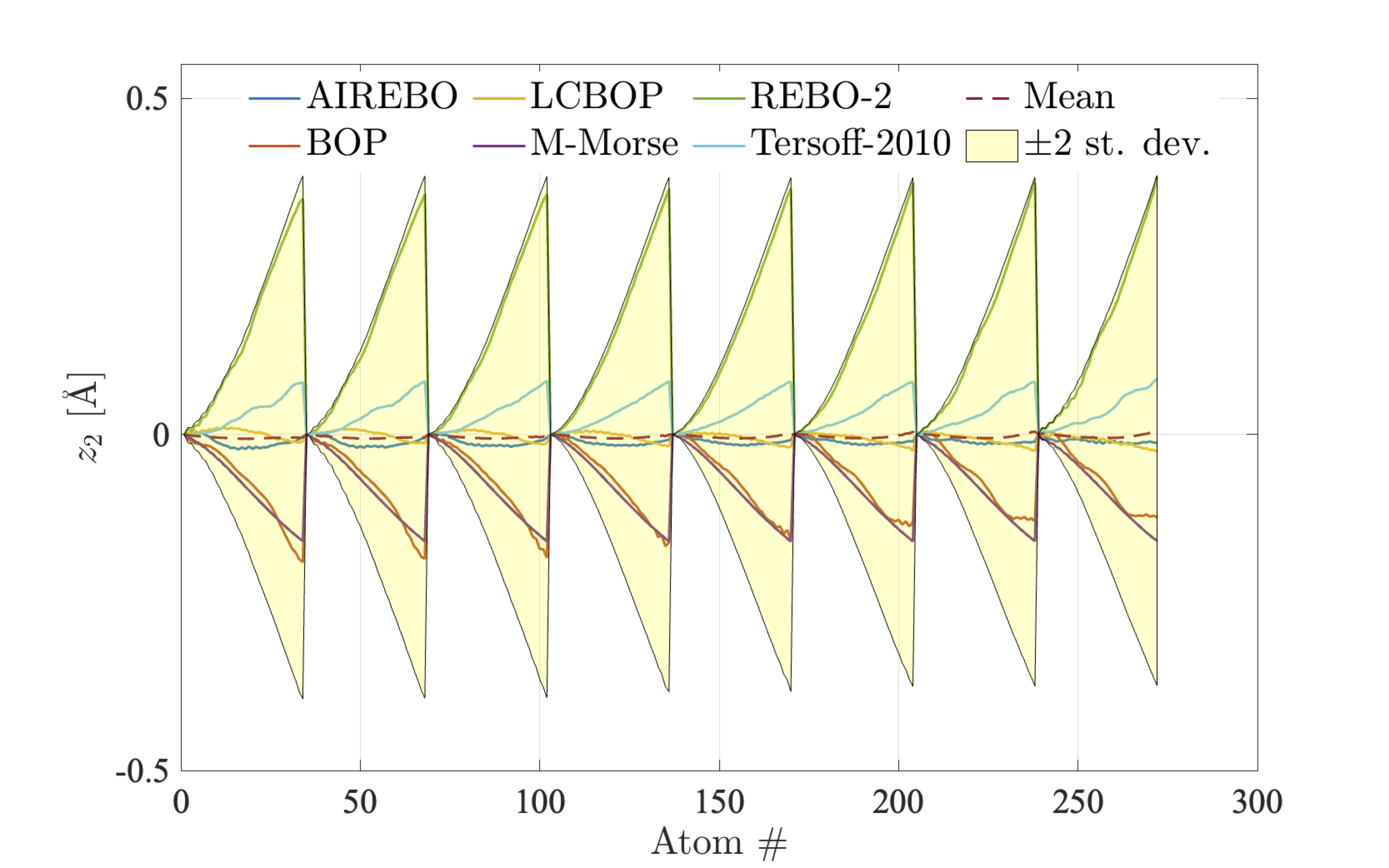}}\;
	\caption{Mean trajectory and confidence interval for the vertical displacement of all atoms at $t = 22$ [ps] (left panels) and $t = 25$ [ps] (right panels). In the top row, the Bop potential is used for all the reduced-order MD simulations, while sample-based selection is carried out in the bottom row.}
     	\label{fig:std_harmonic}
\end{figure}

\begin{figure}[htbp]
	\centering
     	\subfloat[$t=22$ {[ps]} (BOP)]{\includegraphics[width = 0.45\textwidth]{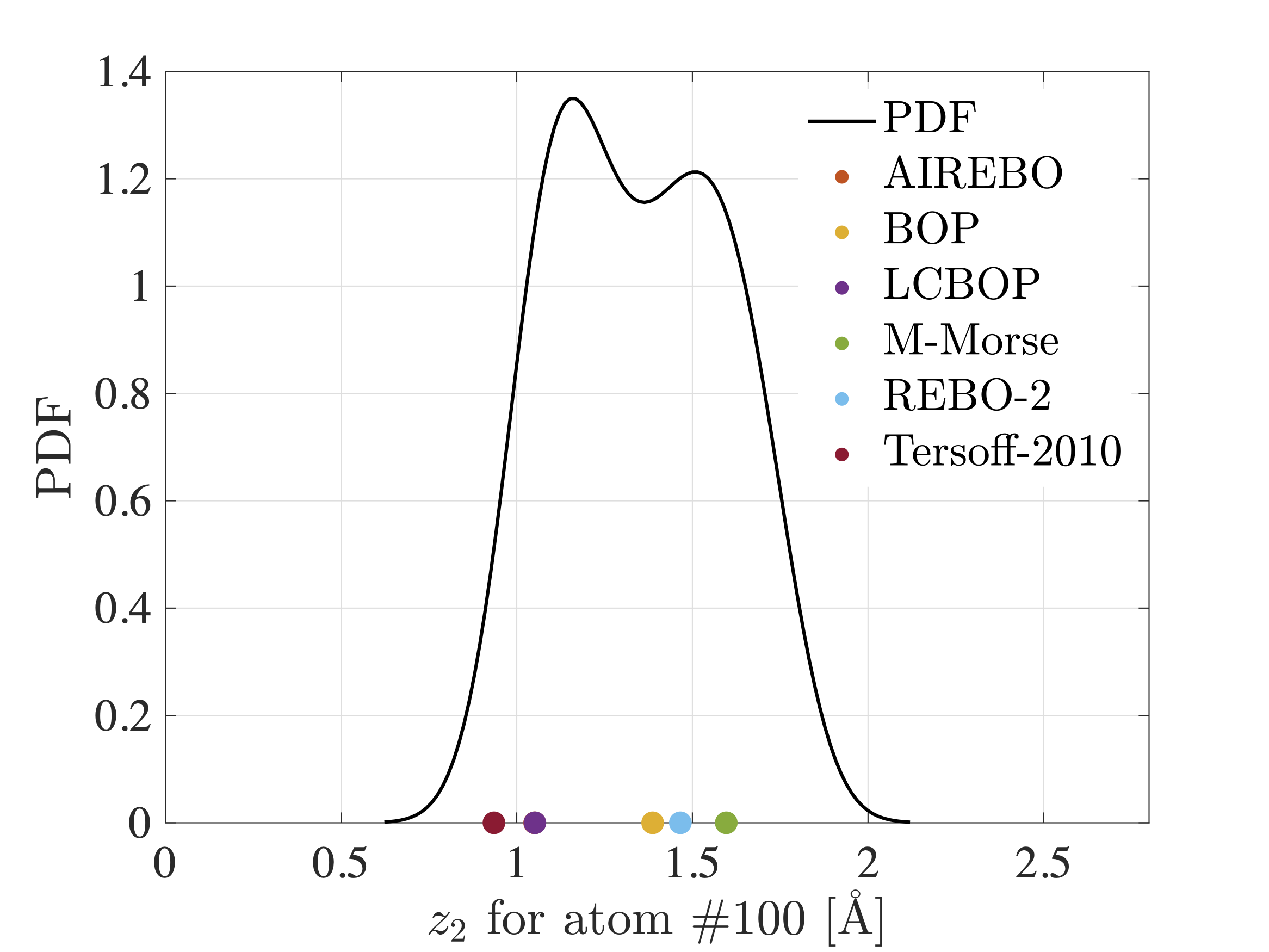}}\;
    	\subfloat[$t=25$ {[ps]} (BOP)]{\includegraphics[width = 0.45\textwidth]{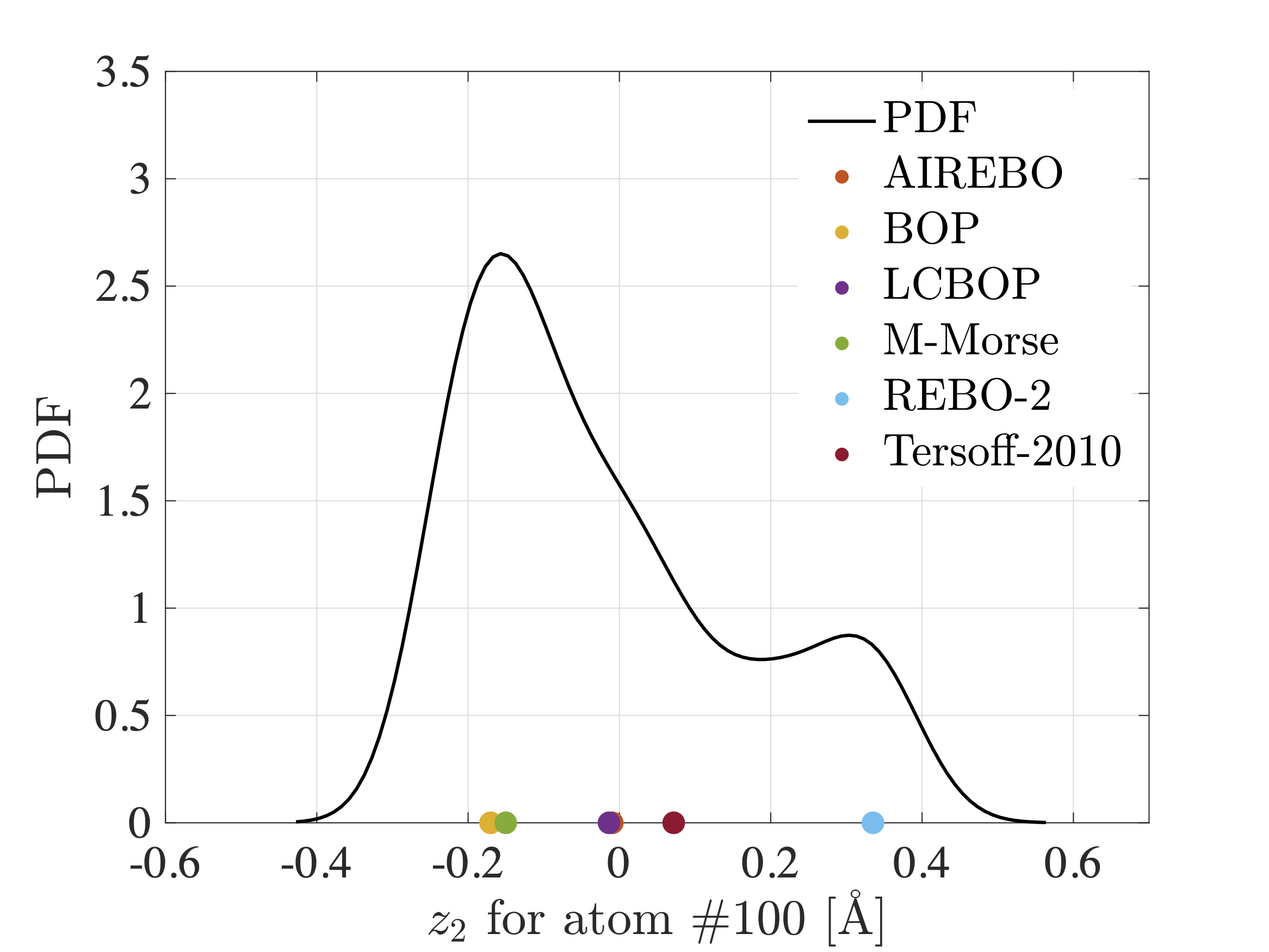}}\;\\
    	\subfloat[$t=22$ {[ps]} (sample-based selection)]{\includegraphics[width = 0.45\textwidth]{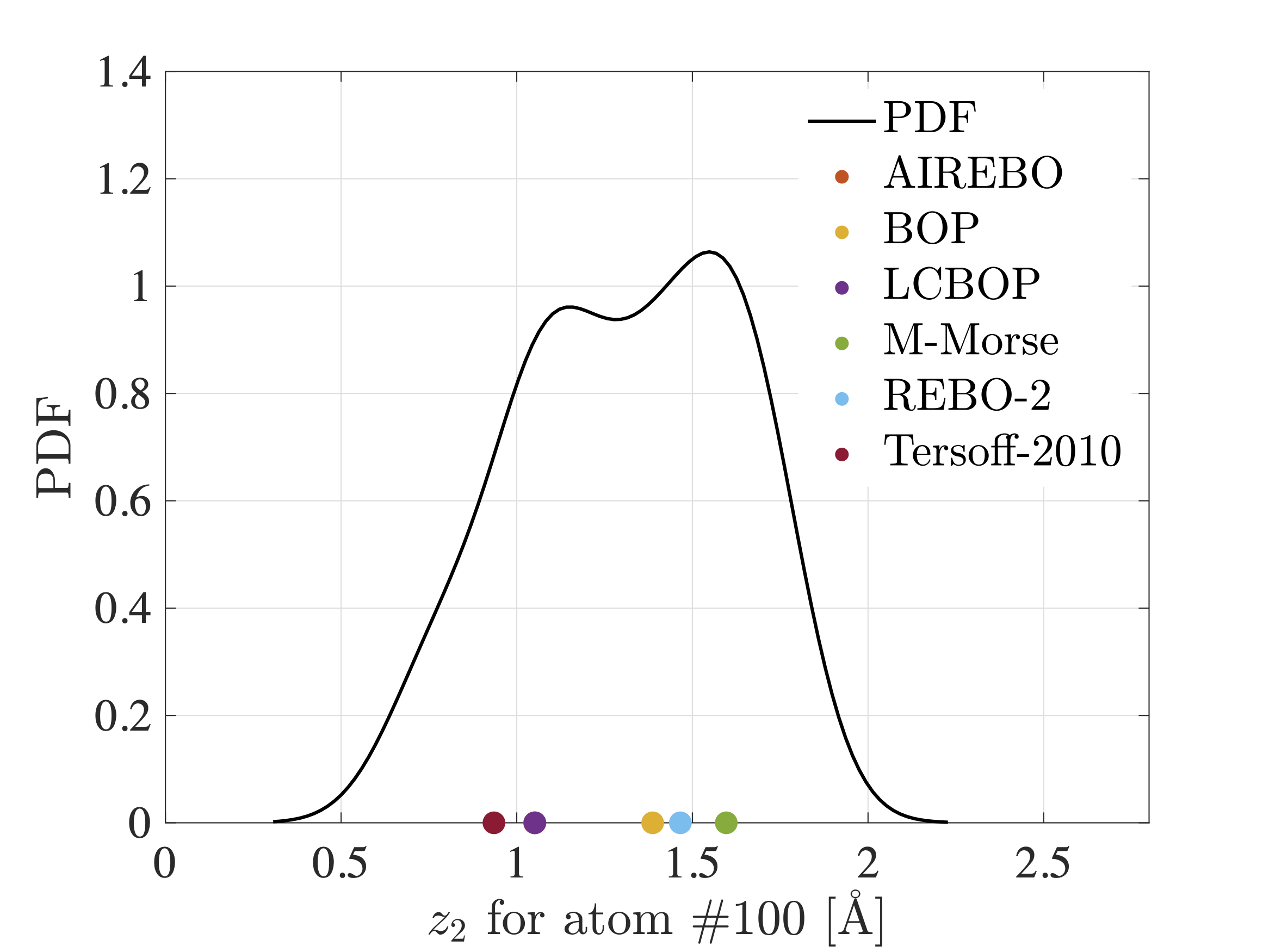}}\;
    	\subfloat[$t=25$ {[ps]} (sample-based selection)]{\includegraphics[width = 0.45\textwidth]{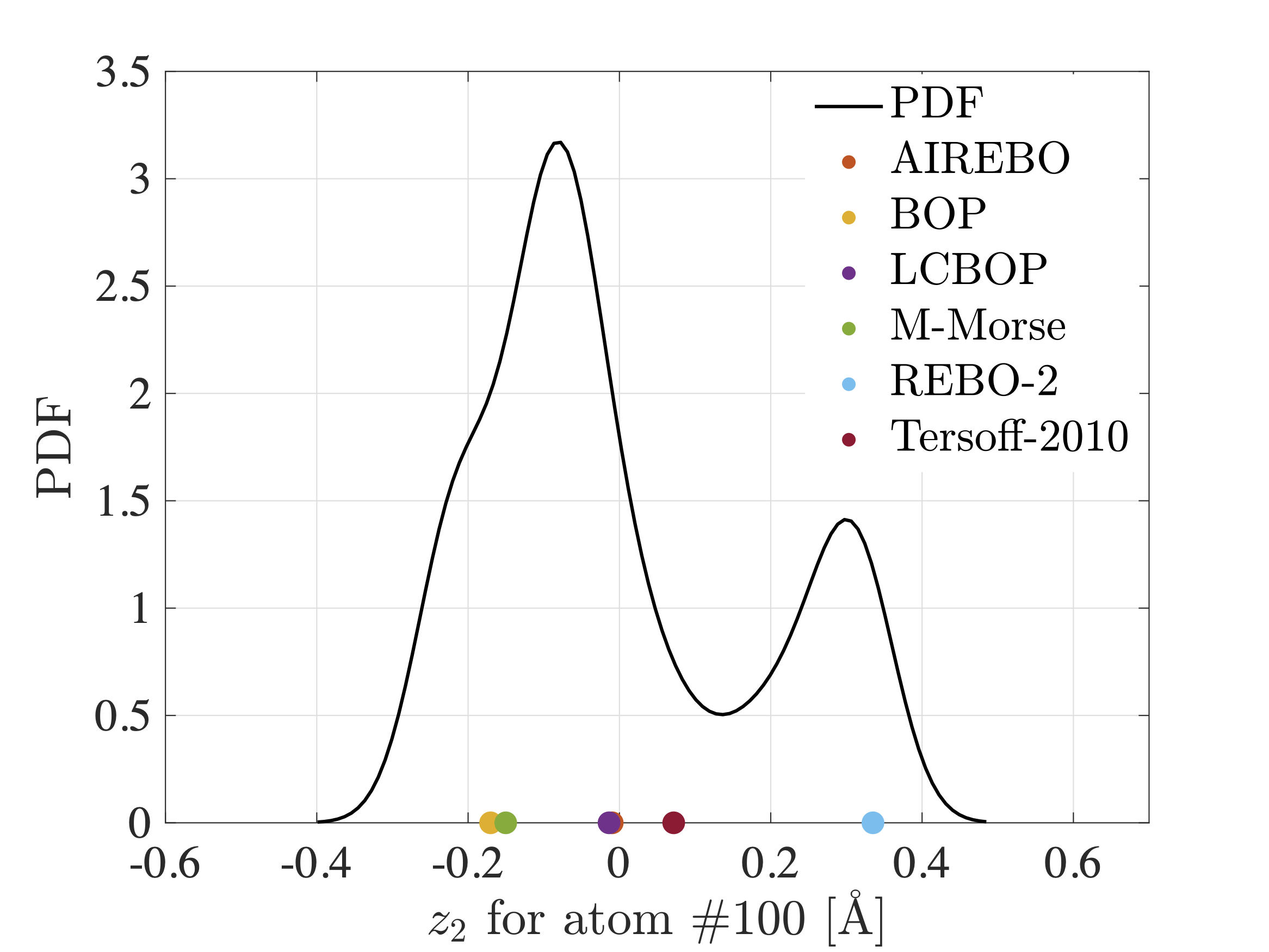}}\;
	\caption{Estimated probability density function (pdf) for the vertical displacement of atom \#100 at $t = 22$ [ps] (left panels) and $t = 25$ [ps] (right panels). Full-order MD simulation results are also reported for the sake of comparison. Note that the markers associated with the Airebo and Lcbop potentials are quite close to one another. In the top row, the Bop potential is used for all the reduced-order MD simulations, while sample-based selection is carried out in the bottom row.}
     	\label{fig:pdf_harmonic}
\end{figure}

Fine-scale uncertainties generated by model error can also be observed using confidence intervals and probability distributions. The mean trajectories and confidence intervals (with a range set to plus-minus two standard deviations) are shown in Fig.~\ref{fig:std_harmonic} for the two selection strategies.
The estimated probability density functions for the vertical displacement of atom \#100 at $t = 22$ and $t = 25$ [ps] are also shown in Fig.~\ref{fig:pdf_harmonic}. 
These results show that the choice of the selection strategies does not significantly impact predictions. It should however be noticed that sample-based selection allows to better differentiate between contributions in the dataset; see, \textit{e.g.}, the peak observed for the Bop potential in the bottom-right figure (as compared to the top-right figure) in Fig.~\ref{fig:pdf_harmonic}. Moreover, this strategy does not generate additional computational cost, and does not rely on \textit{a priori} selection. For these reasons, the sample-based selection approach will be used in subsequent calculations and in particular, in the multiscale results presented in Section \ref{sec:multiscale}.

\subsection{Single Graphene Sheet Subjected to Tension}
\label{sec:multiscale}
\subsubsection{System Description}
\begin{figure}[htbp]
	\centering
    	\includegraphics[trim = 0cm 6cm 0cm 5.5cm, clip, width = 0.7\textwidth]{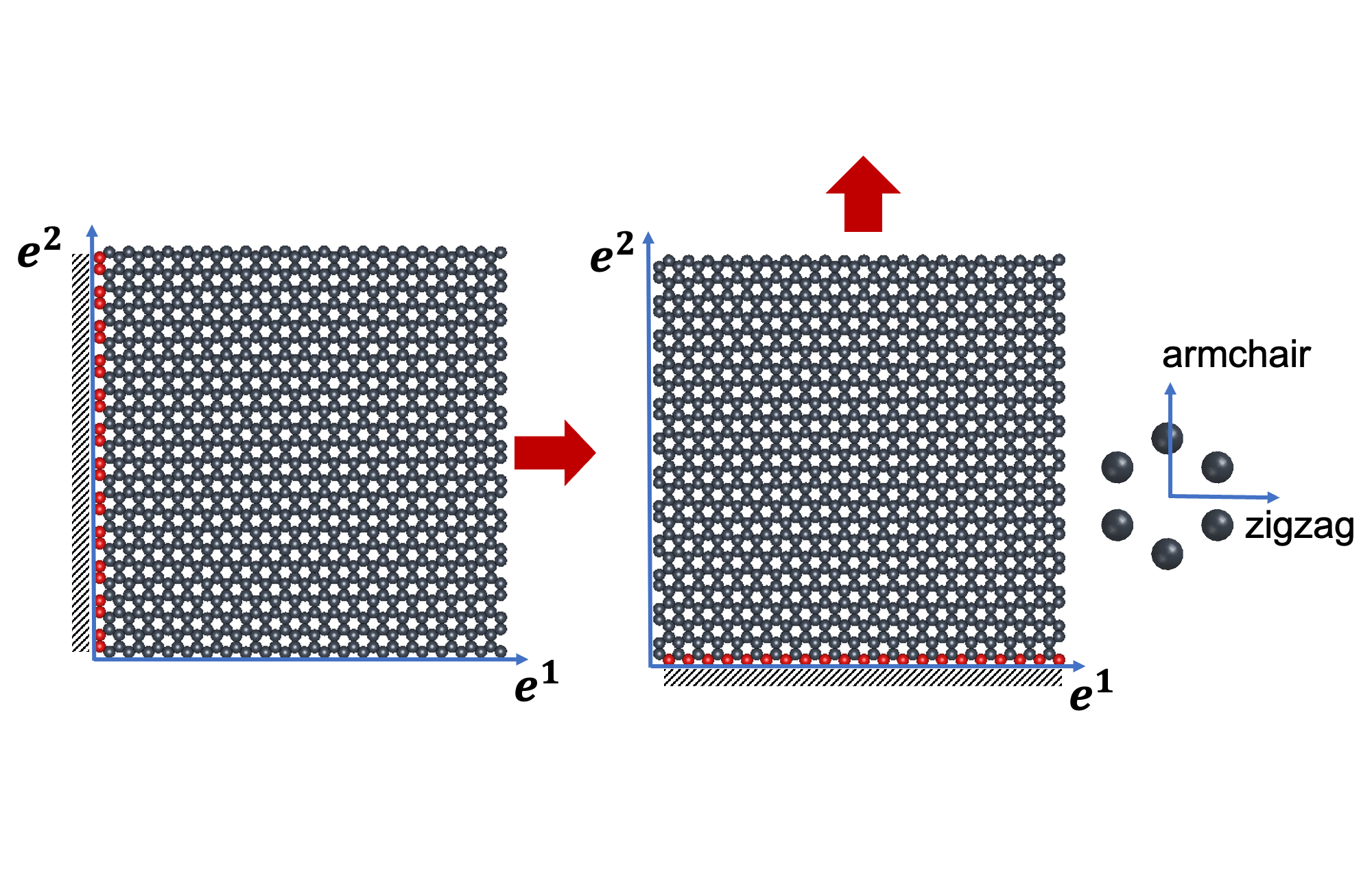}
    \caption{Single-layer graphene sheet with 1,008 carbon atoms. Tensile test is conducted in zigzag (left subfigure) and armchair (middle subfigure) directions. A zero Dirichlet boundary condition is applied on one edge of the sheet and a driving force is applied to the atoms located on the opposite edge.}
    \label{fig:graphene_1008atoms}
\end{figure}

We finally model and quantify the impact of model-form uncertainties in both fine- and coarse-scale predictions on a graphene sheet under tension. The graphene sheet is composed of 1,008 carbon atoms with an overall in-plane size of $50.17 \times 49.74$ [$\angstrom$], a size that is large enough to produce size-independent coarse-scale tensile test results (see \cite{size}). The carbon bond length is selected as $1.418$ [$\angstrom$], in accordance with \cite{bond}. The tensile test is conducted in both zigzag and armchair directions, as shown in Fig.~\ref{fig:graphene_1008atoms}. 
In each virtual tensile test, a zero Dirichlet boundary condition is applied to the atoms located on one edge of the sheet (\textit{e.g.}, to the atoms satisfying $x_1 = 0$ for the zigzag direction) and a stretching force is applied to the atoms on the opposite edge. The time step is set to $1$ [fs], with a total simulation time of $20$ [ps] for both directions. Load stepping is used with increments prescribed every 100 time steps in order to ensure proper relaxation. The loads are specifically defined such that the largest engineering strain rate is equal to 0.22 for the tensile test in the zigzag direction, and to 0.19 in the armchair direction.

\subsubsection{Deterministic Forward Simulations}
The uncertainty resulting from the selection of the interatomic potential is evaluated using full-order MD simulations and the AIREBO, BOP, LCBOP, REBO-2, and Tersoff-2010 potentials (see Section \ref{subsubsec:harmonic-system-desc}). 

Two quantities of interest are considered. First, a fine-scale characterization is obtained by analyzing the displacements along the $\bs{e}^1$ and $\bs{e}^2$ directions. Second, the impact on a coarse-scale property, namely the apparent strain energy, is illustrated. For the sake of comparison, results obtained with the continuum-mechanics-based model presented in \cite{prl} are also reported as complementary reference. The relationships between the strain energy and the engineering strain, denoted by $u$ and $\epsilon$ respectively, are given by
\begin{equation}
    \begin{aligned}
         u_{zz} & = \frac{1}{2}\frac{E}{1-\nu^2}\epsilon^2 + \frac{1}{6}C_{111}\epsilon^3\,,\\
         u_{ac} & =  \frac{1}{2}\frac{E}{1-\nu^2}\epsilon^2 + \frac{1}{6}C_{222}\epsilon^3\,,
    \end{aligned}
\end{equation}
where the subscripts ``$zz$'' and ``$ac$'' refer to the zigzag and armchair directions, $E=312$ [N/m] denotes the Young's modulus, $\nu=0.31$ is the Poisson ratio, $C_{111}=-1689.2$ [N/m] and $C_{222}=-1487.7$ [N/m] are the elastic constants \cite{prl}. 

\begin{figure}[htbp]
	\centering
    	\includegraphics[width = 0.9\textwidth]{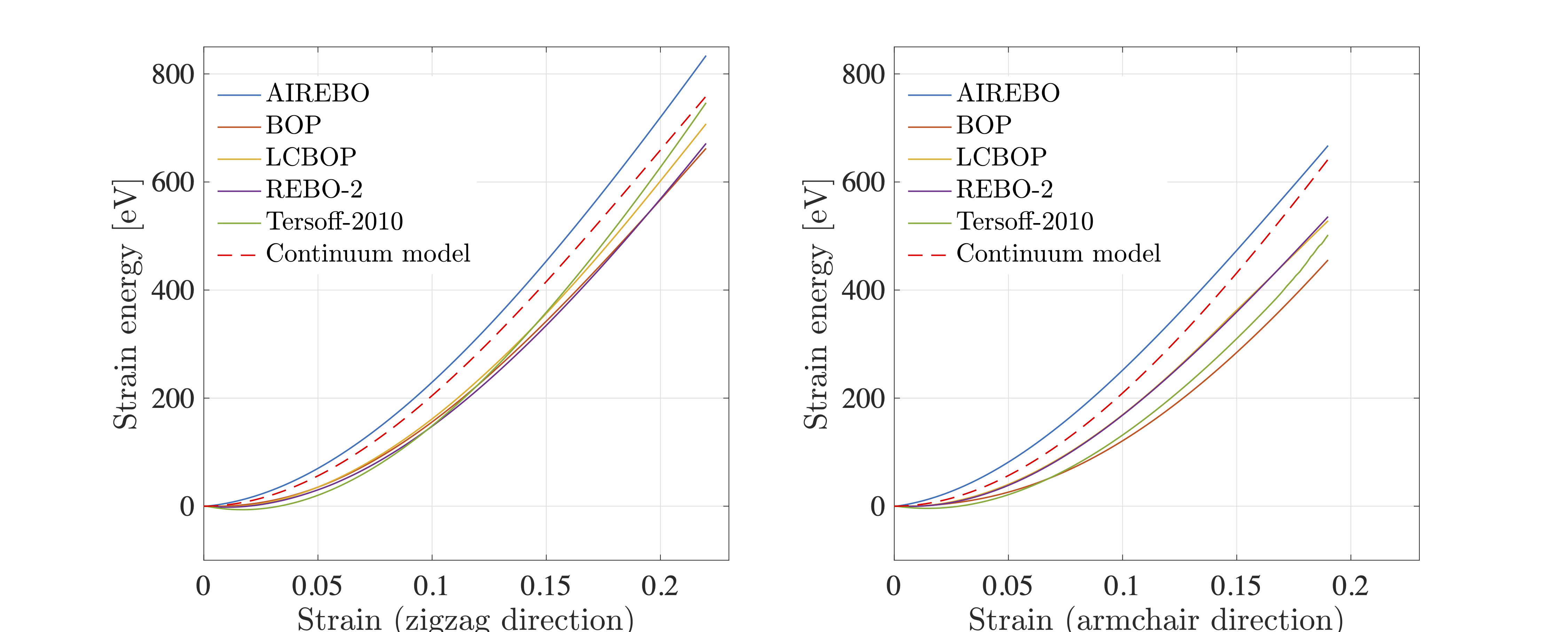}
    	\caption{Evolution of the strain energy (in [eV]) as a function of applied strain, for all potentials and the continuum-mechanics-based model, in the zigzag (left) and armchair (right) direction.}
    	\label{fig:strain-energy5}
\end{figure}

The evolution of the strain energy in both directions and for all potentials is shown in Fig.~\ref{fig:strain-energy5}.
Very large variations induced by model-form uncertainties can be observed. Specifically, the discrepancy in strain energy for a 15\%-strain is 26.3\% in the zigzag direction, and 40.6\% for the armchair direction. Such discrepancies may generate substantial fluctuations when propagated through multiscale approaches \cite{fish}, which  underpins the need to properly capture such variability and perform uncertainty quantification within MD simulations.  

\subsubsection{Stochastic Modeling and Forward Propagation of Model Uncertainties}\label{subsubsec:sto-model-tension}
In order to apply the proposed modeling framework, a total number of 500 snapshots are collected for each tensile direction and all interatomic potentials. The five reduced-order bases $\{[W^{(1)}],\dots, [W^{(5)}]\}$ (associated with AIREBO, BOP, LCBOP, REBO-2, and Tersoff-2010 potentials, respectively), together with the global reduced-order basis $[W]$, are then calculated using the POD approach. Selecting $n = 10$ modes leads to a truncation error that is less than $10^{-4}$ for all candidates, so we consider sampling on $\mathbb{S}_{3024,10} \subset St(3024,10)$. 

Model-form uncertainties can then be propagated using the modeling strategy summarized in Section \ref{subsec:summary}, combined with a Monte Carlo approach. The concentration parameters are determined by solving the quadratic programming problem given by Eq.~\eqref{eq:quadprogpb} to ensure that Fr\'{e}chet mean of the generated samples are close to the global ROB, which is the base point to define the tangent space. These coefficients are found to be 
\begin{equation}\label{eq:alpha-zz}
    \bs{\alpha}_{zz} = (0.89,0.80,0.45,0.78,1.08)^T
\end{equation}
for the zigzag direction, and
\begin{equation}
    \bs{\alpha}_{ac} = (0.82,0.80,1.14,0.62,0.62)^T
\end{equation}
for the armchair direction.

\begin{figure}[htbp]
	\centering
	\subfloat[Fine-scale displacement]{\includegraphics[width = 0.45\textwidth]{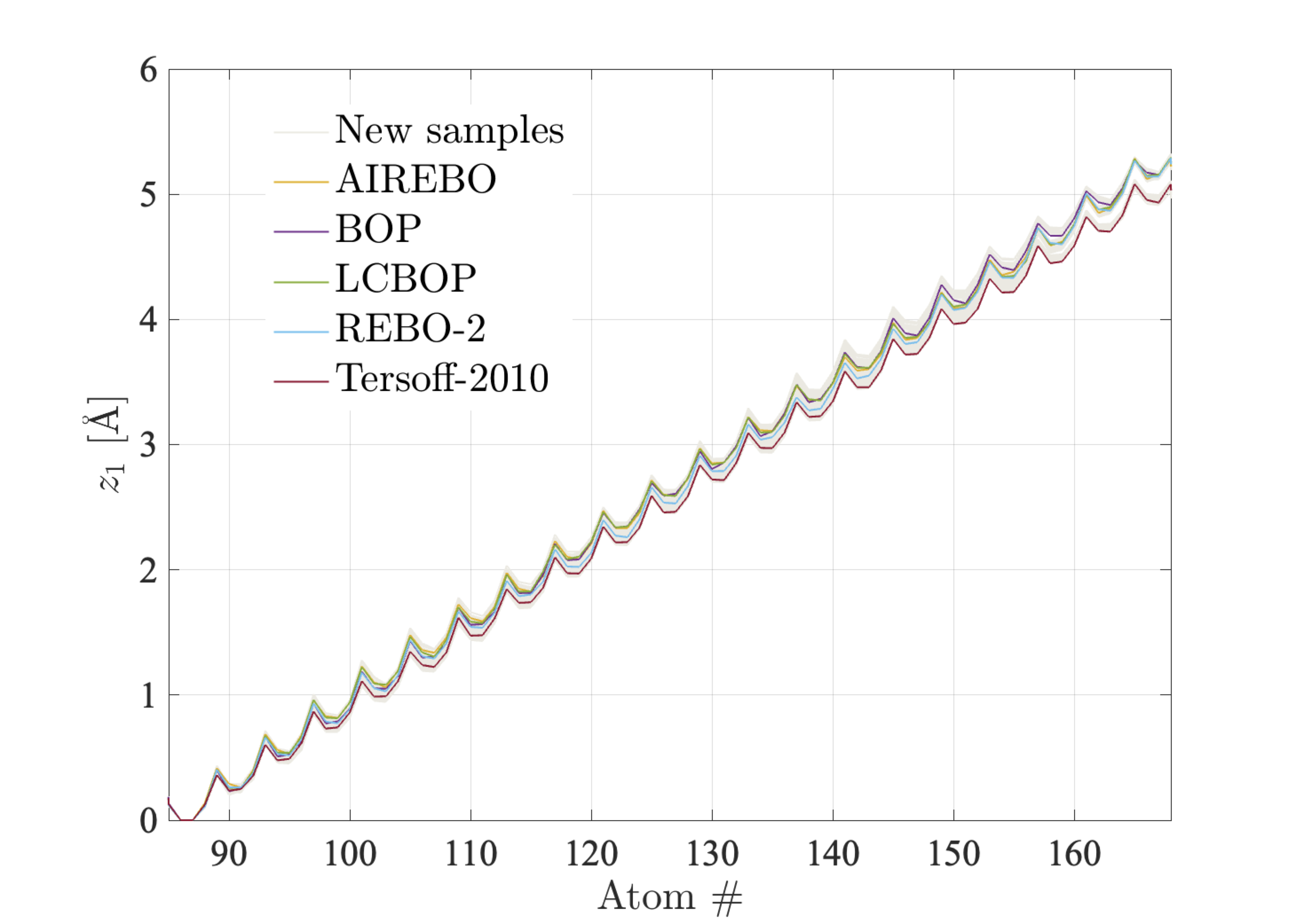}}\;
    	\subfloat[Strain energy at 0.11 strain]{\includegraphics[width = 0.45\textwidth]{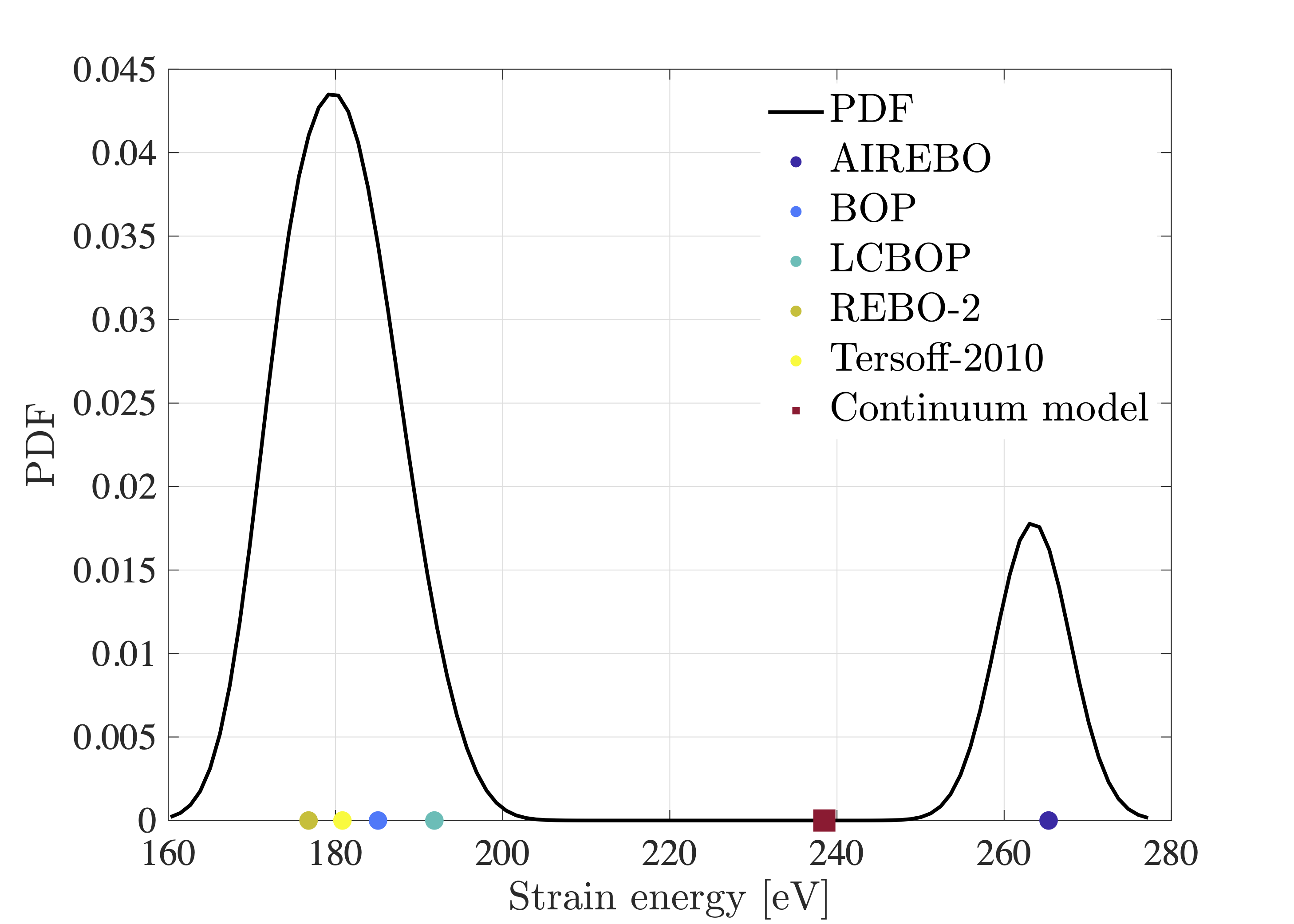}}\;
	\caption{Fine-scale and coarse-scale scale predictions for tensile test in the zigzag direction, obtained for $c = 1$ (stochastic Riemannian convex combination) and the concentration parameters given in Eq.~\eqref{eq:alpha-zz}. The interatomic potential in physical space is chosen using sample-based selection.}
    	\label{fig:fine-coarse-scales-c1}
\end{figure}

Fine-scale and coarse-scale scale stochastic predictions for the tensile test in the zigzag direction are shown in Fig.~\ref{fig:fine-coarse-scales-c1}, using Riemannian convex combinations ($c = 1$) and 200 samples.
It is seen that while the sampled trajectories are evenly distribution within the region defined by the potential candidates (Fig.~\ref{fig:fine-coarse-scales-c1}, left subfigure), the distribution of the coarse-scale properties is limited to a small region around the original full-order-model results (Fig.~\ref{fig:fine-coarse-scales-c1}, right subfigure). In order to increase the range of coarse-scale fluctuations (if required based on the application), scaling of the fluctuations can be performed in the tangent space; see Section \ref{subsubsec:scaling-tangent-space}. A simple and natural way to identify the additional parameter $c$ is to impose that the range of observed values for a given coarse-scale quantity of interest is included in the confidence region predicted by the stochastic model. Other strategies to solve statistical inverse problems can also be deployed, depending on the availability and nature of coarse-scale data.
 
\begin{figure}[htbp]
	\centering
	\subfloat[Zigzag direction ($c=8$): Displacement $z_1$]{\includegraphics[trim = 14cm 13cm 14cm 17cm, clip, width = 0.45\textwidth]{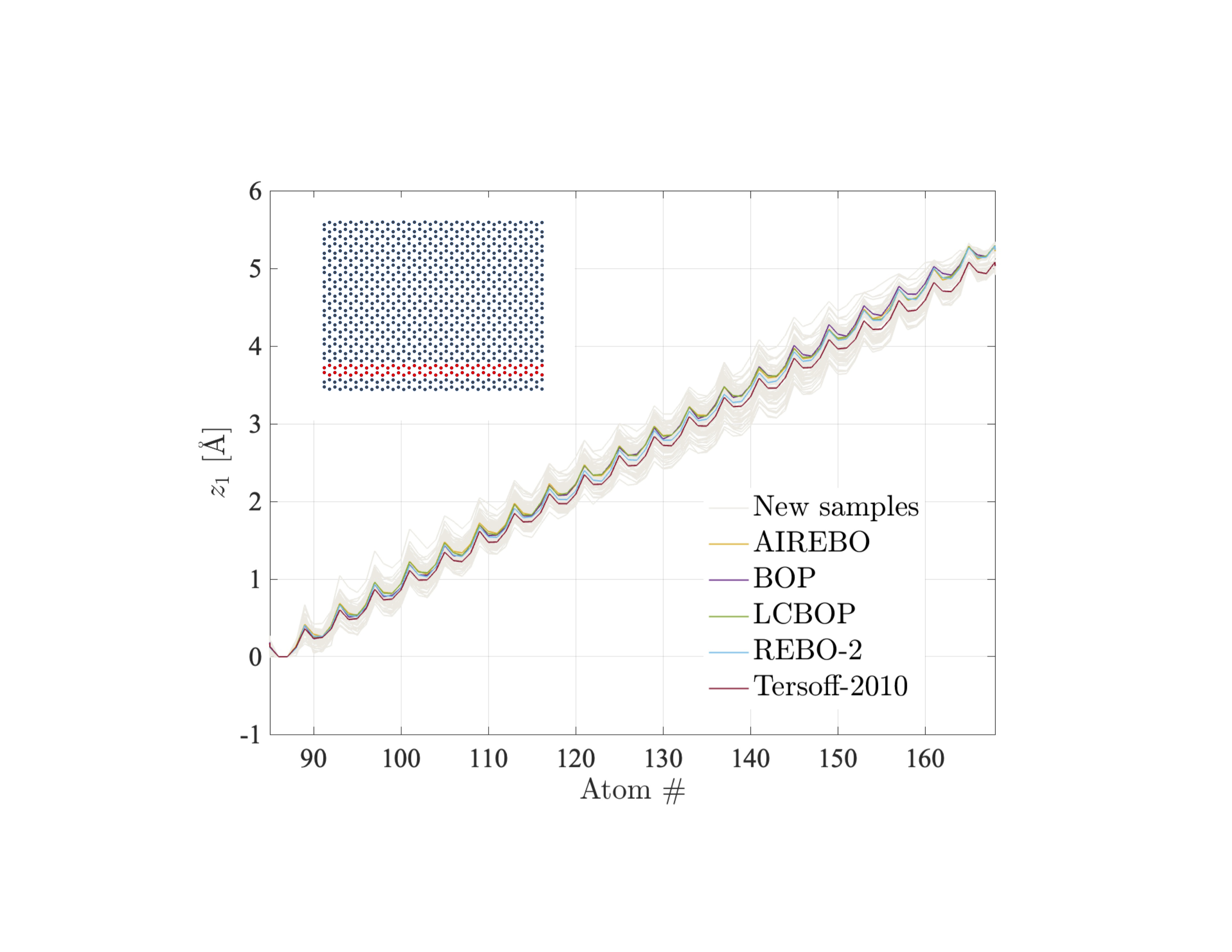}}\;
    	\subfloat[Zigzag direction ($c=8$): Displacement $z_2$]{\includegraphics[width = 0.45\textwidth]{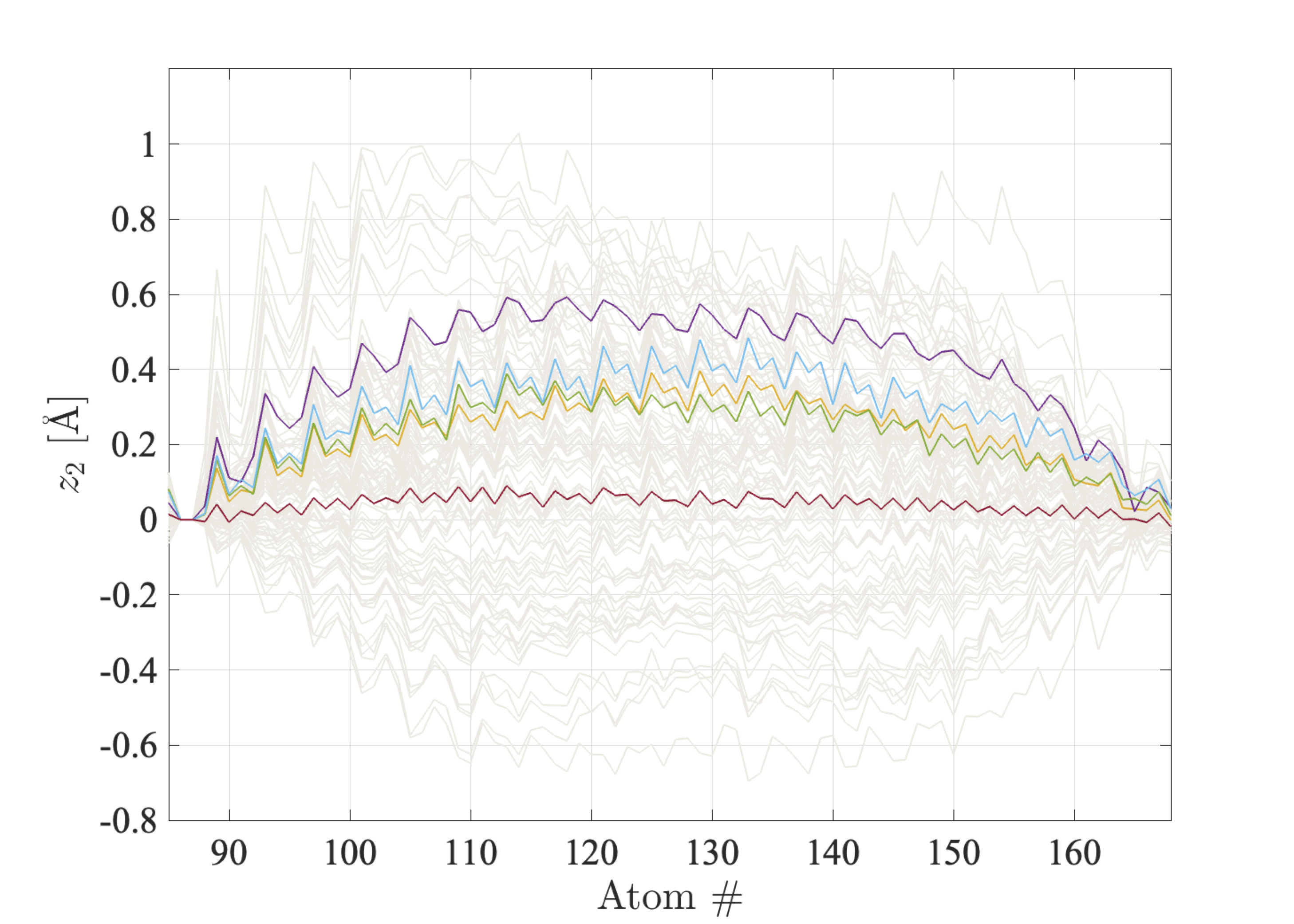}}\;\\
    	\subfloat[Armchair direction ($c=4$): Displacement $z_1$]{\includegraphics[width = 0.45\textwidth]{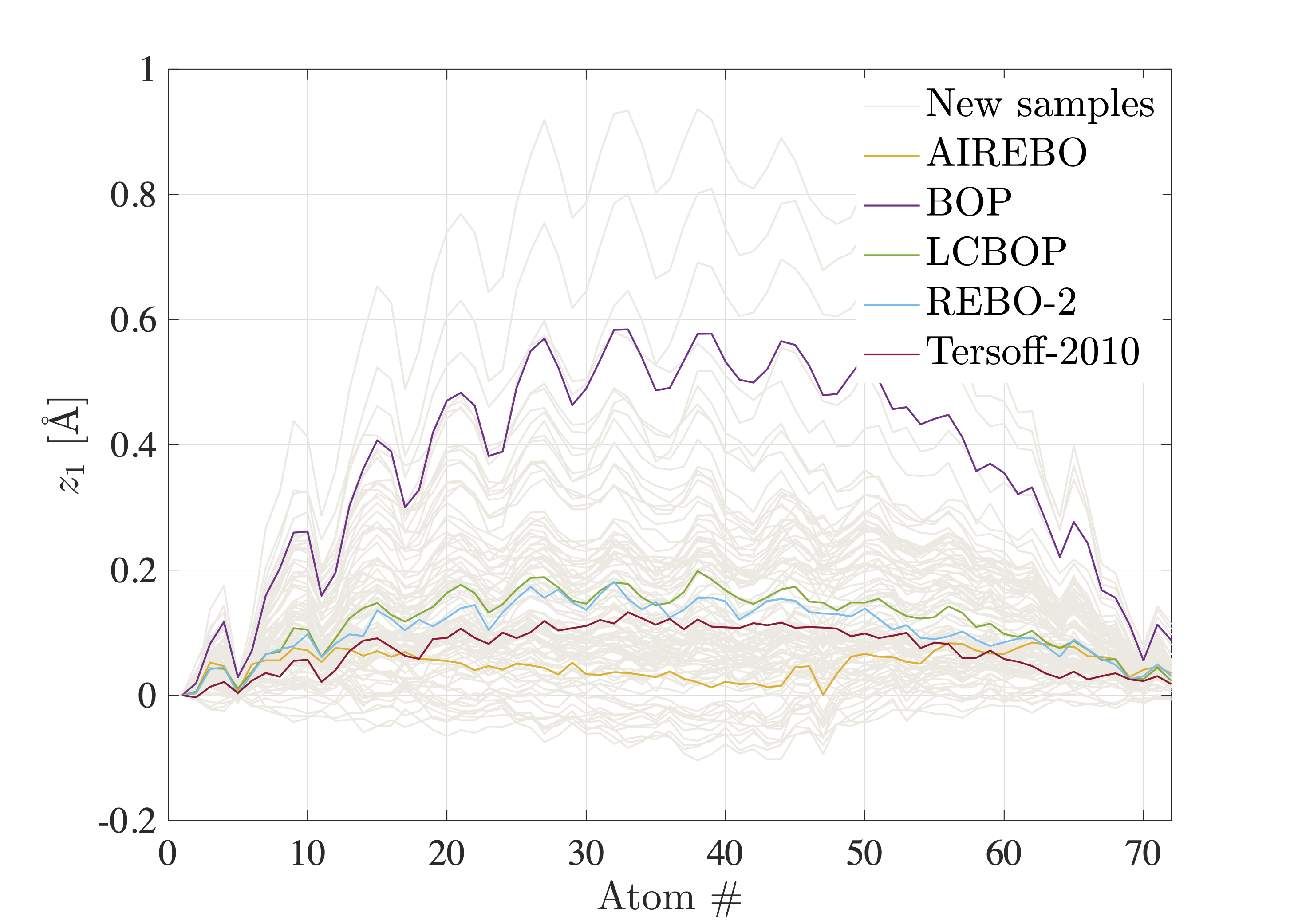}}\;
    	\subfloat[Armchair direction ($c=4$): Displacement $z_2$]{\includegraphics[trim = 14cm 13cm 14cm 17cm, clip, width = 0.45\textwidth]{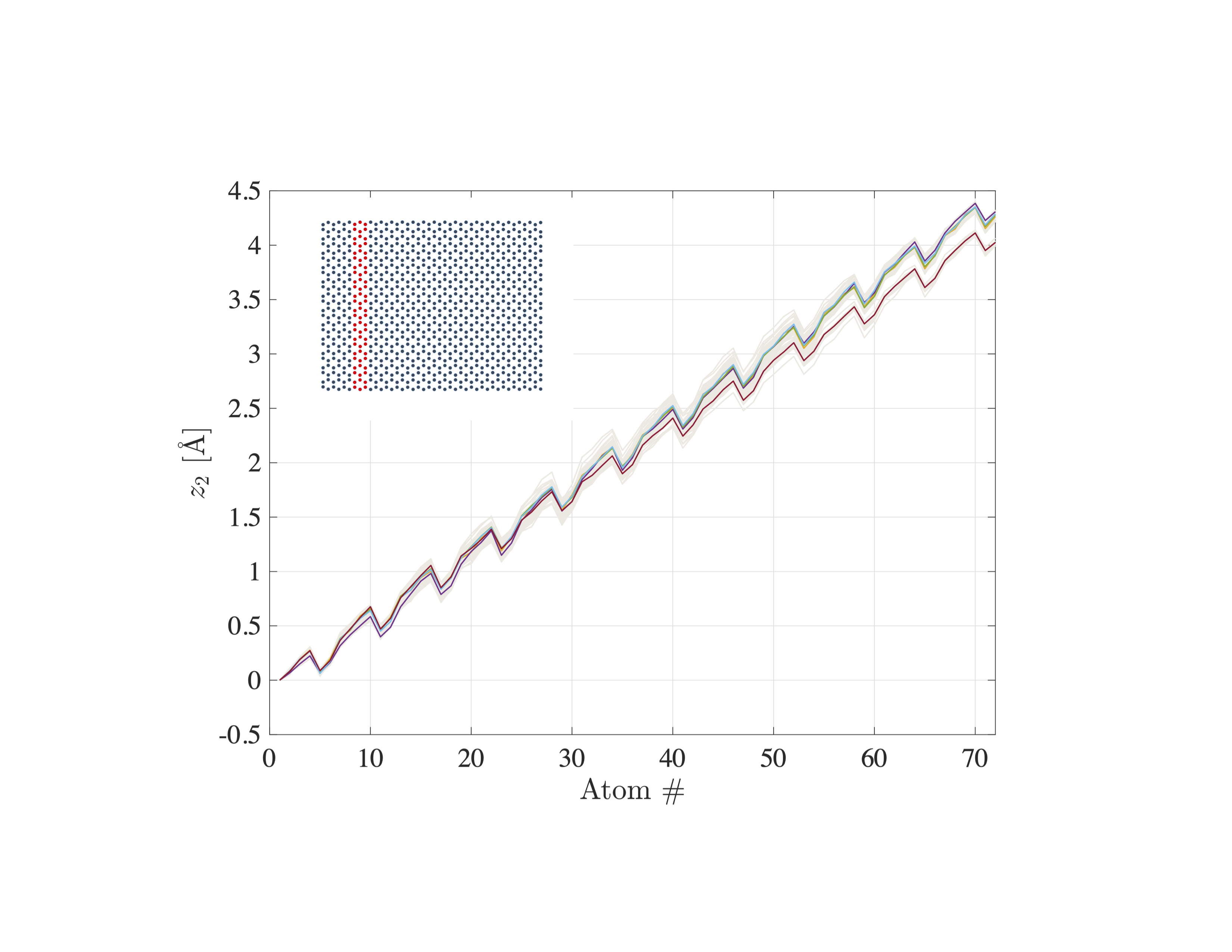}}\;
	\caption{Atom displacements for tension in the zigzag direction (first row, with scale factor $c=8$) and armchair direction (second row, with scale factor $c=4$). Selected atoms are displayed in red in the top-left and bottom-right subfigures. Grey curves correspond to samples obtained with the stochastic reduced-order MD simulations, while the colored curves are associated with full-order MD simulations.}
     	\label{fig:traj_zigzag_armchair}
\end{figure}

A total number of 200 reduced-order basis samples are generated on $\mathbb{S}_{3024,10}$ for both the zigzag and armchair directions. The scaling factor is set to $c=8$ for the zigzag direction, and to $c=4$ for the armchair direction. Fine-scale results, in the form of displacements for a few selected atoms, are first shown in Fig.~\ref{fig:traj_zigzag_armchair} (using sample-based selection for the potential in physical space). 
Displacement along $\bs{e}^1$ and $\bs{e}^2$ are collected at simulation time $ t = 10$ [ps]. As expected, it is observed that the range of displacements becomes much larger as $c$ increases, hence highlighting the sensitivity to this parameter.

\begin{figure}[htbp]
	\centering
	\subfloat[Tensile test in zigzag direction]{\includegraphics[width = 0.45\textwidth]{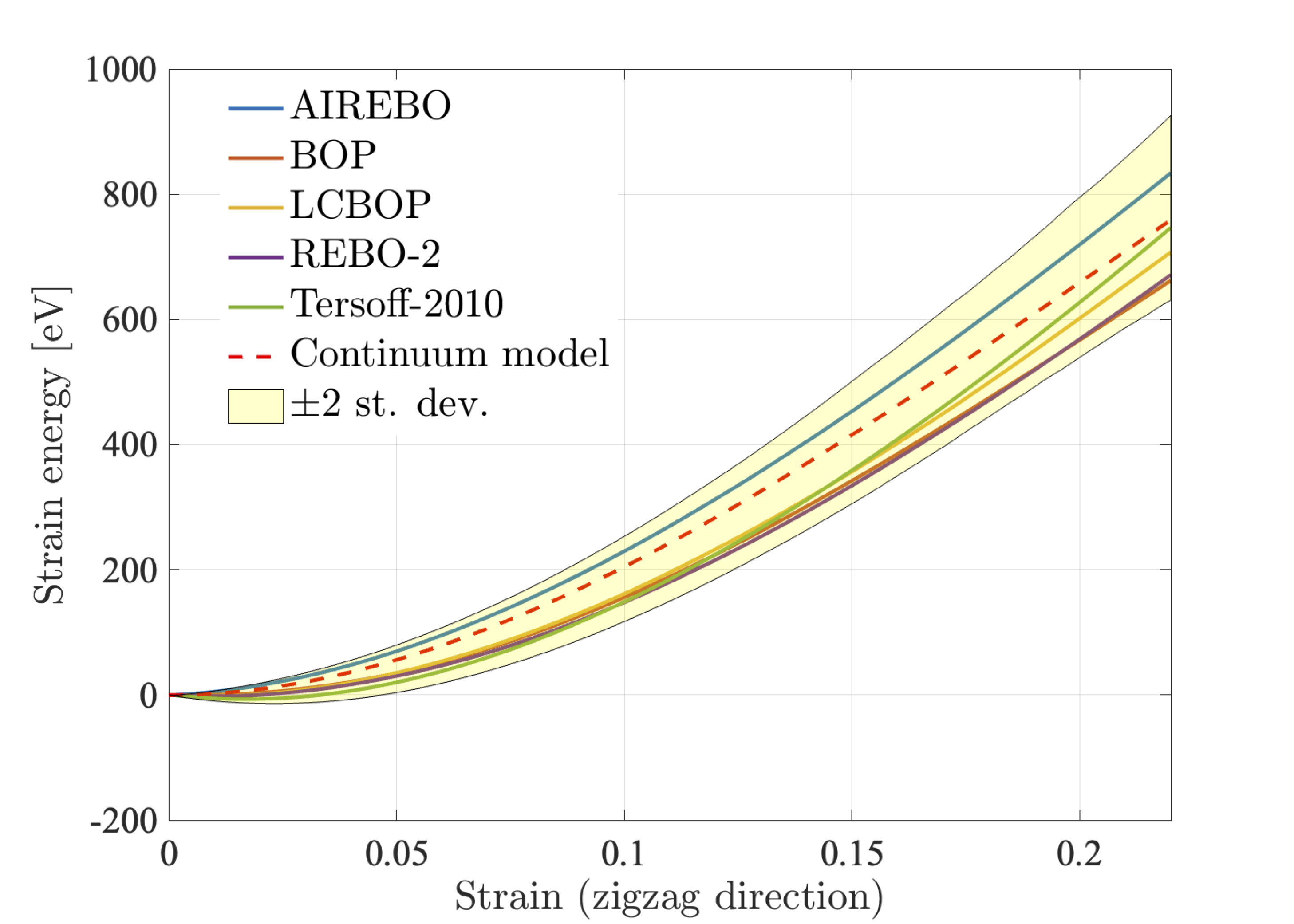}}\;
    	\subfloat[Tensile test in armchair direction]{\includegraphics[width = 0.45\textwidth]{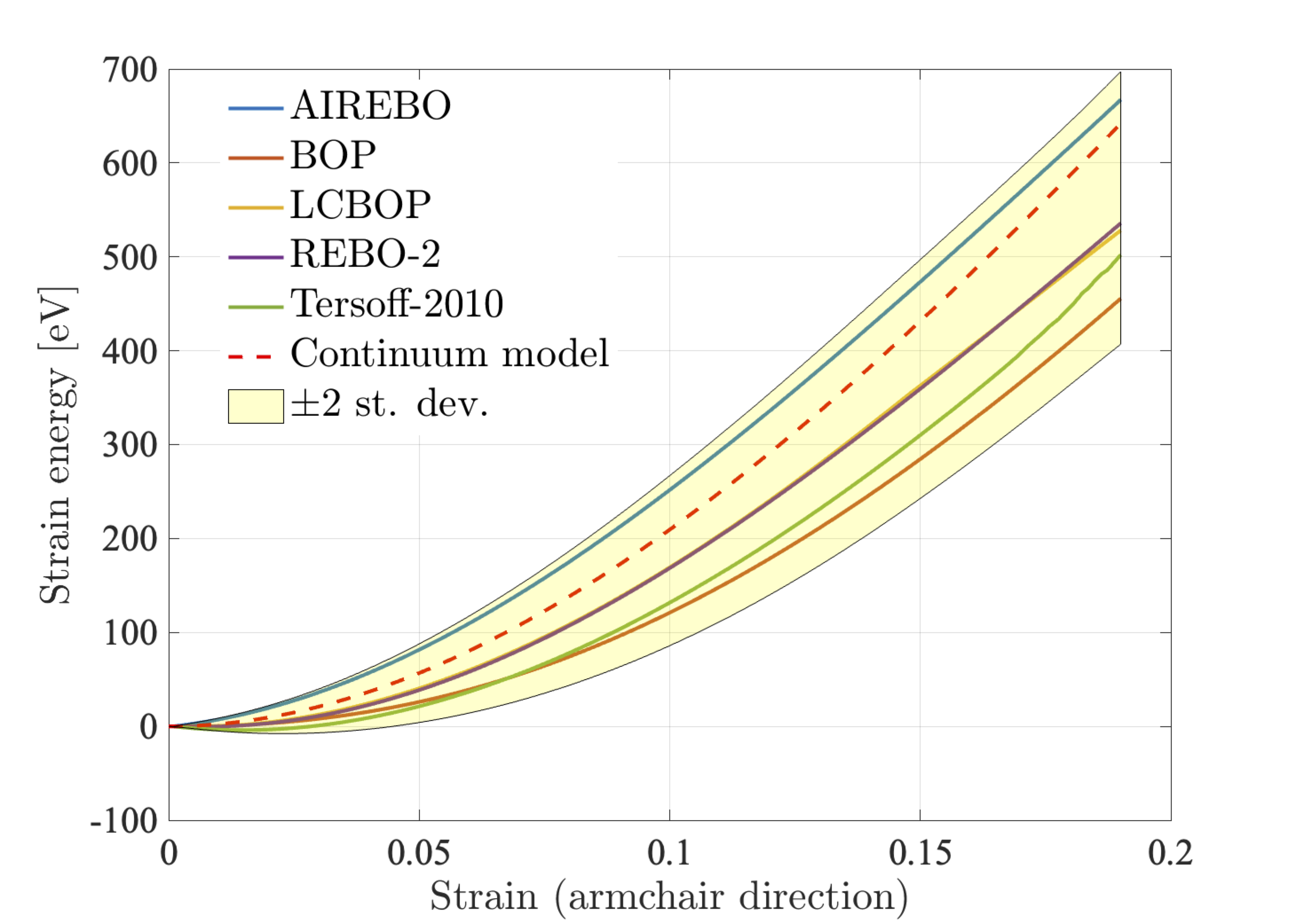}}\;
	\caption{Confidence interval (defined by plus-minus two standard deviations) for the coarse-scale strain energy, together with the strain energy obtained with the full-order MD simulations and the continuum model.}
     	\label{fig:std-tensile-test}
\end{figure}

\begin{figure}[htbp]
	\centering
	\subfloat[Zigzag direction, 0.11 strain]{\includegraphics[width = 0.45\textwidth]{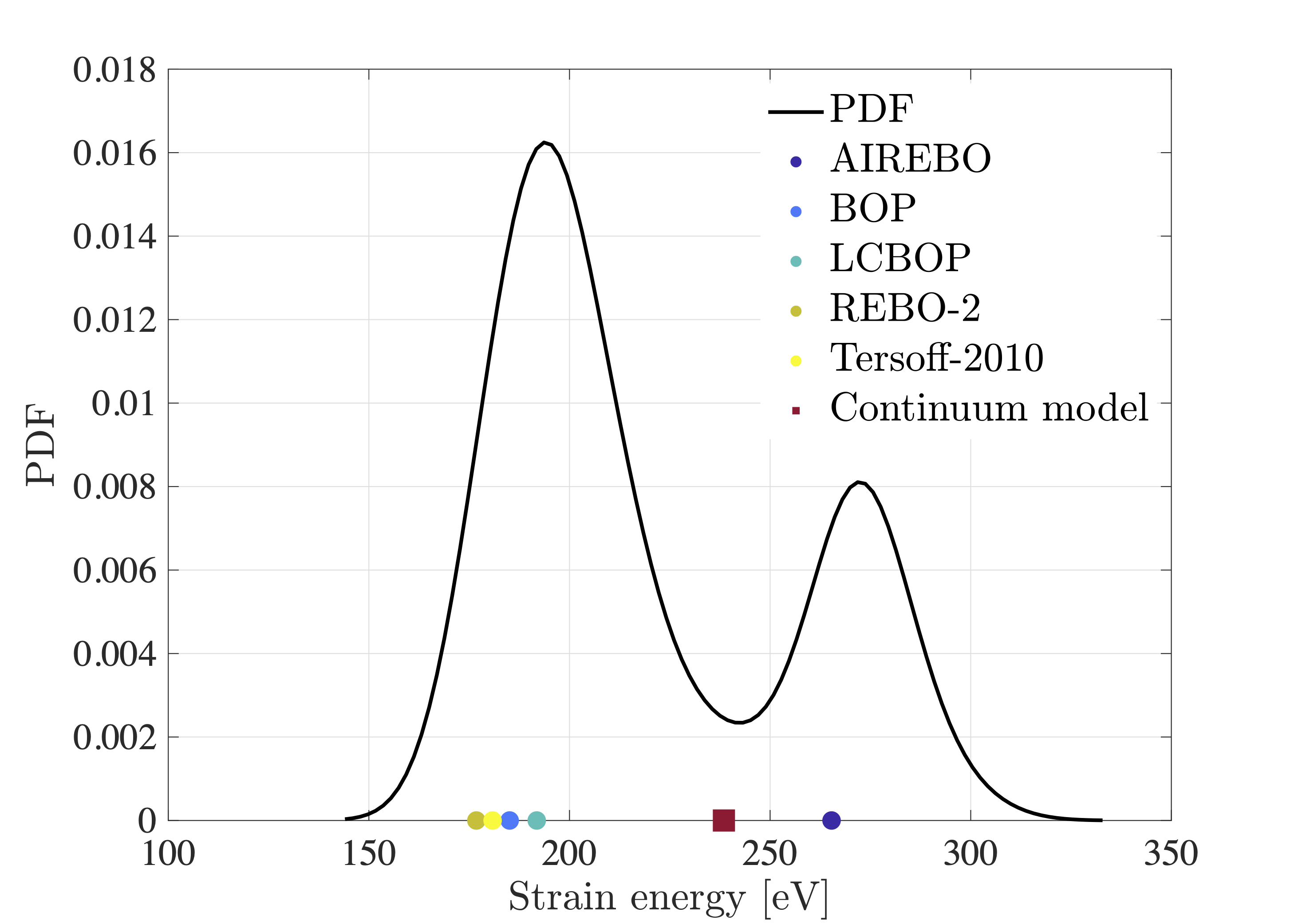}}\;
    	\subfloat[Armchair direction, 0.10 strain]{\includegraphics[width = 0.45\textwidth]{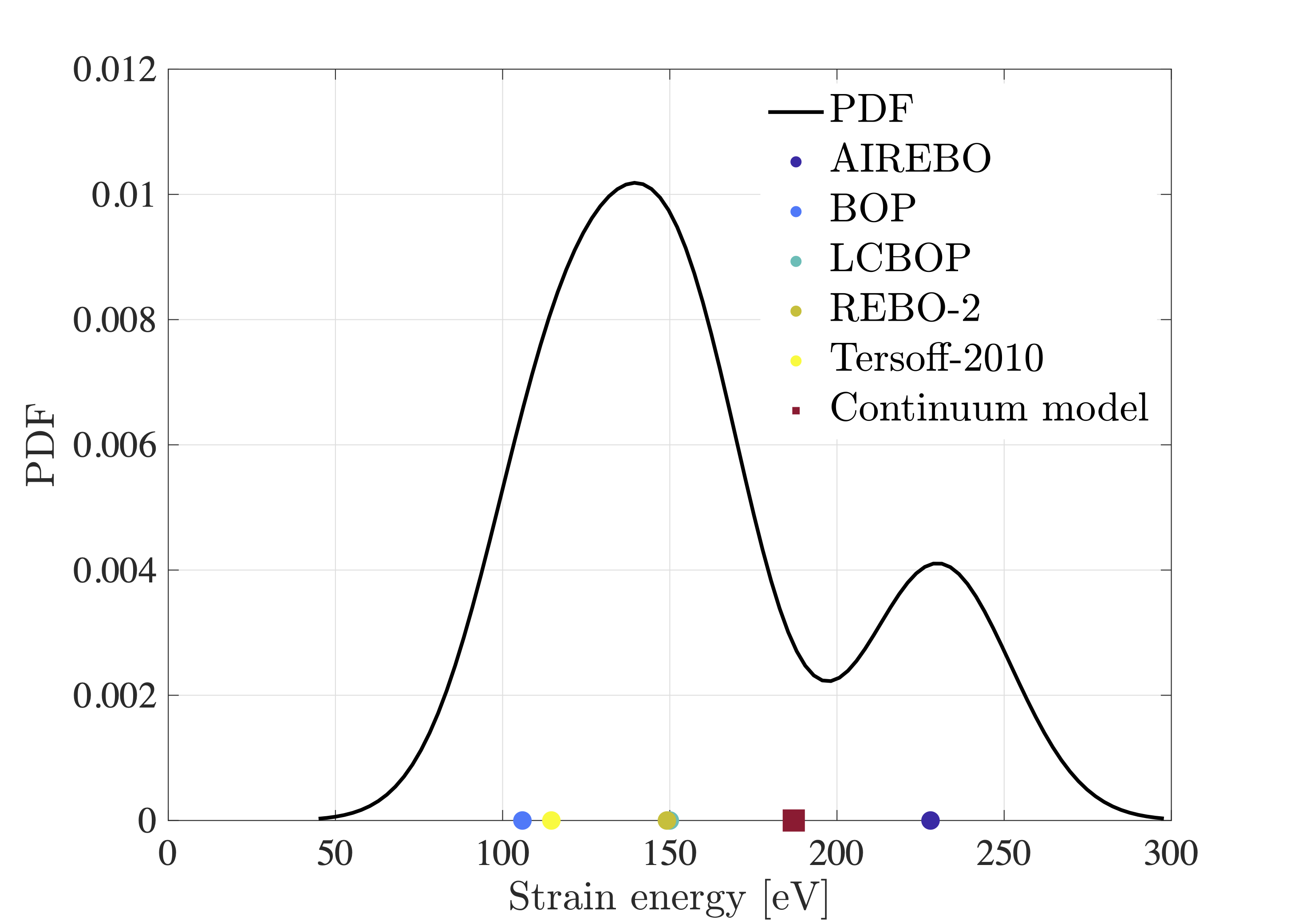}}\;
	\caption{Estimated probability density function for the strain energy in zigzag direction (left) and armchair direction (right). Values obtained with full-order MD simulations (with all potential candidates) and the continuum-mechanics-based model are also reported.}
     	\label{fig:pdf-tensile-test}
\end{figure}

The impact of model uncertainties can also be quantified on the distribution of the coarse-scale strain energy. Fig.~\ref{fig:std-tensile-test} shows the confidence interval of the strain energy ($\pm 2$ standard deviations) with regard to the engineering strain in the two stretching directions. 
The estimated probability density function for the strain energy is shown in Fig.~\ref{fig:pdf-tensile-test} for the two different stretching directions. In these figures, the strain is chosen as 0.11 for the zigzag direction, and as 0.10 for the armchair direction.
These results demonstrate the capability of the proposed formulation to capture model-form uncertainties at fine scale and to propagate them on a coarse-scale quantity of interest. Such uncertainties can be properly encoded into the ROB samples such that the distribution of the quantity of interest can be analyzed in a multi-scale pipeline.

\section{Conclusion}
\label{sec:conclusion}
A Riemannian stochastic representation of model-form uncertainties in molecular dynamics was proposed. The approach relies on a stochastic reduced-order model, defined through a randomized projection basis on a subset of the Stiefel manifold. It was shown that the use of Riemannian projection and retraction operators allows linear constraints, relevant to Dirichlet boundary conditions for instance, to be preserved. This fundamental property enables the consideration of convex Riemannian combinations on the tangent space. The proposed formulation offers several advantages, including a simple and interpretable low-dimensional parameterization, the ability to constraint the Fr\'{e}chet mean solving a quadratic programming problem, and ease of implementation and propagation through stochastic collocation methods. The relevance of the proposed modeling framework was finally demonstrated on various applications, including sampling on the unit sphere and multiscale simulations on graphene-based systems.

\section*{Acknowledgments}
The work of the J.G. was supported by the National Science Foundation, Division of Civil, Mechanical and Manufacturing Innovation, under award CMMI-1942928.

\appendix
\section{Computation of the Fr\'echet Mean}\label{app:frechet}
The algorithm to compute the Fr{\'e}chet mean on a Stiefel manifold, denoted by $\mathbb{M}$, based on a set of samples is given in Alg.~\ref{alg:gdm} (see \cite{frechet_mean_alg}).
\begin{algorithm}[htbp]
\caption{Calculation of Fr{\'e}chet Mean Based on a Gradient Descent Method}
\begin{algorithmic}[1]
\Require 
    set of samples $\{[Y^{(i)}] \in \mathbb{M}\}_{i=1}^{q}$, stepsize $t$,
    convergence threshold $\epsilon$, algorithms to compute the projection and retraction operators 
    \State Choose a initial guess, denoted by $[Y_{(0)}] \in \mathbb{M}$, for the Fr{\'e}chet mean, and set Err = Inf and $k=0$
    \While{Err $> \epsilon$}
     \State Calculate: $\nabla h([Y_{(k)}]) = -\sum_{i = 1}^{q} P_{[Y_{(k)}]}([Y^{(i)}])$
     \State Calculate the error: $\textnormal{Err} = \| h([Y_{(k)}])\|_2$
     \State Update the Fr{\'e}chet mean: $[Y_{(k+1)}] = R_{[Y_{(k)}]}(-t^k \nabla h([Y_{(k)}]))$
     \State $k+1 \gets k$
    \EndWhile
\Ensure Fr{\'e}chet mean: $[\underline{Y}] = [Y_{(k)}]$
\end{algorithmic}
\label{alg:gdm}
\end{algorithm}

\bibliography{mybibfile}
\end{document}